\RequirePackage{fix-cm}
\documentclass[smallextended]{svjour3}       
\smartqed  
\usepackage{graphicx}
\usepackage{comment} 
\usepackage{natbib}
\setcitestyle{aysep={}}
\usepackage{xcolor}
\usepackage{lscape}
\usepackage{cancel} 
\usepackage{tabularray}
\usepackage{amssymb}
\usepackage{amsmath}
\usepackage[colorlinks=true,linkcolor=blue,citecolor=blue,urlcolor=blue]{hyperref}
\newcommand{\msun}{\ensuremath{\mathrm{M}_\odot}}

\newcommand{\msub}{\ensuremath{{\rm sub}{\mathrm{M}_{\rm Ch}}}}{}

\journalname{Astron Astrophys Rev }
\begin{document}

\title{Type Ia supernova progenitors: a contemporary view of a long-standing puzzle}




\author{Ashley Jade Ruiter         \and
        Ivo Rolf Seitenzahl 
}

\authorrunning{A.J. Ruiter and I.R. Seitenzahl} 

\institute{A.J. Ruiter\at
              School of Science \\
              University of New South Wales \\
              Australian Defence Force Academy \\
              Canberra, ACT, Australia\\
              and\\ 
              OzGrav: The Australian Research Council \\
              Centre of Excellence for Gravitational \\
              Wave Discovery, Hawthorn VIC 3122, Australia\\
              and\\
              ASTRO-3D: The Australian Research Council \\
              Centre of Excellence for All-Sky Astrophysics \\
              in 3 Dimensions \\
              and\\
              Heidelberger Institut f\"ur Theoretische Studien\\
              Schloss-Wolfsbrunnenweg 35\\
              Heidelberg, 69118, Germany\\     
    \email{ashley.ruiter@gmail.com} \\
           \and
           I.R. Seitenzahl\at
              Heidelberger Institut f\"ur Theoretische Studien\\
              Schloss-Wolfsbrunnenweg 35\\
              Heidelberg, 69118, Germany\\ 
}
\date{Received: April 2024 / Accepted: 2024}

\maketitle

\begin{abstract}
Type Ia supernovae (SNe Ia) are runaway thermonuclear explosions in white dwarfs that result in the disruption of the white dwarf star, and possibly its nearby stellar companion. SNe Ia occur over an immense range of stellar population age and host galaxy environments, and play a critical role in the nucleosynthesis of intermediate-mass and iron-group elements, primarily the production of nickel, iron, cobalt, chromium, and manganese. Though the nature of their progenitors is still not well-understood, SNe Ia are unique among stellar explosions in that the majority of them exhibit a systematic lightcurve relation: more luminous supernovae dim more slowly over time than less luminous supernovae in optical light (intrinsically brighter SNe~Ia have broader lightcurves). This feature, unique to SNe~Ia, is rather remarkable and allows their peak luminosities to be determined with fairly high accuracy out to cosmological distances via measurement of their lightcurve decline. Further, studying SNe~Ia gives us important insights into binary star evolution physics, since it is widely agreed that the progenitors of SNe~Ia are binary (possibly multiple) star systems. 
In this review, we give a current update on the different proposed Type Ia supernova progenitors, including descriptions of possible binary star configurations, and their explosion mechanisms, from a theoretical perspective. We additionally give a brief overview of the historical (focusing on the more recent) observational work that has helped the astronomical community to understand the nature of the most important distance indicators in cosmology.  
\end{abstract}

\setcounter{tocdepth}{3} 
\tableofcontents

\section{Introduction}
\label{intro}

\subsection{Importance to astrophysics}

Type Ia (or thermonuclear) supernovae are extremely energetic stellar transients that quickly evolve to outshine 100,000,000 stars before rapidly fading in brightness. Though now understood to be a heterogeneous population, it is agreed that a Type Ia supernova (SN Ia) stems from a runaway nuclear reaction in a degenerate (and assumed to be rich in carbon and oxygen) white dwarf star when that star has achieved physical conditions that enable explosive nuclear burning to take place  \citep{thielemann1986a,khokhlov1991a}. 

Type Ia supernovae occur only under special circumstances involving interacting stars with initial masses below ${\sim} 10$ \msun, and under completely different physical conditions than core-collapse supernovae. Core-collapse SNe involve the collapse of a stellar core to a compact object, which can only occur in stars with a massive stellar core (e.g. stars whose total mass on the Zero-Age Main Sequence was on the order of {$\sim$} $8-10$ \msun\ or heavier, assuming they do not produce [pulsational] pair-instability supernovae \citep{rahman2022a}).  
The explosion mechanism in core-collapse supernovae, of which there are a growing number of observationally-classified sub-types \citep{galyam2017a} is an ongoing debate \citep{fryer2012a,mueller2017a}. Type Ia supernova explosions on the other hand are fueled by a nuclear energy source -- on the order of $10^{51}$ erg with typical bolometric luminosities reaching $10^{43}$ erg/s \citep{maguire2017a}. SNe~Ia involve the explosion of a carbon-oxygen\footnote{but see also \citet{kirsebom2019a}.} white dwarf that has obtained sufficiently high density to initiate sub-sonic and/or super-sonic burning, likely brought on via matter accretion from a close stellar companion. We discuss in more detail the various explosion mechanisms for Chandrasekhar mass (approximately 1.4 \msun) and sub-Chandrasekhar mass explosion models in Sects.~\ref{sec:Chandra} and \ref{sec:subChandra}, respectively, after first giving a more holistic overview of the field. For a modern overview of stellar evolution, we encourage the reader to look up `Understanding Stellar Evolution' by \citet{lamers2017a}.

In particular, SNe~Ia are famous for the characteristic that it is possible to `standardize' a large number of their lightcurves (i.e. stretch or compress them to fit a template), which then allows us to use them as our most important distance indicators on the astrophysical distance ladder \citep{rust1974a,pskovskii1977a,phillips1993a,hamuy1996c,phillips2017a}. 
While observed properties of some massive-star (core-collapse) supernovae such as Type II-P \citep{hamuy2002c,poznanski2009a,maguire2010a} and superluminous supernovae \citep[SLSNe,][]{inserra2014a} have allowed some core-collapse supernovae to be regarded as useful cosmological distance indicators, the typically larger luminosities of SNe~Ia, as well as their birthrates \citep{li2011a,prajs2017a}, have rendered SNe~Ia as fundamental cosmological tools over the decades. 
The technique of SN~Ia standardization has grown in complexity over the decades, and requires lightcurve fitting to models in a variety band passes and spectra. 
For an excellent review on the `rise' of SNe~Ia as trusted standardizable candles in cosmology, we refer the reader to \citet{kirshner2010a}.

It is observational data of SNe~Ia that enabled the Supernova Cosmology Project and High-z Supernova Search teams to determine that the rate of expansion of the Universe is actually accelerating (Nobel Prize in Physics, 2011). This fundamental discovery firmly rejuvenated the importance of a cosmological constant -- the seedling idea having been postulated by \citet{einstein2017a} as a universal constant permeating the Universe (then denoted by $\lambda$ in the Einstein Field Equations), though its effects are negligible over small ($\sim$~galactic) scales. Since this profound discovery, there has been significant research dedicated toward understanding the concept of `dark energy' \citep[for a review, see][]{frieman2008a}.

In addition to their importance in the advancement of cosmological studies, and thus our understanding of our Universe's previous and future evolution, SNe~Ia are also the most important source of iron-group elements (e.g. V, Cr, Mn, Co, Fe, Ni) in the Universe. Though core-collapse supernovae (types II, Ic, Ib and their sub-types) exceed SNe~Ia by number among young stellar populations, per event a typical SN~Ia produces around 0.5\,\msun\ of iron \citep[e.g.][]{stritzinger2006a,mazzali2007a,scalzo2014a,bora2022a}, and therefore about 12 times more iron than a typical Type II core-collapse SN \citep{rodriguez2021a}. Stripped envelope core-collapse SNe (e.g. SNe~IIb, SNe~Ib/c) produce more iron than their Type II SN cousins, but event-by-event, the mean over-production of iron by SNe~Ia relative to core-collapse SNe that also include stripped envelope events is still a factor of about 9 \citep{rodriguez2022a}. This results in about just over half of the iron at the current epoch originating from SNe~Ia \citep[see e.g. Fig.~3 of][]{maoz2017a}.
Iron is the most critical element for spectroscopic measurements of metallicity in stellar populations because iron is used as a tracker of chemical enrichment in studies of galaxy evolution, especially for our own Milky Way \citep{sharma2022a}. In addition to elements near the iron-peak (which also includes notable production of Cu and Zn), SNe~Ia also produce a significant amount of intermediate-mass elements, especially the even atomic number elements Si, S, Ar, Ca, and the transitional element Ti, which sits at the border between intermediate-mass elements (IMEs) and iron-group elements (IGEs). 

`Catching' a supernova in the act of exploding, or even capturing data in those first hours or days after explosion where crucial physics can be constrained, is not easy. It can take days for the new SN to be recognized by the community as an astrophysical transient. This was especially true prior to the age of synoptic sky surveys \citep[e.g. SkyMapper,][]{scalzo2017a}. Amateur astronomers have also had a substantial contribution toward the discovery of new SNe \citep{galyam2013a}.

Around 15-20 days after explosion, the SN~Ia will reach its maximum value in bolometric luminosity, and it is around this time when the SN~Ia be can classified further (see below). The SN~Ia optical lightcurve is powered by the radioactive decay of $^{56}$Ni (see Sect.~\ref{sec:specandLC})\footnote{$^{56}$Ni (with a half-life of 6.075 days) decays to the radioactive isotope $^{56}$Co (with a half-life of 77.236 days), which then eventually decays to the stable isotope $^{56}$Fe.}. The time evolution of Type Ia supernova lightcurves around maximum light is shaped by the interplay of the ongoing yet decreasing energy input (due to the exponentially decaying activity from the radioactive decay law) and the diffusion/propagation of the photons through the SN ejecta towards the stellar photosphere. It is this delaying effect of the diffusion through the initially opaque ejecta that leads to a maximum in the lightcurve (as opposed to a purely exponentially decaying function) in the first place \citep[][see Sect.~\ref{sec:ejecta-masses}]{arnett1982a}.

In simple terms, more radioactive $^{56}$Ni in the centre of the supernova tends to yield more luminous events, and more mass surrounding it leads to greater diffusion times and broader lightcurves with a reduced luminosity at maximum. Whether a SN~Ia's lightcurve evolves more quickly than average (having a narrow lightcurve) or more slowly than average (broader lightcurve) already allows us to speculate that these two types of events are intrinsically different -- especially when accounting for the fact that the faster events tend to be found among regions with less active star formation. 

Taking together SN~Ia lightcurve behaviour in combination with a notable variation in spectral properties near maximum light, and occasionally other physical properties too, it has become possible to delineate most SNe~Ia into various ``sub-classes'' \citep{taubenberger2017a}: the sub-luminous, fast-declining SNe~Ia -- the most well-known of which are the 1991bg-likes \citep{leibundgut1993a}; the `super-luminous' SN~Ia such as the broad-lightcurve 1991T-likes \citep{phillips1992a} or the so-called `Super-Chandrasekhar' events \citep[][see also Sects.~\ref{sec:ejecta-masses} and \ref{sec:prog-MCh}]{silverman2011a}; Ca-rich (or `gap transient') faint events that are more often associated with old quiescent stellar populations, sometimes relatively far from their host galaxies \citep{kasliwal2012a,de2020a}; the `Type~Iax' supernovae (see Sect.~\ref{sec:companion_search}) comprises a large, rather diverse group, of which the most infamous is the 2002cx-like sub-population \citep{li2003a} -- a type of `peculiar' supernova. Finally, the ``normal'' SNe~Ia \citep{branch1993a} refer to those Type Ia supernovae whose lightcurves are well-behaved in the sense that it is possible to fit them to a template that is useful for cosmological studies. 

This is by no means an exhaustive list of sub-classes. We also note that these sub-class labels are tied directly to observational features, namely peak luminosity and lightcurve `speed', with no initial link to theoretical predictions, though over the years certain explosion scenarios have emerged as arguably more favourable over others for certain sub-classes \citep[e.g. a `failed' explosion of a Chandrasekhar mass white dwarf for the case of SN~Iax event SN 2020udy,][see also Sect.~\ref{sec:faileddef}]{maguire2023a}. We refer the reader to Sect.~2 of \citet{galyam2017a} for a discussion on various SN~Ia sub-classes with reference to specific spectral properties \citep[see also][for a brief overview of the different sub-classes]{ruiter2020a}.

There is widespread agreement that different progenitor scenarios result in the synthesis of chemical elements in different proportions \citep[e.g.][]{roepke2012a}. Therefore, not knowing the nature of SN~Ia progenitors, or what fraction the different sub-classes of progenitors contribute among different stellar environments, makes for quite a challenging puzzle for galactic chemical evolution studies, which rely on yields from the explosive nucleosynthesis in SNe~Ia as critical simulation input. Append the fact that different SN~Ia progenitors have a variety of feedback-timescales, or `delay time distributions'\footnote{The Delay Time Distribution (DTD) is the distribution of times over which SNe~Ia explode following a (hypothetical) burst of star formation assuming all stars were born at time $t=0$.} (see Sect.~\ref{sec:Rates}), and we have a truly challenging problem if we are to understand the process of chemical enrichment in galaxies in detail. 

Even though SNe~Ia have famously served as extremely useful cosmological tools, this review focuses on the role SNe~Ia play in astrophysics from a perspective of understanding stellar and binary evolution of low- and intermediate-mass stars, while highlighting the critical role that SNe~Ia play in nucleosynthesis, and thus chemical evolution (see Sect.~\ref{sec:chem}).

In this section, we give a brief overview of what we know about SNe~Ia in terms of their progenitor configuration, with an emphasis on historically significant pioneering theoretical works. In Sects.~\ref{sec:DirObs} and \ref{sec:InDirObs}, 
we provide an overview of observational works that have made substantial contributions to understanding the nature of SN~Ia progenitors through direct and indirect methods, respectively. 
In Sects.~\ref{sec:Chandra} and \ref{sec:subChandra}, we shift our focus to theoretical work pertaining to the study of Chandrasekhar mass and sub-Chandrasekhar mass explosion models, respectively, that were published this century \citep[for a review on earlier works, see][]{hillebrandt2000a}. 

\label{sec:1}

\subsection{Progenitor scenario: star types} 
\label{sec:progenitors}

Historically, SNe~Ia were seen as residing in one of two silos: single degenerate (SD) or double degenerate (DD). Both categories involve a white dwarf that ultimately makes up the bulk of the exploding mass that had reached ignition conditions (i.e. critical density/temperature), but the star that is donating this crucial extra mass (the `donor') could be either a regular star still undergoing nuclear burning in its core or in a shell(s), or it could be a degenerate (e.g. another white dwarf) star.  

Such a catagorization (SD or DD), though elegant in some ways, ignores the implied explosion mechanism -- which is thought to be predominantly governed by the exploding white dwarf mass --  and thus the nucleosynthetic implications of the progenitor become washed out. This silo catagorization also leaves little room for `grey' areas in terms of how one defines degeneracy, as some donor stars might exist in the realm of `semi-degenerate' \citep[see][]{nelemans2001a,iben1991a}. 

Delineating the progenitor types by exploding white dwarf mass rather than donor star type gives a simpler, more physically-motivated line of reasoning in discussing SN~Ia origin, since it is widely believed that Chandrasekhar mass white dwarfs undergo a thermonuclear explosion through a different mechanism than white dwarfs that explode well below the Chandrasekhar mass limit \citep[][Sect.~4]{hillebrandt2013a}. It turns out that both `silos' of progenitors (SD and DD) can each harbour both Chandrasekhar mass and sub-Chandrasekhar mass SNe~Ia. Additionally, DD SNe Ia can occur at nearly any stellar age, or delay time, from ${\sim} 50$ Myr post-star formation, and the situation is similar for SD progenitors, especially when including progenitors with helium-burning donors  \citep{yungelson2000a,wang2009a,ruiter2011a,claeys2014a}. For a schematic picture of several progenitor evolutionary channels, we refer to reader to Fig.~2 from \citet{liu2023a}. 

\subsubsection{White dwarf with non-white dwarf: SD scenarios in brief} 
\label{sec:SD}

The canonical view of the SD scenario \citep{whelan1973a} consists of a white dwarf rich in carbon and oxygen that approaches the Chandrasekhar mass limit via stable\footnote{In this context, `stable' Roche-lobe overflow (RLOF) means mass is transferred to the accretor on either a nuclear or thermal timescale, but is not necessarily conservative.} mass transfer from a non-degenerate star (but see Sect.~\ref{sec:Chandra} regarding the `core-degenerate' scenario). The donor is usually assumed to be a main sequence star or a red giant, but could be another type (see below). Binary systems with a massive white dwarf accreting from a non-degenerate companion are indeed observed in nature, which has given some credence to this proposed scenario compared to others.\footnote{The dearth of observed white dwarf binaries having short orbital periods has been one (poor) argument against the plausibility of double degenerate binaries contributing to the SN~Ia population; see Sect.~\ref{sec:dd}.} 

For the case of a Chandrasekhar mass explosion that results in a Type Ia supernova, the favoured explosion scenario involves a sub-sonic deflagration followed by a super-sonic detonation that explodes the white dwarf star. In a nutshell, the central density of the carbon-oxygen-rich white dwarf becomes high enough for carbon-burning to begin. After some time on the order of 1000\, years \citep[e.g.][]{piro2008b} of quiescent, convective carbon-burning referred to as the `simmering phase' \citep{piro2008c}, explosive carbon-burning is ignited as a flame \citep[e.g.][]{garcia1995a, kuhlen2006a}, which, in most if not all cases, completely disrupts the white dwarf star. In some cases, a carbon detonation may not be successfully achieved in a Chandrasekhar mass white dwarf, resulting in a `failed detonation', often termed a `failed deflagration', since the sub-sonic deflagration did not successfully transform into a super-sonic detonation. These weaker explosions are currently the favoured mechanism in explaining some peculiar supernovae that exhibit SN~Ia-like features but with lower luminosities and lower ejecta velocities \citep{jha2017a} referred to as SN~Iax (see Sect.~\ref{sec:companion_search} for a bit more discussion on this sub-class). We refer to Sect.~\ref{sec:Chandra} for details on modelling approaches for Chandrasekhar mass explosions. In fact, the failed deflagration explosions are not only the favoured model to explain the fainter SN~Iax supernovae, but it is also suspected that the donating star in these cases was not a red giant or main sequence companion, but a star that had previously lost its hydrogen envelope through binary interactions but was still burning helium -- imagine an AGB star without its envelope. The helium-burning star channel in the context of Chandrasekhar mass SNe~Ia has been found to be a promising formation scenario to explain SNe~Iax events \citep{maguire2023a}, given their affinity for being found among young stellar environments, e.g. at short delay times \citep[][]{ruiter2009a,wang2009a,takoro2020a}.  
Single degenerate SNe~Ia could also occur in sub-Chandrasekhar\ mass white dwarfs through the `classical' double-detonation scenario in which a star that is rich in helium 
donates matter to a white dwarf through RLOF at relatively low accretion rates \citep{woosley1994b,yungelson2000a}. The star donating helium rich matter could be a white dwarf (see Sect.~\ref{sec:dd}) or a star that is still undergoing significant nuclear burning, or could be partially degenerate. The star would have lost its hydrogen-rich envelope in previous binary interactions, as wind mass loss rates in low- and intermediate-mass stars are generally not high enough to cause any significant loss of the hydrogen-rich envelope \citep{iben1983a}. A hydrogen-stripped, helium-burning star is rather compact in size compared to a red giant. Therefore, when such a helium-burning donor fills its Roche lobe, the separation from the accreting white dwarf would be much smaller compared to the separation in the the red giant donor case \citep[see][for equations characterizing the Roche lobe radius]{eggleton1983a}. When accretion from the helium-rich star begins, it may proceed on a thermal timescale and thus initially have a high rate of mass transfer (${\sim} 10^{-5}$ \msun/yr), but later decrease to a lower value as the binary system's orbital separation increases over time. We note that the mass transfer may not necessarily be conservative \citep{piersanti2014a}. For such binaries containing rather compact stars where it may be assumed that mass transfer is driven by gravitational wave radiation, one can estimate the mass transfer rate as

\begin{equation}
\dot{M} = \left(\frac{5}{6} + \frac{\beta}{2} - \frac{M_{2}}{M_{1}}\right)^{-1}\frac{M_{2}}{\tau_{g}}\, ,
\end{equation}

\noindent where $M_{1}$ and $M_{2}$ are the accretor and donor masses respectively, $\tau_{g}$ is the timescale of orbital decay as a consequence of gravitational wave emission, and $\beta$ is a dimensionless factor related to the masses of both stars (see \citealt{savonije1986a}, Sect.~2; see also \citealt{paczynski1971a} for a general description of common assumptions that are made when modelling mass transfer in close binaries). When accretion rates are rather low ($< 2 \times 10^{-8}$ \msun\ yr$^{-1}$), helium-burning is not able to take place and rather a shell of helium-rich material accumulates on the white dwarf, which can eventually detonate 
\citep{taam1980a,iben1987b,livne1995a,fink2007a}. This initial detonation in the helium-rich shell drives shock-waves through the white dwarf surface layers and core, which converge off-centre on the opposite side of the ignition region. The convergent shocks result in compression and drive up the temperature and density of the white dwarf in a region off-set from the star's (carbon-oxygen-rich) centre, resulting in a subsequent, second detonation \citep[thus there is a `double-detonation';][]{livne1990a,livne1991a,fink2010a,woosley2011a}\footnote{Double-detonation SNe~Ia explosions can stem from non-degenerate, helium burning donor stars as discussed here, but they can also plausibly stem from semi- or fully-degenerate helium-rich white dwarf donors.}. It is this second detonation which unbinds the (sub-Chandrasekhar mass) white dwarf star. It was pointed out by \citet{perets2010a} that the low-luminosity, calcium-rich transient SN2005, labelled initially as a Type~Ib supernova, was a prime example of a helium detonation explosion\footnote{The reader is referred to \citet{frohmaier2018a} or \citet{de2020a} for a recent overview of a search for these transients with the Zwicky Transient Facility.}. Given SN 2005E's low ejecta mass, large amounts of intermediate mass elements and low amounts of Fe-peak elements observed in the spectra, in addition to its apparent old stellar population age, SN 2005E remains a promising example of an explosion that involved a helium detonation of some kind \citep[see also][]{dessart2015a}. In Sect.~\ref{sec:subChandra}, we discuss in more detail models of sub-Chandrasekhar mass explosions. 

Both of these `sub-categories' of single degenerates -- Chandrasekhar mass explosion and sub-Chandrasekhar double-detonation -- require different criteria in terms of chemical composition (donor and accretor) and rate of accretion to end up as a Chandrasekhar mass explosion, a sub-Chandrasekhar mass explosion, or a nova eruption, with not all studies in total agreement regarding the physical conditions necessary to result in the various outcomes \citep{hashimoto1983a,starrfield1985a,isern1991a,prialnik1995a,nomoto2007a,shen2009a,jose2020a}. 
Different assumptions of how accretion proceeds regarding retention efficiency of the accreted material \citep{bours2013a,toonen2013a,ruiter2014a,piersanti2014a} and how this varies with composition, and from which location in the binary system the angular momentum is removed from the orbit, are just some factors that 
should be included in assessing the relative numbers of plausible progenitor scenarios in binary population synthesis (BPS) studies \citep[see][for a comprehensive example of binary star evolution modelling in population synthesis]{han2020a,riley2022a}. These physical considerations of course matter not only during the final RLOF phase preceding the SN~Ia explosion, but also during any mass loss/transfer episodes that occur over the course of binary evolution.

\subsubsection{White dwarf with white dwarf: DD scenarios in brief} 
\label{sec:dd}

The double degenerate scenario involves the merger of two white dwarfs that typically are assumed to merge as a consequence of orbital angular momentum loss brought about by gravitational radiation emission \citep{peters1964a,paczynski1967a,iben1984a,webbink1984a}. The time between the last mass-exchange (i.e. a common envelope or a stable RLOF phase) and the merger has a large range; it can be anywhere from just a few years to many Gyrs after the two white dwarfs are formed. The time delay $t$ between formation of a binary system on the Zero-Age Main Sequence and the time when the resulting two white dwarfs merge can be estimated if we assume that the main mechanism for loss of orbital angular momentum is gravitational wave emission \citep[see][Sect.~3]{ruiter2009a} and is found to be fairly well-represented by a power-law $t^{-s}$ with $s {\sim} 1$ \citep{maoz2012b,castrillo2021a}. An observationally-derived delay time distribution (DTD) from Type Ia supernovae in elliptical galaxies was measured by \citet{totani2008a} and the resulting DTD was found to align very well with the one predicted to arise from white dwarf mergers (see Sect.~\ref{sec:Rates} for further discussion on the SN~Ia DTD).

One must realise though that there are large uncertainties associated with deriving an observational DTD because in most cases, one must make important assumptions about the star formation history of the stellar population (e.g. \citealt{maoz2010a}, but see also \citealt{maoz2011a,strolger2020a} for alternative approaches). Mass exchange interactions within binaries can significantly alter the nuclear burning lifetimes of the stars, which also affects the delay time. The \citet{totani2008a} observations sparked a rejuvenated interest in the double degenerate scenario of SNe~Ia, launching a stream of new theoretical work in this area. In the case of degenerate donors, it could be a helium white dwarf or possibly a `hybrid' He-CO white dwarf. Both types of helium-rich WDs are only made through binary star interactions \citep{iben1985a}\footnote{We note that the term `hybrid' is also used to refer to white dwarfs that are of carbon-oxygen-neon composition. We do not discuss those heavier white dwarfs here but refer the reader to \citet{denissenkov2013a}.}. We discuss such models in some more detail in Sect.~\ref{sec:dyn}. 

\begin{figure*}[ht]
\centering
    \includegraphics[width=\textwidth]{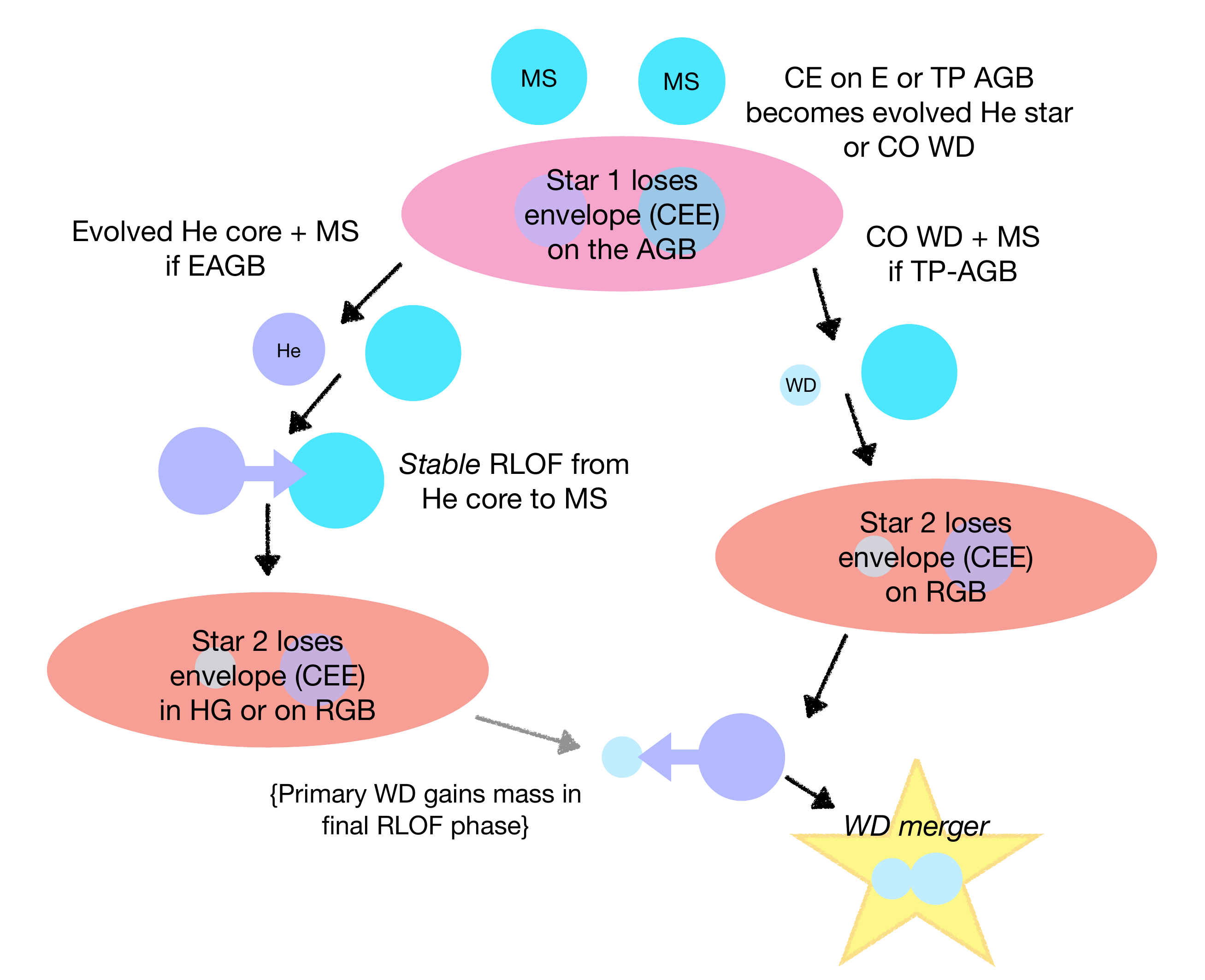}
\caption{Schematic cartoon showing two channels of white dwarf mergers that can plausibly lead to SNe~Ia from the {\sc StarTrack} \citep{belczynski2008a} binary population synthesis code. The two channels shown are those more likely to occur when the first mass transfer episode is unstable (i.e. a common envelope), which often results in a second CE later on in the evolution. We note however that double white dwarf merger progenitors can also easily form through undergoing just one CE event. {\it Left channel}: When the first CE occurs on the early-AGB (EAGB), the binary will often undergo a stable mass transfer event before the second CE takes place when the initially more massive star is a slightly-evolved, H-stripped, He-burning star. Then, another CE event occurs when the secondary is either in the Hertzsprung Gap or on the RGB. 
{\it Right channel}: When the first CE occurs later in the AGB's evolution, when it is thermally-pulsating (TPAGB), the emerging core is a white dwarf, and the next mass transfer episode to take place is unstable. {\it Both channels}: In both channels, there is a stable mass transfer episode predicted to occur later in the evolution of the binary (after both CEs in this case) whereby the slightly-evolved, hydrogen-stripped, helium-burning star donates mass to the first-formed white dwarf. Such a phase is crucial for building up the mass of the primary white dwarf to be on the order of ${\sim} 1$ \msun \, \citep{ruiter2013a}, but further detailed studies should be carried out to determine the thermal response of the stars to assess whether this phase indeed leads to continued, stable accretion and eventual formation of two white dwarfs, or whether a `premature' merger is likely to occur during this phase. The grey arrow indicates the phase is encountered less frequently.}
\label{fig:dwd-cartoon}      
\end{figure*}

\citet{badenes2012a} used radial velocity measurements to estimate the merger rate of double white dwarfs in the Galaxy. Their (lower velocity resolution) sample was limited to systems with small separations between $0.001 - 0.05$ AU. They found that the total Galactic DWD merger rate -- including WD mergers whose total mass would be below the Chandrasekhar mass limit -- was consistent with the estimated specific rate of SNe~Ia in the Galaxy from \citet[][their Fig.~4, in other words 0.1 SNuM, or ${\sim} 1 \times 10^{-13}$ SNe~Ia yr$^{-1}$ \msun$^{-1}$]{li2011c}.   
Expanding the sample, \citet{maoz2018a} were able to probe binaries with separations out to $4$ AU to set more stringent limits on the Galactic merger rate of white dwarfs. They found the double WD merger rate to be even larger than their previous estimate: 6 times higher than the expected Galactic SN~Ia rate. In other words, if most SNe~Ia do arise from white dwarf mergers, ${\sim} 15$ per cent of Galactic double WD mergers will eventually produce a SN~Ia (see Sect.~\ref{sec:Rates} for more discussion on SN~Ia rates). For a recent study on detection of close white dwarf binaries and their properties, we refer the reader to \citet{shahaf2024a}. 

To date, very few double white dwarf binaries have been directly detected \citep[][39 systems in Table~2]{napiwotzski2020a}. However, such low detection rates are expected within a limited volume given their faint nature. This is not a concern for the double white dwarfs as a leading progenitor scenario for SNe~Ia, since current technology does not even enable us to easily detect large numbers of double white dwarfs. Unlike the `textbook' SN~Ia progenitors that involve a white dwarf stably accreting from a non-degenerate star -- occasionally exhibiting thermonuclear outbursts -- detached double white dwarf binary systems are electromagnetically `quiet' \citep[see][who analysed the probability of detecting plausible double degenerate progenitors with current facilities]{rebassa-mansergas2019a}. It is expected that with the launch of the first space-based gravitational wave observatory, anticipated to be the the Laser Interferometer Space Antenna \citep[LISA, see Living Review by][]{amaro-seoane2023a}, we will finally be able to detect the presence of ${\sim} 60,000 $ double white dwarf binaries in the Galaxy \citep{korol2018a,korol2022a}, some of which we can expect to be likely progenitors of SNe~Ia. Though it is quite unlikely that LISA will `see' directly the merger of a double white dwarf over the lifetime of the mission,\footnote{LISA will be sensitive to double white dwarfs within a certain frequency band that corresponds to orbital periods on the order of a few hours or less, thus the very wide (and more numerous) systems will not be observable. Further, LISA will only be able to detect close double white dwarf binaries in our own Galaxy and nearby, which limits the total detection volume, but see \citet{yoshida2021a} regarding observations of post-WD-merger remnants in the decihertz band.} detection of many thousands of white dwarf binary pairs that will merge many Myr from now will give us the opportunity to understand pre-merger white dwarf system parameters in an unprecedented way. This will be useful not only for solving the progenitor problem \citep{maoz2012b}, but will give insights into the formation of other, non-explosive transient sources also presumed to be formed via white dwarf mergers, such as hydrogen-deficient carbon (HdC) stars \citep[R Coronae Borealis, `dustLess' HdC stars, and extreme helium stars,][]{tisserand2022a}. In Fig.~\ref{fig:dwd-cartoon}, we show a schematic picture of just two plausible formation channels leading to the formation of a merging double CO WD binary. 

 In terms of the explosion mechanism, double degenerate systems possessing the right physical properties are now thought to readily lead to explosions in sub-Chandrasekhar mass white dwarfs, likely through a double-detonation (see Sect.~\ref{sec:subChandra}). Even so, the more traditional idea of a white dwarf merger leading to the disruption of a larger, less-massive white dwarf that forms an accretion torus around the more massive primary, which subsequently accretes mass until the Chandrasekhar limit is approached, still remains a possibility \citep[see a detailed hydrodyamical study by][]{neopane2022a}. In this case, the explosion mechanism would proceed similarly to what occurs in the single degenerate Chandrasekhar mass explosion (see Sect.~\ref{sec:Chandra}).  

\subsubsection{Early history -- observed SNe~Ia}

In the 19th century a supernova occurred in the nearby spiral galaxy Andromeda: SN 1885A \citep{devaucouleurs1985a}, also called S Andromedae, thought to have been a SN~Ia based on historical records. Observations were not made easier by the fact that the supernova appeared relatively close to M31's nucleus. Going back further and closer to home, the Galactic remnant SN 1181 AD has gained much recent attention. This supernova event has been identified to have properties that render it a poor match to `normal' SNe~Ia that are used for cosmological studies, but is a reasonable candidate to link to another SN~Ia sub-class. It has been speculated by \citet{schaefer2023a} that this was a SN~Iax event that could have been triggered from the merger of a CO WD and an ONe WD \citep[see also][]{Fesen2023a}. 

The earliest known supernova thought to have been of Type Ia observed by and recorded by humans dates back as early as 185 BC (SN 185, well documented by Chinese court astronomers, see also \citet{broersen2014a}), followed by the more well-known SN Ia events SN 1006, SN 1572 (Tycho) and SN 1604 (Kepler) \citep[][table 1]{hamacher2014a}. However, we know based on supernova remnant observations that there have indeed been Type Ia supernovae exploding over the last ${\sim}$2 millennia, but only a handful have present-day lasting records. For example, SN remnants SNR 0509-67.5, 0519-69.0, and N103B in the Large Magellanic Cloud (LMC) are estimated to have ages of $400\pm120$, $600\pm200$, and $\approx$860 years, respectively, from observations of their light echoes \citep{rest2005a}.
These events were likely observable from communities in the Southern Hemisphere, and ongoing collaborations with Indigenous elders are examining possible records of these events in oral tradition  \citep[cf.][]{hamacher2022a}. The lack of any record of the supernova that gave rise to SNR 0509-67.5 seems particularly puzzling since supernova remnant evolution models that take the position of the reverse shocked ejecta emitting in [Fe\,\textsc{xiv}]5303 \AA\
into account place the explosion into the early 18th century \citep{seitenzahl2019a, arunachalam2022a}.\footnote{Reverse shocked ejecta in SNR 0509-67.5 were detected in MUSE data as forbidden highly-ionized `coronal' lines, so named because the same optical lines are seen in the solar corona \citep{seitenzahl2019a}.} Perhaps the most plausible explanation is that the supernova simply exploded in the (southern hemisphere) winter, when the LMC was below the observable horizon for most localities frequented by humans at the time. 
A summary of historical supernovae is out of the scope of this review, but the interested reader is encouraged to see \citet{pagnotta2020a}. 

\subsubsection{Type Ia supernovae: the last 100 years}
\label{sec:20thcent}

Edwin Hubble is famous for having made observations of far-away galaxies and is credited with the discovery of the expanding Universe \citep{hubble1929a}, though it should be noted that the same redshift-distance relation was discovered by Georges Lema\^itre even earlier \citep{lemaitre1927a}. Though the presence of something we now refer to as dark energy was predicted by Einstein's field equations, the notion that the rate of expansion of our Universe is increasing with time would not be confirmed until near the turn of the 20th century. 
Hubble used Cepheid variable stars to demonstrate the Universe is expanding. It was possible to use Cepheids as `standard candles' owing to their their well-known pulsation period--luminosity relationship, first uncovered by Henrietta Leavitt \citep{leavitt1912}. As instrumentation capability and technology grew, it became possible to detect other, more luminous and thus distant `standard candles': Type Ia supernovae (more accurately referred to as `standardizable candles' since lightcurve-fitting must first be applied; see Sect.~\ref{sec:hosts}). 

\citet{burbidge1957a}, famously known as B$^{2}$FH, and independently \citet{cameron1957b} first outlined rather comprehensively the plausible mechanisms for synthesis of the elements. B$^{2}$FH showed that elements from Li to U were created in stars, not synthesized in the Big Bang. The fact that the B$^{2}$FH paper offered testable predictions -- namely through chemical evolution of galaxies -- makes it a profound discovery paper. More than two decades earlier, it was reported by \citet{lundmark1932a} through careful consideration of historical records available at that time that there indeed appeared to be two different classes of explosion sites capable of polluting the interstellar medium based on their peak luminosity -- common novae (today known simply as novae), and supernovae \citep[see also][]{baade1934b,baade1934a,lundmark1935a,baade1938a,minkowski1939a}. It was noticed even around that time that these supernova explosions occur more frequently (per unit stellar mass) in galaxies that host primarily young stellar populations \citep{zwicky1942a}. 

The concept that core-collapse supernovae likely arise from massive stars and Type Ia supernovae might originate from degenerate cores of evolved stars was pointed out by 
\citet{hoyle1960a}:

\begin{quote}
``\emph{...there appear to be two distinct conditions that can lead to a major stellar
explosion: (1) A catastrophic implosion of the core. This condition is necessary when the
nuclear fuels are non-degenerate. We shall find this to be the case in massive stars
($M > 30 \, $\msun). (2) Degenerate nuclear fuels are inherently unstable. Explosion can take
place during normal evolution -- i.e., without a catastrophic implosion being necessary.
We shall find this to be the case in stars with mass somewhat greater than \msun. The
existence of two distinct conditions for explosion suggests an association with the two
types of supernovae identified by observers.}''.
\end{quote}

The idea that the dominant abundance of $^{56}$Fe among the stable iron-group{\footnote{For some general guidance regarding what we refer to as iron-group elements, we refer the reader to \citet{woosley1973a}, Figs.~2 and 6.}} isotopes in the solar system is linked to its radiogenic origin from the doubly-magic $^{56}$Ni likely goes back to German nuclear physicist Otto Haxel \citep[for a historical review see][]{clayton1999a}. In his PhD thesis, \citet{pankey1962a} made the important connection between the amount of $^{56}$Ni synthesized and the peak luminosity: 

\begin{quote}
``\emph{It is also obvious that any large scale formation of
iron by this method, in the explosion of a super nova, would lead to a subsequent decay of activity that would be linear for approximately 100 days, and then would follow the exponential beta decay of Co-56 to Fe-56}''.
\end{quote}

With a new value (but at the time uncertain) for the binding energy of $^{56}$Ni available, \citet{clifford1965a} showed, by numerically solving the equations of nuclear statistical equilibrium, that for nuclear matter of nearly equal numbers of protons and neutrons, under many conditions $^{56}$Ni will indeed be the preferred nucleus synthesized. \citet{truran1967a} then showed with time-dependent nuclear reaction network calculations that supernova shock waves in material composed of self-conjugate intermediate mass nuclei, such as $^{12}$C, $^{16}$O, or $^{32}$S, leads to copious amounts of $^{56}$Ni produced and relative isotopic ratios of the stable iron-peak elements that can explain those found in the solar system. 

\begin{figure*}[ht]
\centering
    \includegraphics[scale=0.8]{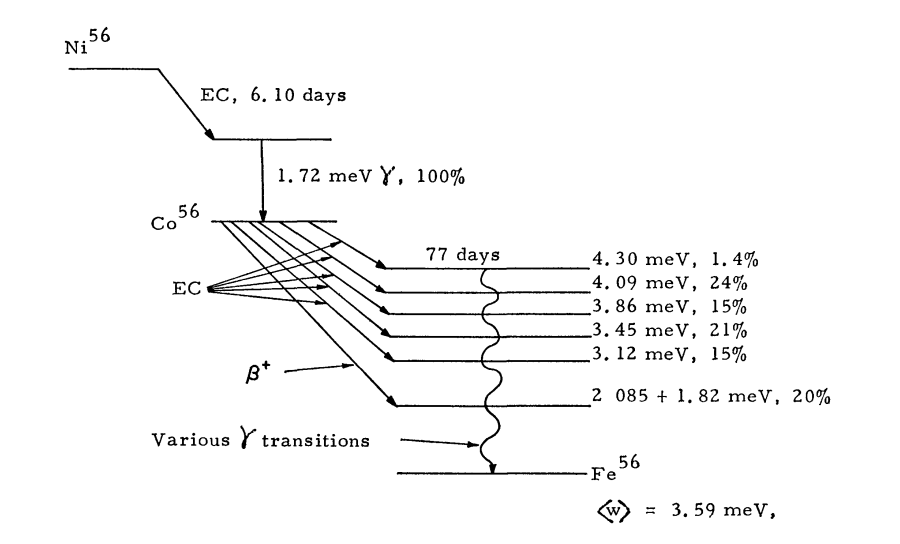}
\caption{Decay scheme of $^{56}$Ni $\rightarrow$ $^{56}$Co $\rightarrow$ $^{56}$Fe, which is long-known to supply the most power to the optical lightcurve of thermonuclear supernovae. Image reproduced with permission from \citet{colgate1969a}, copyright by AAS. 
}
\label{fig:decay}      
\end{figure*}

\citet{colgate1969a} further investigated explosion models with subsequent radioactive decay, concluding that radioactive decay of $^{56}$Ni may indeed play a significant role in providing the bulk of optical luminosity in supernova lightcurves (see Fig.~\ref{fig:decay}). \citet{seitenzahl2009d} extended the theory of the source terms of radioactively powered supernova lightcurves by including the $^{57}$Ni and $^{55}$Co decay chains, which owing to their longer half-lives dominate the heating at late times \citep[observational evidence of this prediction can be found in][]{graur2016a,tucker2022a}.

The influential papers of \citet{hansen1969a} and \citet{arnett1969a} first proposed detonations in massive white dwarfs powered by the explosive thermonuclear fusion of $^{12}$C $+$ $^{12}$C as a model for SNe~Ia. While the 1.37\,\msun\ carbon-detonation model of \citet{arnett1969a} and the 1.42\,\msun\ model of \citet{hansen1969a}  were soon after rejected as viable models for SNe~Ia, the key ideas that SNe~Ia are powered by detonations in massive WDs, fusing lighter carbon (and e.g. oxygen) nuclei into more tightly bound nuclei, including large amounts of radioactive $^{56}$Ni, has survived the test of time. The general idea that lower- and intermediate-mass stars may be the main contributors to such thermonuclear explosions was gaining ground, as supported by \citet{paczynski1970a}: ``\emph{Stars of 3, 5 and 7 $M_{\odot}$ ignited carbon in the centre at the density of $3 \times 10^9$ g/cc. This will probably lead to the type of thermonuclear supernova explosion suggested by Arnett.}". 

Later on, it would be agreed that the specific \citet{arnett1969a} direct-detonation model of an accreting near-Chandrasekhar mass WD fails because the direct initiation of a central detonation -- where the white dwarf eludes a pre-expansion phase -- is unlikely \citep[e.g.][]{niemeyer1997b};
the thermonuclear flame should rather ignite as a sub-sonic deflagration slightly off-centre \citep[e.g.][]{nonaka2012a}. 
Moreover, and more importantly, the incineration of a near-Chandrasekhar mass WD by a super-sonically advancing detonation leads to very high explosion energies and the production of predominantly iron-group elements (mostly radioactive $^{56}$Ni) -- and only small amounts of intermediate mass elements such as silicon or sulfur are produced, which is in conflict with observations \citep{arnett1971a}.

But even in the 1970s, it was still not clear what the progenitors of various types of supernovae (thermonuclear or core-collapse) in general were, though the dichotomy was becoming increasingly apparent in that core-collapse supernovae were being found among more massive, younger stellar populations while `type I' supernovae originate in lower mass stars \citep{oke1974a,tinsley1977a}. It was already suggested by \citet{finzi1967a} that such `Type I' supernovae could arise from electron captures resulting in the implosion of heavy white dwarfs near the Chandrasekhar mass limit.  
\citet{arnett1979a} postulated that Type I supernovae may arise from the core-collapse of an evolved star that underwent mass-loss of its H-rich envelope, the pre-SN progenitor being a helium-rich star in the mass range $1.5 - 4$ \msun. 
In terms of progenitor configuration, \citet{whelan1973a} discussed the likelihood of type I supernovae originating from near-Chandrasekhar mass white dwarfs accreting from a low-mass, highly-evolved giant companion. This picture would become the favoured scenario configuration for SN~Ia progenitors for the next ${\sim}$ 4 decades \citep[see also][]{hansen1969a}. 

After the pioneering work of \citet{arnett1969a}, hydrodynamical studies on detonation and deflagration physics started in the 1970s \citep{bruenn1971a,buchler1974a,ivanova1974a,nomoto1976a}  
\citep[see also the early work of][in the context of rejuvenvation of helium cores]{nomoto1977a}. 
Progress on theoretical work on explosive nuclear burning in degenerate matter and nucleosynthesis calculations paved the way for more plausible connections with problems in stellar astrophysics. 
The 1980s brought the development of more sophisticated burning models for both Chandrasekhar mass and sub-Chandrasekhar mass white dwarfs: \citep{taam1980a,nomoto1982a,hashimoto1983a,starrfield1985a,nomoto1984a,nomoto1984b,nomoto1985a,branch1985a,thielemann1986a,mueller1986a,woosley1986b,hernanz1988a}.
 
 The most successful and famous model of the early explosion models is the eminent W7 model of \citet{nomoto1984b}, a 1D explosion model of a 1.38\,\msun\ CO WD where the nuclear burning occurs behind a parameterized fast deflagration flame. While initial variants of the W7 model nucleosynthesis \citep{thielemann1986a, iwamoto1999a} overproduced certain stable iron-group isotopes (e.g. $^{54}$Fe, $^{58}$Ni), implementation of the revised electron capture rates on pf-shell nuclei \citep{langanke2000a} removed much of this problem \citep{brachwitz2000a}. Such 1D models like W7, although un-physical in the sense that the buoyancy and hydrodynamical instabilities (especially the Rayleigh-Taylor instability) that accelerate the flame are suppressed, resulted in an explosion profile that in many ways matched characteristics such as spectra and lightcurves \citep[e.g.][]{hoeflich1996a} and nucleosynthesis \citep[][]{khokhlov1991b} of ``normal" \citep{branch1993a} SNe Ia. More realistic simulations of deflagrations in near-Chandrasekhar mass white dwarfs that modelled the thermonuclear flame propagation and hydrodynamical evolution, including the growth of hydrodynamical instabilities in 3D, however failed to create sufficient amounts of fast intermediate mass elements (such as S, Si) and explosions strong enough to resemble ``normal" SNe Ia, \citep[see e.g.][]{niemeyer1996a, niemeyer1997b, reinecke1999b}.

The situation was thus that for near-Chandrasekhar mass explosion models, on the one hand an initial period of sub-sonic burning was required to expand parts of the star to lower density to avoid the pitfalls of the \citet{arnett1969a} pure detonation model. On the other hand, after some period of expansion the burning had to accelerate again to a supersonic combustion front to produce sufficient amounts of intermediate mass elements and sufficiently energetic explosions. A particular variant of this ``delayed detonation" is the pulsational delayed detonation model, where a detonation is triggered via an interplay of the turbulent combustion at the flame front and the strong pulsation of the white dwarf \citep[the explosion was not strong enough to disrupt the star fully,][]{ivanova1974a, khokhlov1991b}.

Another, more widely accepted and elegant solution to this problem invokes the transition of the subsonic turbulent deflagration to a supersonic detonation (deflagration-to-detonation-transition -- DDT), a physical phenomenon that is also observed in terrestrial combustion \citep[see e.g.][]{zeldovichbook}. \citet{khokhlov1991a} first proposed DDTs as a solution to obtaining explosions in near-Chandrasekhar mass white dwarfs that produced $^{56}$Ni and intermediate mass elements in the right proportions to explain normal SNe Ia. The literature on how to model the physics of unconfined DDTs in the context of turbulent combustion in SNe~Ia is vast and often rather technical, and we therefore refer the interested reader to the review by \citet{roepke2017a} as a first point of entry. 

The apparent success of the (un-physical) 1D near-Chandrasekhar mass models in reproducing observable characteristics, such as spectra and lightcurves, and the at the time perceived high-degree of homogeneity among observed SNe~Ia led to a strong acceptance in the community of near-Chandrasekhar mass explosion models as the natural explanation for SNe~Ia. Explosion models with lower mass primary WDs were certainly discussed and investigated, but the influential work by \citet{woosley1994a} investigated double-detonation models with rather low mass primaries and quite massive helium shells (${\sim} 0.2$ \msun). The detonation in the massive helium which produced nucleosynthesis yields contradicting the solar system values and the low core mass resulted in sub-luminous events (see Sect.~\ref{sec:subChandra}). Sub-Chandrasekhar mass primaries thus fell out of favour for an extended period of time in comparison with their near-Chandrasekhar mass ``competitors". For a comprehensive review on the state of the art of Type Ia supernova models up until the year 2000 see \citet{hillebrandt2000a}.

The ground-breaking discovery that Type Ia supernovae could be used as cosmological distance indicators sparked a whole new global effort to try to better-understand the nature of their progenitors, thus launched a rich era of observational supernova research. 
Toward the end of the 1990s, SNe~Ia were being discovered at rather large distances with the Hubble Space Telescope. 
The discovery of the accelerating Universe by these teams led to the Nobel Prize in Physics (2011) being awarded jointly to Saul Perlmutter (50\%) and Adam Riess (25\%) and Brian Schmidt (25\%) ``{\em for the discovery of the accelerating expansion of the Universe through observations of distant supernovae}”\footnote{\url{https://www.nobelprize.org/prizes/physics/2011/press-release/}} \citep{riess1998a,schmidt1998a,perlmutter1999a}. 

In the next sections, we discuss in some more detail the observations and theoretical models that have guided our understanding of SNe~Ia over the course of this century so far. 

\section{The 21st century: direct observations that constrain the explosion} 
\label{sec:DirObs}

It is almost haunting how in all of their cosmological prowess, 
the lack of understanding of what make SNe~Ia has not been a major obstacle in our ability to use them as effective cosmological tools. However, this lack of understanding profoundly effects our knowledge about the yield/origin of the elements, particularly those in and near the iron-group (Ti, V, Cr, Mn, Fe, Co, \& Ni). 
There have been a number of direct SN~Ia observations where spectra and lightcurve information have contributed to a deeper understanding of the physics behind these explosions, and in some cases the nature of the progenitor system. 
We refer the reader to the relatively recent review papers by \citet{jha2019a} and \citet{maoz2014a} for an overview of observational properties of Type Ia supernovae.  
The SN~Ia observables are of course affected by the physics of the explosion mechanism, which is affected by stellar structure (namely WD mass but also ignition and other factors) as well as, to some degree though indirectly, the nature of the donor star. So far, relatively nearby SNe~Ia have been used to calibrate or `standardize' SNe~Ia and the (emperical) relationship between intrinsic brightness and lightcurve width is mapped and applied to more distant supernovae, assuming that high-redshift supernovae `behave' the same way as those that are closer to us. 

Though more luminous SN~Ia have historically been observed among younger stellar populations \citep{branch1996a,umeda1999a}, recent photometric data (IR and optical) confirm previous cosmological studies that, after applying cosmological standardization fits to the data, the brighter SNe~Ia are those occurring among more massive host galaxies \citep{ivanov2000a,hamuy2000a,uddin2020a}. 
We discuss more about SNe~Ia in the context of their hosts in Sect.~\ref{sec:hosts}. 

\begin{figure*}[ht]
\includegraphics[width=\textwidth]{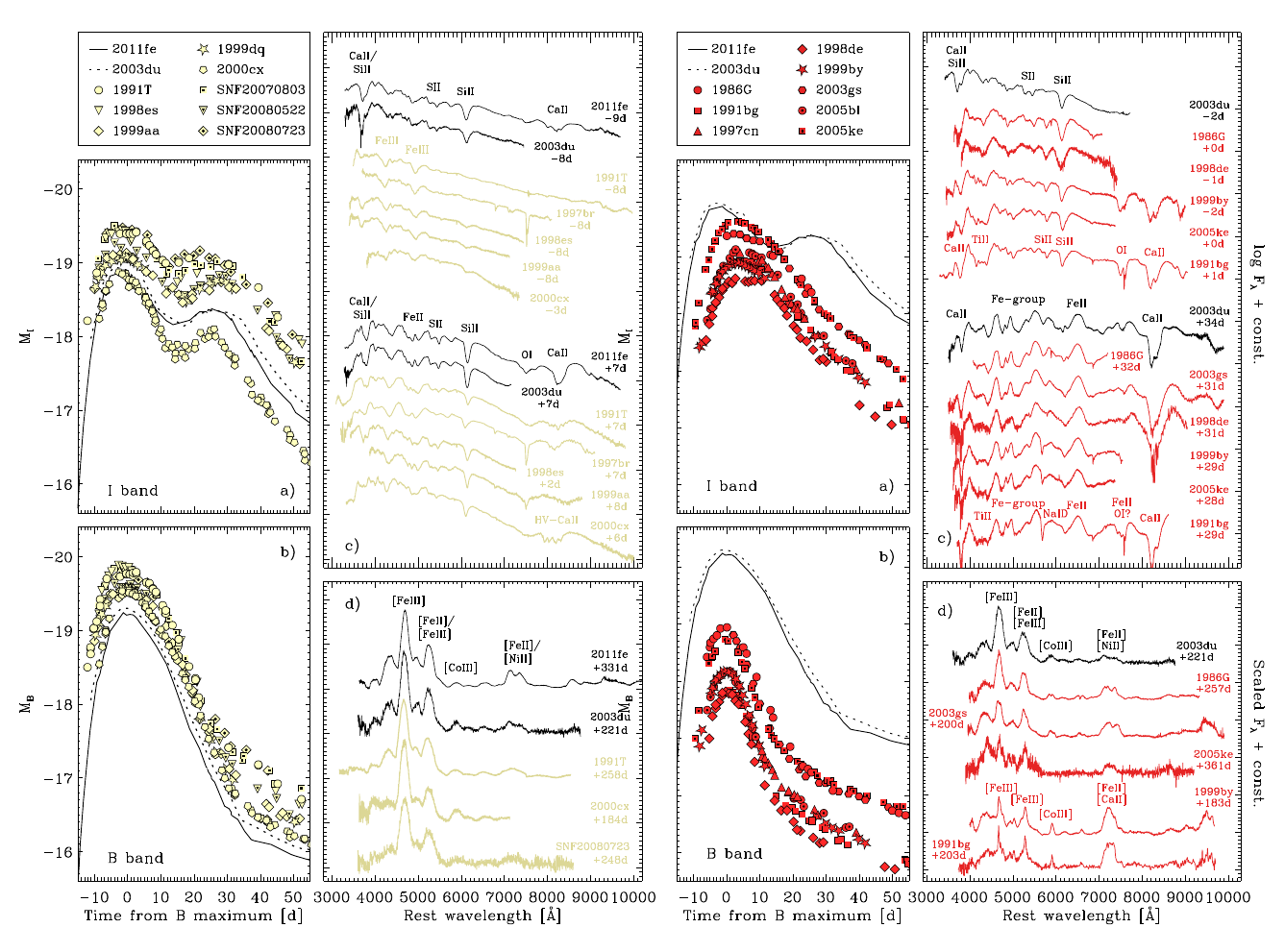}
\caption{Lightcurves and spectra of canonically-luminous SNe~Ia (shown in yellow; left) and canonically sub-luminous SNe~Ia (shown in red; right). In both (yellow and red) examples: lightcurves are in B-band (lower left) and I-band (upper left), while photospheric (top right) and nebular (bottom right) spectra are shown for comparison. In all figure panels, the lightcurve or spectrum of a normal SN~Ia (2011fe and 2003du in this case) is plotted for comparison. Images reproduced with permission from \citet{taubenberger2017a}, copyright by Springer.}
\label{fig:TheTaub}      
\end{figure*}

The large range in maximum luminosity within the observed SN~Ia population has been known for a few decades \citep[][]{phillips1993a,filippenko1997a,phillips1999a}. 1991T-like supernovae, characterized as being very luminous, can be ${\sim} 3 \times$ more luminous at peak in B-band than the fainter, faster-declining 1991bg-like supernovae \citep{spyromilio1992a}. This is nicely illustrated in Fig.~\ref{fig:TheTaub}, adapted from \citet{taubenberger2017a}, which shows B-band and I-band lightcurves and spectra of 1991T-like supernovae alongside those of 1991bg-like supernovae. 
While normal SNe~Ia make up the majority of SNe~Ia, there is a substantial fraction, on the order of ${\sim 15-35}$\%, that do not fit into this category \citep{moeller2016a,graur2017a}.  

Indeed, a large fraction of non-normals are more likely to be missed observationally owing to the fact that most peculiar SN~Ia events are sub-luminous compared to normal SNe~Ia. 
SN2011fe and SN2003du are shown for comparison with black lines in all panels. We note as well that through photometric studies, \citet{yang2022a} found the sub-class of 1991T/1999aa-like supernovae to also be very useful distance indicators. Going in the other direction, \citet{graur2024a} found that a nearby sample of sub-luminous 1991bg-like SNe function as very promising standardizable candles when taking into account correlations between their peak absolute magnitudes and the `colour-stretch' parameter s$_{\rm BV}$.

With the more recent benefit of deep transient surveys, the true diversity among SNe~Ia has become even more apparent than is illustrated here \citep[see][especially their Fig.~1]{taubenberger2017a}. Fainter, more exotic sources thought to be thermonuclear in nature are being found in surveys and can more readily be followed-up spectroscopically. 
Spectroscopy is indeed what is needed to learn more about the detailed physics of the explosion as it gives us an opportunity to test the explosion + nucleosynthesis + radiative transfer predictions from detailed models. 

\subsection{Spectra and lightcurves that constrain the explosion mechanism} 
\label{sec:specandLC}

Lightcurves of SNe~Ia directly probe how much radioactive material ($^{56}$Ni) is synthesized in the explosion, which is often used as a proxy for how much mass was in the exploding white dwarf \citep{leibundgut2000a}. 
Spectra are the most useful observational asset and act a decoder for the explosion, because from spectra we can learn about the nucleosynthetic composition of the explosive process and, possibly to some extent, the nature of the stars involved in the explosion. 
For example, for pure deflagration models that fail to unbind the star, less carbon-oxygen fuel is consumed compared to delayed-detonation models, so we would expect a large amount of unburned carbon and oxygen to be present even deep within the ejecta structure, and this may give rise to certain observables even at late times \citep{blondin2012a}. 
Ejecta line velocities give us information about the energy of the explosion, while measured velocities from material in the vicinity of the supernova can also potentially help to reveal the progenitor evolutionary channel, because such material is indicative of circumstellar non-accreted material originating from the donor \citep[e.g. blueshifted, time-varying Na~\textsc{i} D absortion features,][]{patat2007a,sternberg2011a}.

We anticipate the ability to put tighter constraints of the explosion mechanism(s) of SNe~Ia to greatly improve as the number and quality of state-of-the-art spectral observations increases. We note that while there are many works that focus on directly constraining SN~Ia explosions at high energies \citep[UV, X-rays;][]{xiaofeng2012a,immler2006a,margutti2014a}, and low energies \citep[radio;][]{chomiuk2016a}, our overview here is by no means complete and focuses on the optical regime, as this is where the bulk of the results thus far have fallen, especially in terms of modelling predictable observables (spectra and lightcurves). Thus far, setting any robust constraints on explosion models by comparing synthetic spectra and lightcurves from different models to real SNe~Ia \citep[e.g.][for SN2011fe]{roepke2012a} has been partly limited due to the dearth of nearby events, in addition to an overall lack of 3D explosion models \citep{pakmor2024a}. Some recent work has already been published using infrared nebular spectra with JWST where a number of lines from species near the iron-peak were measured \citep{kwok2023a}. However, one should remain cautious in the interpretation of observational data when making comparisons with models. Since a supernova explosion is inherently 3D -- particularly important when considering Chandrasekhar mass explosions or double degenerate mergers in which both white dwarfs explode \citep{pakmor2022a} -- it is not currently possible to accurately infer progenitor structure by comparing 1D explosion models to nebular spectra alone. Therefore, it is necessary to compute and carry out full analyses, both of the explosion and radiative transfer simulations, in 3 dimensions to make any significant progress as a scientific community in understanding SN~Ia progenitor structure from observed nebular spectra. 

\subsubsection{Early interaction (early excess)} 

Obtaining very early-time spectra of SNe Ia can be quite a challenge. However, early-time spectra can reveal crucial information about the explosion mechanism. Interaction between the companion and the supernova ejecta at early times is predicted to result in excess emission in the blue and in the UV bands \citep{kasen2010a}, though such signatures may be difficult to observe for certain viewing angles. 

A strong reverse shock is expected to form when the supernova ejecta smacks into the companion, thus causing surrounding material to heat up. As such, thermal radiation can be seen as an excess in the blue region of the lightcurve. Such an excess is expected to appear rather early, and may quickly fade a few days after explosion \citep{cao2015a}. Observational evidence of early shock interaction has not been ubiquitous. This is not surprising given the short timescales over which the early excess signature appears, though pushing for shorter cadence intervals (than the more typical 3-day cadence) in synoptic surveys could help to rectify this \citep{magee2022a}. \citet{hayden2010a} analyzed lightcurves of 108 supernovae from the SDSS-II Supernova Survey in search for signatures of early interaction, but no such candidates were found in the data. Now, a number of years later and owing to statistical analysis of improved early observations  
\citep[e.g.][]{magee2022a}, it is becoming apparent that a significant fraction (about one in five) of SNe~Ia shows an early excess bump in the lightcurve, with the sub-class of 91T/99aa SNe showing the greatest (about one in two) prevalence \citep{deckers2022a}. Some well-studied individual examples that cover a range of sub-types and interpretations include iPTF14atg \citep{cao2015a}, SN2017cbv \citep{hosseinzadeh2017a}, SN2020hvf \citep{jiang2021a}, SN2021zny \citep{dimitriadis2023a}, SN2023bee \citep{hosseinzadeh2023a,wang2024a}, 
SN2023ywc \citep{srivastav2023a}, and SN2022xkq \citep{pearson2024a}; see also works related to the Young Supernova Experiment \citep{jones2021a,aleo2023a}.

An overarching trend is that early-time lightcurves indicate that these supernovae probably did not originate from symbiotic-like (wide WD + RG) binaries \citep[307 lightcurves from TESS,][]{fausnaugh2023a}.

We highlight here one interesting example: SN2019yvq, a slightly under-luminous SN~Ia, exhibited an excess in the early lightcurve in the blue and UV \citep[see][]{miller2020a}. It is one of the rare events known to exhibit a blue early flash but also additionally calcium emission in the nebular spectra \citep{siebert2020a}. Emission of [Ca~\textsc{ii}] in particular is consistent with this event having originated from a helium-shell detonation \citep[see also][]{jiang2017a}, but no single progenitor scenario seems to be able to explain this event without introducing new discrepancies \citep{tucker2021a}.  
 
Single degenerate scenarios are not the only formation scenario expected to give rise to early-time emission, and in fact the excess emission could be due to interaction with circumstellar material \citep{piro2016b}. Though typically the presence of early interaction emission is taken to mean we can exclude the possibility of a double degenerate progenitor, it turns out that mergers cannot be completely ruled out. Interaction between a tidal tail produced in a white dwarf merger interacting with the interstellar medium could give rise to a gas shell; such an interaction is predicted to produce spectral features, potentially across a wide range of wavelengths and timescales, with an early shock breakout phase giving rise to optical/UV emission \citep{raskin2013a}.  
Though useful for setting some limits on plausible progenitor scenarios \citep{hosseinzadeh2017a}, it appears very challenging to fully disentangle the possible SN~Ia sub-classes with early excess emission \citep{jiang2018a}, as photometric data alone offer only modest constraints on progenitor scenarios at best \citep[see radiation hydrodynamical study by][]{noebauer2017a}. 

\subsubsection{Photospheric velocities} 

SNe~Ia explosions give rise to a myriad of spectral lines in the optical regime. Some rather intriguing correlations exist, for example, \citet{garnavich2004a} showed that the depth of the 580 nm spectral feature, which has potential contributions from both Si~\textsc{ii} and then Ti~\textsc{ii} in cooler photospheres, correlates with luminosity for sub-luminous SNe Ia. 

Before and around maximum light, the SN spectrum is characterized by broad emission and absorption line complexes. The observed photons are originating from a photosphere that moves inward in a Lagrangian sense. Since we are observing the part of the photosphere that is rapidly moving towards us (the view of the receding far side is obstructed by the optically thick core), the (blue-shifted) Doppler shifts of the lines can be measured. Such ``expansion velocities" around maximum brightness are most prominently determined for Si~\textsc{ii} $\lambda 6355$ \AA\ (but also others, such as e.g. Ca~\textsc{ii} H\&K).

 \begin{figure*}[ht]
 \centering
    \includegraphics[scale=0.4]{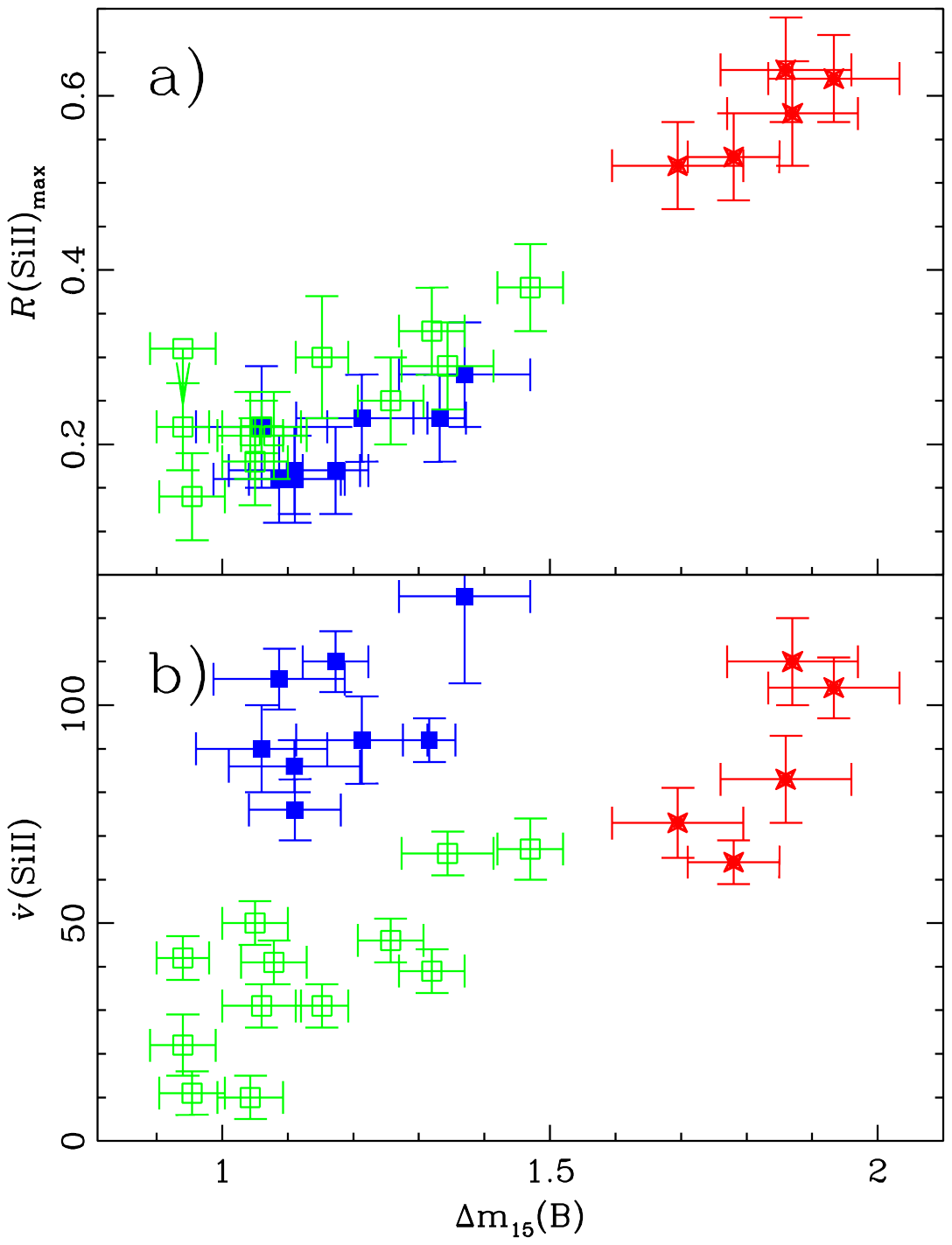}
\caption{Open symbols show LVG SNe, filled squares show HVG SNe, and starred symbols show FAINT SNe from \citet{benetti2005a}. It was later shown by \citet{blondin2012a} that these groupings were less distinct when a stricter, more consistent definition of velocity gradient is used \citep[see Sect.~5.2 of][]{blondin2012a}.}
\label{fig:Benny}      
\end{figure*}

\begin{figure*}[ht]
\centering
\includegraphics[width=\textwidth]{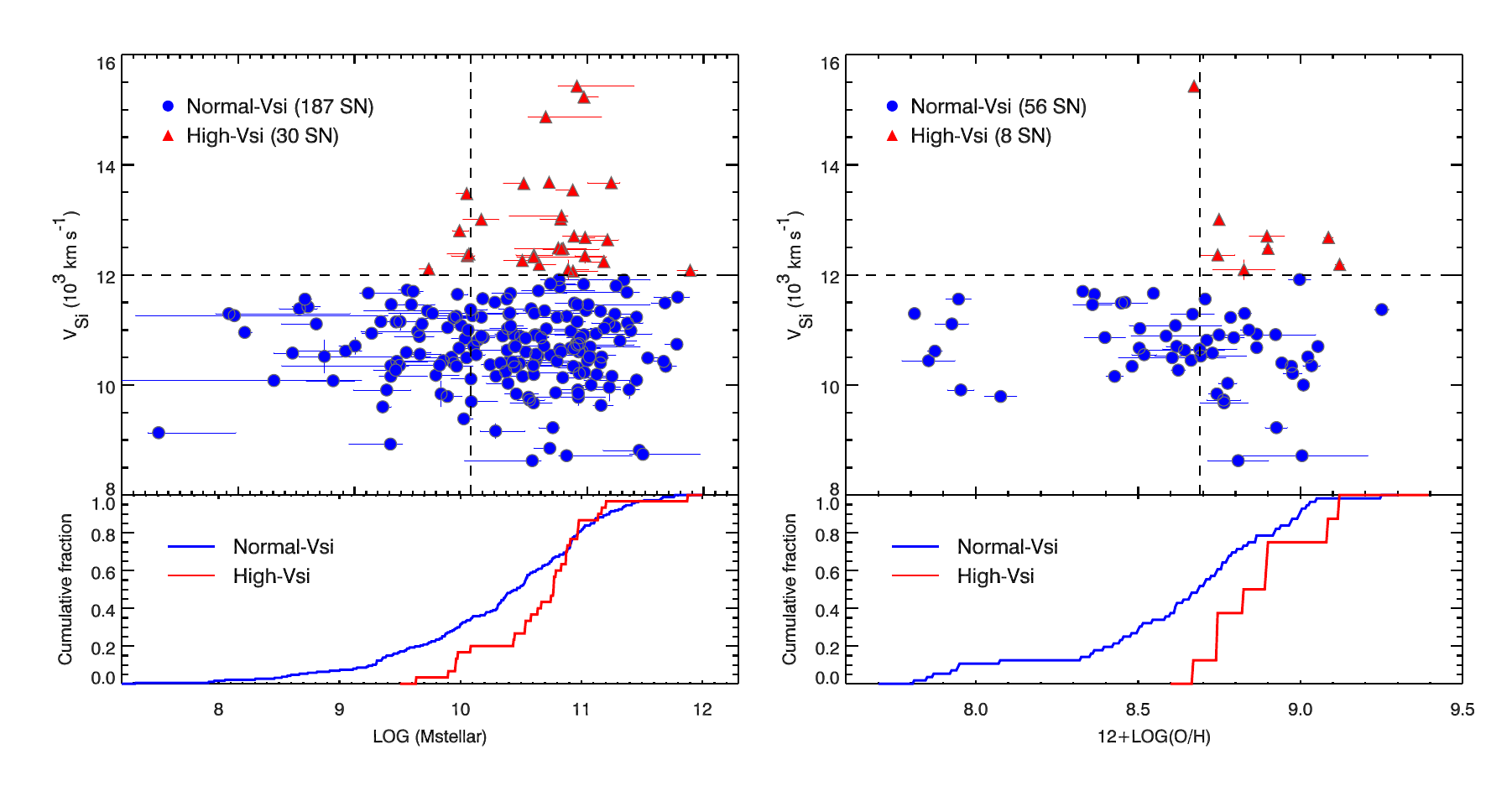}
\caption{Measured Si~\textsc{~\textsc{ii}} $\lambda6355$ \AA\ velocities as a function of host-galaxy stellar mass. The red triangles are the so-called high-velocity SNe~Ia, and tend to be found among massive host galaxies, while the lower-velocity SNe are found over the entire range of galaxy masses. Image reproduced with permission from \citet{pan2020a}, copyright by AAS.}
\label{fig:Pan}      
\end{figure*}

\citet{benetti2005a} classified the diversity among SNe~Ia by examining the time evolution of the Si~\textsc{ii} expansion velocity (the velocity gradient) and concluded that a population of 26 SNe could be broken into 3 `groups': FAINT, high velocity gradient (HVG) and low velocity gradient (LVG), with the split occurring around $70$ km s$^{-1}$ {\rm day}$^{-1}$ (see lower panel of Fig.~\ref{fig:Benny}). The FAINT group, which include the well-known 1991bg event, tend to originate in older stellar populations (early type galaxies) and decline rather quickly in comparison to the other two groups, though were found to have velocity gradients similar to those of the HVG group in \citet{benetti2005a}. Those authors speculated that HVG SNe could be delayed-detonation explosions in which variations in density at the time of deflagration-to-detonation transition is what may be causing the slight diversity among observed properties for this group (see Sect.~\ref{sec:Chandra} for a description of various explosion models in Chandrasekhar mass WDs). Speculation about the explosion mechanism for LVG SNe, which include the well-known SN 1991T event, was however less straightforward, though it was proposed this group is simply an extension of the HVG SNe in which ejecta mixing or circumstellar interaction could play some larger role. There is also evidence that explosions with high-velocity ($> 12,000$ km/s) Si~\textsc{ii} $\lambda6355$ \AA\ are found among more massive galaxies on average, with metallicity possibly being an affecting parameter \citep[][see Fig.~\ref{fig:Pan}]{pan2020a}. 

Searching for correlations between observational properties, e.g. lightcurve decline rate and strengths of various absorption features at maximum light, has been explored with the aim of disentangling SN~Ia physical processes. Some correlations have been found to give rise to a `spectral sequence' \citep{nugent1995a}. Though some SNe Ia may show nearly identical lightcurve evolution, they have been known to exhibit a rather large dispersion in velocity gradient for certain features (i.e. Si~\textsc{ii} $\lambda 6355$ {\AA} line). 
The ratio of the depth of two absorption features against ${\Delta} {\rm m}_{15} {\rm (B)}$ is shown in the top panel of Fig.~\ref{fig:Benny}: $R({\rm Si \,~\textsc{ii}})_{\rm max}$ represents the ratio of the depth of Si~\textsc{ii} $5972$ \AA\ and $6355$ {\AA} (rest wavelength) lines at maximum light \citep[see also][where the term `core-normal' originated from based on observations of 26 SNe Ia]{branch2006a}. In this parameter space, the FAINT SNe tend to show a clear separation from both LVG and HVG SNe. However when ${\Delta} {\rm m}_{15} {\rm (B)}$ is plotted against the expansion velocity evolution gradient $\dot{v}$ (Fig.~\ref{fig:Benny}, bottom panel), the three groups are clustered in distinct regions, with the FAINT group more plausibly representing a different explosion pathway than the other two (LVG, HVG) groups, noting that the LVG SNe have lower Si~\textsc{ii} velocity gradients than the other two groups \citep[see also][]{wang2009d}. The $\dot{v}$ parameter ($-\Delta v / \Delta t$) introduced by \citet{benetti2005a} measures the rate of decrease of the expanding ejecta, typically from the silicon line at $6355$ {\AA} which is usually prominent in the first ${\sim}$month after explosion. However, \citet{blondin2012a}, who performed a detailed study of 462 SNe~Ia, found that this definition of velocity gradient is not necessarily consistent since the value will depend on the interval of time over which  the measurements are taken (see their Sect.~8). Nonetheless, when in possession of large amounts of high-quality data, it is useful to carefully explore potential underlying physical correlations that may hold tangible significance. 

\citet{maeda2010b} investigated the LVG and HVG trend further, and came to the conclusion that the LVG and HVG grouping can be simply explained by a geometrical effect, and likely supported evidence for the existence of asymmetric (Chandrasekhar mass) SNe~Ia explosions. 
They argue that viewing angle relative to the deflagration sparks, assumed to be initiated off-centre, will determine the amount of red- and/or blue-shifting of nebular lines [Fe~\textsc{ii}] (7155 \AA) and [Ni~\textsc{ii}] (7378 \AA), which trace the deflagration ash \citep[also see the review by][section 1, for recent discussion]{jha2019a}. In this context, LVG SNe can be explained as being viewed from an angle in which the deflagration was ignited on the nearside, and thus the deflagration ash, once visible, will display line-of-sight velocities that are blue-shifted; vice-versa for HVG events  \citep[see Figs.~2 and 4 of][for details]{maeda2010b}. 

With the well-known idea that the mass of the exploding white dwarf is likely connected to the SN~Ia luminosity at peak brightness \citep{sim2010a}, to first order, one can assume that SNe~Ia exploding from more massive WDs will exhibit lower ejecta velocities owing to having higher gravitational binding energies. Such a relationship -- in this case increasing [Si~\textsc{ii}] velocity with M$_{\rm B}$ -- was uncovered in the double-detonation simulations of \citet[][{see also \citet{zingale2013a}}]{polin2019a}, who simulated double-detonations for different sub-Chandrasekhar mass models in 1D with 0.01 \msun\ shell masses (see their Figs.~11 and 12). \citet{polin2019a} indeed found that a sub-set of the data from \citet{zheng2018a} followed this Si~\textsc{ii} -- M$_{\rm B}$ relationship that would be anticipated for sub-Chandrasekhar events with a range of masses, while another group of SNe~Ia clustered around $v \sim 11,000$ km/s and M$_{\rm B} \sim -19.3$. The clumping group then are more likely to be arising from explosions of a common (Chandrasekhar) mass \citep[see also][for context regarding Ca-rich transients]{polin2021a}.

\subsubsection{The importance of spectra in the nebular phase}

After $\gtrsim 3$ months after explosion, the supernova is entering the nebular phase -- where the expanding ejecta have decreased substantially in density such that photons can stream freely. This phase is extremely interesting since it allows us to `see' both the near-side and far-side of the ejecta, as well as the low-velocity ejecta core. Since the outer ejecta are optically thin, we can directly probe the inner ejecta (species and their velocity red/blueshifts), which helps to constrain ignition model mechanism and geometry \citep{maeda2010b}. As the supernova enters the nebular phase, a large fraction of the radioactive nickel has decayed to cobalt, and it is the radioactive decay of 
$^{56}$Co $\rightarrow$ $^{56}$Fe that powers the lightcurve at this stage. The spectrum transitions from one of continuous emission (with absorption features) to a spectrum dominated principally composed of emission lines from iron peak elements \citep{shingles2022a}, which allows for the determination of, among other properties, synthesized nickel mass \citep[][see also Sect.~\ref{sec:ejecta-masses}]{childress2015a}. 
It has been recently suggested that the near-infrared plateau observed during the nebular phase of some SNe~Ia may be correlated with their peak luminosity, thus further similar studies could potentially give some insight into progenitor origin \citep{graur2020a,deckers2023a}.

An interesting way to probe the progenitor scenario is through the double-peaked velocity profile in the decay products of $^{56}$Ni that is visible in some optical nebular spectra \citep{vallely2020a}. 
It has been argued that, given the ${\sim}$ few thousand km/s spacing between the two velocity peaks, each peak could originate from an exploding white dwarf in the case of a double white dwarf collision \citep{dong2015a} -- though the same could be said for mergers, which are statistically more common than head-on collisions of white dwarfs (see Sect.~\ref{sec:coll}). In such a scenario, to create the distinct velocity peaks, both stars would need to explode before a single, more massive (and more dense) white dwarf is formed.

\subsubsection{H$\alpha$, circumstellar medium, and high-velocity features}

It was noted by \citet{leonard2007a} that should the donor star be hydrogen-rich, one should be able to detect H$\alpha$ emission in the nebular spectra of SNe~Ia \citep[see also][and simulations by \citealt{marietta2000a} and \citealt{pakmor2008a}]{graham2017a}. If a firm connection can be made between observed H$\alpha$ emission and the circumbinary material, then this would substantiate the assumption that the SN~Ia was produced by a Chandrasekhar mass white dwarf. A direct hydrogen signature, however, has only been found for a rare sub-class dubbed ``Ia-CSM", perhaps the most famous example is PTF-11kx \citep{dilday2012a}, but many other examples exist \citep[e.g.][]{silverman2013a, sharma2023a}. Most SNe~Ia however lack any sign of H$\alpha$ \citep{maguire2016a, graham2017a, tucker2020a}.

It has been known for almost 20 years that many (if not most) SNe~Ia exhibit so-called ``high-velocity features" (HVFs) in their early, pre-maximum light spectra \citep{mazzali2005a}. These HVFs are mostly  Si~\textsc{ii} and Ca~\textsc{ii} absorption lines that are at least several thousand km/s offset (faster) than the corresponding photospheric features. These HVFs allow us to study the high-velocity outer layers and the CSM (at length scales of ${\sim}10^{15}$ cm) and there is hope that correlations with other SN properties yield further valuable insights into the explosion mechanism and environment \citep{childress2014b,silverman2015a}. 

The CSM on even longer length scales (${\sim}10^{17}$ cm) surrounding the SN explosion site can also be probed with high-resolution spectroscopy, which sometimes reveals (mostly blue-shifted) time-varying Na~\textsc{i} D absorption features (and also Ca~\textsc{ii}) \citep[e.g.][]{patat2007a,sternberg2011a}. These are typically interpreted as coming from prior outflows of the pre-supernova system, yet while some correlations with other SN properties (e.g. photospheric velocities) 
exist, a consistent picture has not yet emerged \citep[e.g.][]{hachinger2017a}. A recent attempt to establish whether there is a connection between the early HVFs and the time-varying Na~\textsc{i} D features also showed that they appear unrelated \citep{clark2021a}.  

\subsection{Polarimetry}

In the context of thermonuclear supernovae, the most important processes that can introduce a net polarization of the light observed on Earth are scattering and dichroic absorption on aligned interstellar dust grains. Both Rayleigh scattering by molecules and Mie scattering by small dust grains, either in the inter-stellar or circumstellar medium, can be significant. Finally, polarization induced by scattering of photons by electrons (Thomson scattering) is the key physical process that can lead to net intrinsic polarization of light emitted by the supernova. A monochromatic light wave, and in fact quantum-mechanically any individual photon, will have a fixed orientation of the oscillating electric field, and hence maximal polarization. These polarizations of the individual light waves (or photons) will however cancel each other out on average for light emitted (randomly) by an extended, spherically symmetric object with perfect symmetry in composition and structure (e.g. temperature, no preferred magnetic field direction). Since the different processes responsible for polarization have different signatures, we can sometimes disentangle whether it is asymmetries in the distribution of the chemical elements (e.g. clumps), overall departure from sphericity, scattering by circumstellar or interstellar dust, or some combination thereof, that caused the polarization. For an introduction to supernova polarimetry see the review by \citet{patat2017a}. 

 As a class, Type Ia supernovae generally exhibit very low levels of polarization, often no significant polarization is detected down to the 0.2\% level and below \citep[e.g.][]{wang1996a}. This implies an overall high degree of spherical symmetry in Type Ia supernovae; for a review see \citet{wang2008a}. Polarization across strong spectral line features (line
polarization) is a particularly useful probe of the inhomogeneity of the distribution of elements in an otherwise spherically symmetric explosion. The observational characteristics of line polarization can thus be compared to the predictions for different explosion model classes. For example, \citet{wang2007a} found a correlation between the lightcurve decline rate ($\Delta m_{15}$) and the polarization of the Si$\,\textsc{ii}\lambda6355$ \AA\ line and they interpreted this result as strong support for the delayed-detonation (Chandrasekhar mass) explosion of Type Ia supernovae. Generally speaking, the low degree of polarization overall of most SNe~Ia is in better agreement with typically more symmetrical near-Chandrasekhar mass explosions than the more globally asymmetric violent merger or double-detonation models \citep[e.g.][]{bulla2016b,bulla2016a,bulla2020a}. 
Interestingly, \citet{cikota2019a} find in a set of 35 SNe~Ia observed with VLT/FORS2 that the linear polarization across the Si$\,\textsc{ii}~\lambda6355$ \AA\ line correlates with the pre-maximum line velocity, with suspected Chandrasekhar mass SNe having lower polarization compared to SNe~Ia that are suspected to be sub-Chandrasekhar in nature (see their figures 14 and 15).

\subsection{Direct imaging}

While a number of works involving photometric surveys of SNe~Ia abound, such as the Supernova Legacy Survey, or SNLS \citep{sullivan2009a}, the Carnegie Supernova Project \citep{krisciunas2017a}, PanSTARRS \citep[e.g.][]{scolnic2018a}, the Zwicky Transient Facility \citep[ZTF][]{bellm2019a}, and the Dark Energy Survey \citep[e.g.][]{abbott2019a},
the primary goal of such surveys has historically been to determine SN~Ia distances (spectroscopic follow-up is often used to determine redshifts) to constrain the dark energy equation of state by setting limits on cosmological quantities such as the matter density parameter $\Omega_{\rm m}$, where $\Omega_{\rm m}$ = 1 -  $\Omega_{\Lambda}$, and $\Omega_{\Lambda}$ is the density parameter for dark energy, in a flat $\Lambda$CDM Universe \citep[][]{guy2010a}. For an up-to-date study on the data release of over 1500 SN~Ia lightcurves with spectroscopic follow-up, see the Pantheon+ study of \citet{scolnic2022a}; see also the recent study by the Dark Energy Survey \citep{DES2024a}. 

\subsubsection{Companion searches}
\label{sec:companion_search}
Though searches for stellar companions of SNe~Ia relate more to constraining the nature of the progenitor system rather than constraining the explosion mechanism directly, a number of works have focused on searching for stellar companions (ex-donors) of SN~Ia exploders decades to 1000s of years after explosion. Though theoretical studies can be used as a guide to constrain expected observables from main sequence or evolved ex-donors \citep{marietta2000a,meng2007a} and helium star ex-donors \citep{liu2021a}, So far, no unambiguous detection of an ex-donor has been achieved with absolute certainty \citep[][see Sect.~\ref{sec:survivors}]{ruiz-lapuente2004a,kerzendorf2009a,schaefer2012a,do2017a}. 

For recent SNe, ex-donor stars that are red giants would be more easily found for nearby supernova events such as SN2011fe or SN2014J, but small, compact donor stars like white dwarfs, low-mass main sequence stars or low-mass stripped helium-burning stars \citep{warner1972a} are more difficult to detect and thus cannot unambiguously be ruled out as a plausible ex-donor \citep[see Fig.~4 of][which considers the plausibility of different donor star types for nearby SNe~Ia]{kelly2014a}. Searches for signs of a stellar companion in the close vicinity (spatially and temporally) of the explosion \citep{li2011b,kelly2014a,graur2014b,graur2019a} have yielded some limits on what type of companion may have been allowed by the observations for a particular supernova. This type of study is well-suited to nearby SNe that are still in the early part of their ejecta-dominated phase. 

With upcoming deep, optical synoptic sky surveys such as the Legacy Survey of Space and Time (LSST), which is proposed to begin collecting 10 years of data with the Vera C. Rubin Observatory, hereafter Rubin \citep{ivezic2019a} in 2025, the number of pre-explosion images of SNe~Ia is only going to grow. For now though, the number of pre-explosion images is relatively few, with one of the best examples being SN2012Z \citep{mccully2014a,liu2015a}, where an intact blue (likely helium-rich) star companion is arguably detected in pre- and post-explosion images \citep{mccully2022a}. SN2012Z is a well-known ``SN~Iax''\footnote{So-called SN~Iax events are a broad sub-class of peculiar thermonuclear supernovae \citep{foley2014a}.} event and is thought to be the best candidate for a weak explosion of a Chandrasekhar mass white dwarf that is likely to have left behind a bound remnant \citep{vennes2017a}. 

\subsubsection{Lensed supernovae}  

Gravitational lensing offers a unique opportunity to observe very distant explosions, including supernovae. An intervening massive galaxy or cluster of galaxies may conspire to `magnify' the supernova's (intrinsic) luminosity such that it is detectable to us. 

While lensed SNe with resolved images are fairly rare \citep[iPTF16geu at $z=0.409$,][]{goobar2017a}, there is a growing list of reported strongly-lensed and/or multiply-imaged SNe~Ia  \citep{patel2014a,rubin2018a,rodney2021a,goobar2023a,golubchik2023a,pierel2024a,pascale2024a}. 

Strong gravitational lensing of an astrophysical object was first discovered in the 1970s \citep{walsh1979a}. `Double quasar' QSO 0957+561 `components' A and B were identified as having nearly identical spectral features and it was concluded that the most likely reason for this was a common origin.  
Though much less-commonly detected than lensed quasars, lensed SNe~Ia are extremely valuable tools to study otherwise-undetectable (due to their high-redshift) explosions that occurred when the Universe was much younger. Since it is still not certainly known whether the nature of SN~Ia progenitors evolves or changes with redshift, and given their critical importance in cosmological studies, these higher-redshift lensed events have become a coveted asset \citep{cano2018a,pierel2022a}. As we enter an era of synoptic sky surveys that go deep, detection of lensed supernovae will become more common, which also calls for the development of software that will enable rapid identification of events  \citep{morgan2022a}. \citet{oguri2010a} predict on the order of 45 lensed SNe~Ia to be detected over the decade-long lifetime of the Rubin LSST \citep[see also][]{huber2019a}. Recent work by \citet{cikota2023a} demonstrates innovative methods that can be used to recover spectra of high-redshift supernovae through strong lensing. 

\section{Indirect constraints of the explosion}
\label{sec:InDirObs}

\subsection{Host environment} 
\label{sec:hosts}

It is well-known that some physical characteristics of SNe~Ia are correlated with properties of their host galaxies. 
As far as various sub-classes go, low star-forming galaxies tend to host 1991bg-like and Ca-rich SNe while the more luminous 1991T-like and ``SN~Ia CSM" (CSM $=$ circumstellar medium) SNe are more often found in active star-forming galaxies  \citep{panther2019a,hakobyan2020a,chakraborty2024a,qin2024a}. 
It was noted decades ago that, at face value, more luminous SNe~Ia (slow-decliners) tend to be found in younger, star-forming stellar populations, such as spiral or irregular galaxies, while fainter SNe (fast-decliners) are preferentially found in hosts with an older stellar population, e.g. ellipticals \citep[e.g.][]{hamuy1995a}. Nowadays, correlations between supernova brightness and host galaxy properties are often quoted \textit{after} having been standardized for luminosity-decline rate, the main reason being that such a process is needed to use SNe~Ia as standardizable cosmological candles \citep[][]{childress2013a}. In the context of cosmology, the term ``Hubble residual" is used, which is simply the difference between the distance modulus of the SN inferred from its lightcurve via the standardization process and the expectation value of this SN at its redshift \citep{kelly2010a}. These Hubble residuals are known to correlate with a number of (often interrelated) host galaxy properties, such as galaxy mass, metallicity, star formation rate, stellar age, and dust \citep[e.g.][]{sullivan2003a,howell2009a,sullivan2010a,childress2013a,rigault2020a,brout2021a,meldorf2023a,wiseman2023a,grayling2024a}.

\begin{figure*}[ht]
\centering
\includegraphics[scale=0.5]{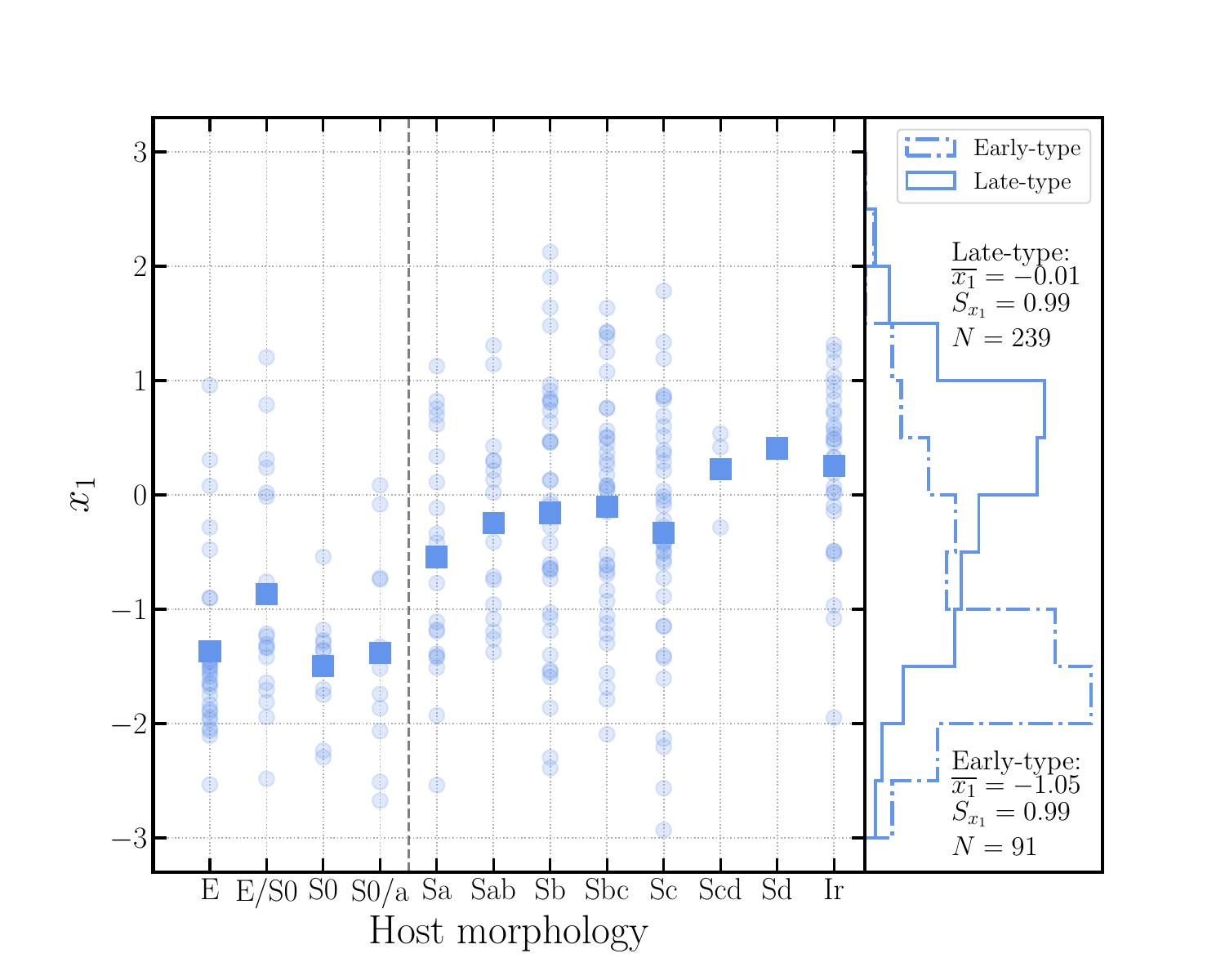}
\caption{SNe Ia from the PANTHEON survey showing {\sc SALT2}$x_1$ stretch parameter vs. galaxy host morphology (elliptical to lenticular to spiral types to irregular). In summary, SNe~Ia in elliptical galaxies have lower stretch values, and in addition generally brighter SNe~Ia but only {\em after} colour corrections are applied. Image reproduced with permission from \citet{pruzhinskaya2020a}, copyright by the author(s).}
\label{fig:Pruz}      
\end{figure*}

In a study by \citet{uddin2017a} of 1338 spectroscopically-confirmed SNe~Ia from 4 surveys, it was determined that, \textit{after standardization}, brighter SNe tend to be found in more massive host galaxies on average (although, of course, these events are intrinsically under-luminous). Further, these trends were not found to vary as a function of redshift. 
Generally speaking: more massive (and generally metal-rich) galaxies host fainter SNe with narrow lightcurves while less massive and star-forming host galaxies tend to harbour more luminous SNe~Ia with slower-declining lightcurves; however, we reiterate that this correlation should not be confused with the seemingly inverse correlations that are found for the Hubble residuals
 \citep[e.g.][see Fig.~\ref{fig:Pruz}]{sullivan2006a,sullivan2010a,meyers2012a,pruzhinskaya2020a}. 

Lower-metallicity intermediate mass stars on the AGB are expected to produce higher-mass stellar cores toward the end of their nuclear burning stage and therefore produce higher mass white dwarfs in single stellar evolution models compared to their solar-metallicity counter-parts, at least when considering stellar models up to approximately twice solar metallicity \citep{karakas2010a,karakas2014b}. 
Thus by the above logic, higher-metallicity (e.g. solar metal mass fraction) stars would form less massive white dwarfs compared to their metal-poor (e.g. higher redshift) counterparts. 
But as was pointed out, exploding white dwarf mass might be the dominant physical parameter that determines SN~Ia peak luminosity (see Sect.~\ref{sec:wdm}), so if we expect to find more massive white dwarfs in binaries in some particular environment, then we would naively expect more luminous SNe~Ia to occur there.  Going in the same direction, \citet{timmes2003a} suggested that, in the context of Chandrasekhar mass explosions, SNe~Ia from lower-metallicity progenitors would produce more $^{56}$Ni, thus lower-metallicity progenitors should be more luminous, which follows the same general expected physically-motivated trend outlined above. 
But the face-value argument for more massive white dwarfs and thus brighter supernovae from lower-metallicity stars is not at all straightforward. Older, lower-metallicity massive white dwarf populations also tend to have lower carbon mass fractions \citep{umeda1999a}, thus plausibly make dimmer SNe~Ia at least in the context of Chandrasekhar mass white dwarfs \citep[see also][]{kang2020a}. In short, nothing is straightforward!  

One crucial parameter to consider in linking SN~Ia probability to environmental properties is the core mass at which the creation of a white dwarf becomes impossible owing to the core collapsing to a neutron star (or black hole) instead. One can investigate the plausible outcomes using detailed stellar evolution models of single stars in the mass range 8--10 \msun. 
A recent study of (single) AGB star nucleosynthesis uncovered an interesting trend: That, as a function of increasing metallicity (initial metal mass fraction) from $Z=0.0025 - 0.05$, ignition of the carbon core to produce a core-collapse supernova would require higher initial stellar masses. However, at $Z \sim 0.04$, the trend reverses, and lower mass stars have a growing capacity to ignite as metallicity increases \citep[see][Fig.~1]{cinquegrana2023a}. 
We do not propose specific solutions to the challenges discussed here, but rather highlight that several factors crucial in stellar evolution, in particular metallicity, play a role in influencing the occurrence rate and host stellar population properties of Type Ia Supernovae.

We finally note that not all SNe~Ia are connected to a known host: a recent example is the fast-declining event KSP-OT-201509b, for which no host galaxy has been discovered \citep{moon2021a}. Upcoming deep surveys that can allow for improved image-stacking capability should be able to shed more light (literally) on host environments of such sources.

\subsection{Chemical evolution} 
\label{sec:chem}
Type Ia supernovae are important sites for the nucleosynthesis of heavy elements. With their unique (but delayed compared to core-collapse supernovae) nucleosynthesis signature they are a crucial ingredient for the chemical evolution of galaxies \citep[e.g.][]{matteucci1986a}. For decades, Type Ia supernovae were approximated as a homogeneous, metallicity independent class of events, with a single yield set (typically W7) and delay time distribution \citep[e.g.][]{timmes1995a, chiappini1997a,kobayashi2006a}. \citet{seitenzahl2013b} relaxed this rigid assumption and considered different nucleosynthesis for different sub-channels of SNe~Ia, thereby allowing for the impact of the diversity of SNe~Ia on the chemical evolution of galaxies. A brief discussion of theoretical yields of iron-peak species can be found in Sect.~\ref{subsubsec:fegroup}.  

\subsubsection{Milky Way}
\label{subsec:MW}
It has been estimated that within the solar neighbourhood, more than half of the iron came from thermonuclear supernovae, with the rest coming from core-collapse events \citep{maoz2017a}. Various works have estimated the relative fraction of different sub-classes of Type Ia supernovae with galactic chemical evolution models -- initially breaking SNe~Ia up into two main categories: Chandrasekhar mass explosions vs. sub-Chandrasekhar mass explosions, each with their own set of yield tables \citep{seitenzahl2013a}. In general, including nucleosynthetic sources arising from products of binary star evolution has not been carried out extensively in the literature owing to the complex parameter space involved \citep{dedonder2004a}. Though nucleosynthetic yields for SNe~Ia -- including sub-Chandrasekhar mass models -- started to be incorporated into galactic chemical evolution models once they were made available in the 1990s \citep{samland1997a}, there has been little, further advancement until recently. In recent years, galactic chemical evolution studies have started to incorporate a wider variety of SN~Ia channel yields in an effort to reflect the observed diversity in observed properties, including incorporating chemical evolution feedback timescales by adopting different theoretical, or observationally-motivated, delay time distributions \citep{lach2020a,eitner2023a,dubay2024a}. The key result of \citet{seitenzahl2013b}, that explanation of the chemical evolution of the elemental ratios in the iron group (especially Mn/Fe, Cr/Fe, Ni/Fe)
requires a mix of both near-Chandrasekhar and sub-Chandrasekhar mass progenitors, still holds, although non-LTE corrections \citep{eitner2020a} and improved calculations of the central ignition density \citep{bravo2022a} now favour a scenario where the sub-Chandrasekhar mass systems are the dominant channel. 
For a more comprehensive overview of Galactic chemical evolution, we refer the reader to \citet{matteucci2021a}.

\subsubsection{Dwarf galaxies} 

\begin{figure*}[ht]
    \includegraphics[width=\textwidth]{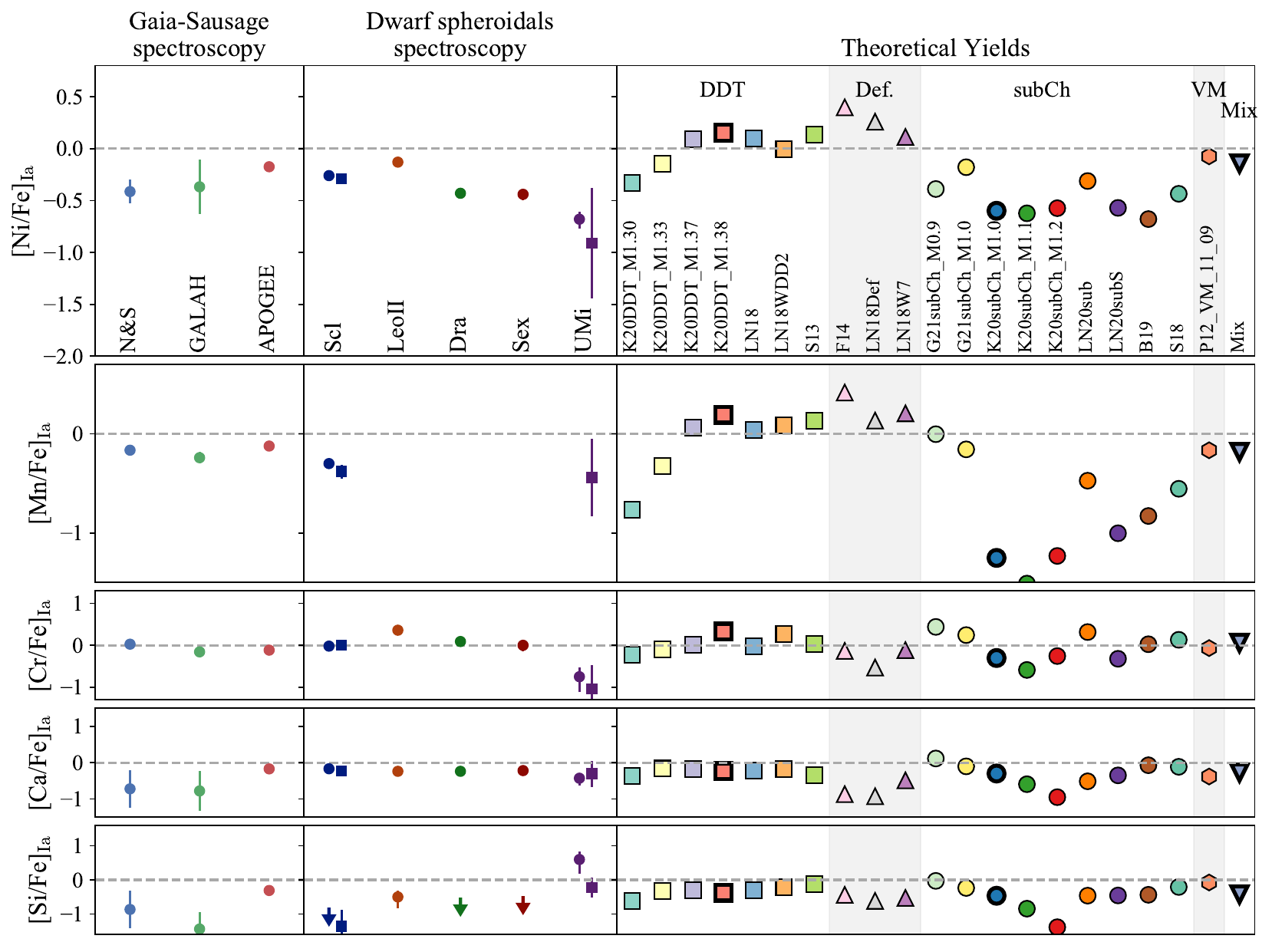}
\caption{Observed and theoretical SN~Ia yields for the Gaia Sausage galaxy (observed; far left), other dwarf spheroidal galaxies (observed; middle column), and numerous SN~Ia nucleosynthesis calculations from different explosion models: Chandrasekhar mass deflagration to detonation transition models (DDT), pure deflagration models (Def), and various sub-Chandrasekhar mass models. All dwarf spheroidals shown here exhibit sub-solar [Mn/Fe] and [Ni/Fe], as do the sub-Chandrasekhar mass theoretical yields. Image reproduced with permission from \citet{sanders2021a}, copyright by the author(s).}
\label{fig:sanders}      
\end{figure*}

The previous Sect.~\ref{subsec:MW} outlines how the chemical evolution of our Milky Way Galaxy can provide meaningful constraints on the explosion mechanisms and progenitor scenarios of SNe~Ia. While the large number of stars in the Milky Way that are bright enough for spectroscopy and atmospheric abundance modelling is an advantage, the Milky Way's complicated star formation history, past in-fall-, accretion-, and merger-events, and different stellar populations (e.g. thick disk, thin disk, bulge, halo) add much complexity. It was recently argued by \citet{sanders2021a} that there must have been a significant population of sub-Chandrasekhar supernovae that contributed to enriching the Gaia Enceladus (or `Sausage') dwarf galaxy, which merged with our Milky Way over 8 Gyr ago.  The support for sub-Chandrasekhar explosions in this (metal-poor) dwarf galaxy is evidenced by relatively low [Mn/Fe] and [Ni/Fe] abundance in stellar spectra (see Fig.~\ref{fig:sanders}). 

Dwarf satellite galaxies of the Milky Way have much fewer stars accessible, but they tend to have much more simple star formation histories \citep{tolstoy2009a}, which offers unique opportunities to constrain SN~Ia explosion models and progenitor channels via their distinct nucleosynthetic signatures and delay times. In the following paragraph, we discuss the prominent example of the element Mn in some detail. 

The chemical evolution of Mn in dwarf satellite galaxies of the Milky Way, such as e.g.~Fornax, Sculptor, Leo I, or Carina, has been known to exhibit systematically different trends than that of our Galaxy \citep[e.g.][]{mcwilliam2003a,north2012a}, which had been attributed to a possible metallicity dependence of the Type Ia supernova Mn yield by these and other works \citep[see also][]{cescutti2008a}. \citet{seitenzahl2013b} suggest the alternative explanation that the different abundance trends of [Mn/Fe] in these dwarf galaxies could also be explained if SNe Ia did not arise from a unique channel \citep[see also][]{kobayashi2015a}. If the more copiously Mn producing SN~Ia channel had a longer delay time, then they might only begin to explode in dwarf galaxies after the gas has been depleted and star formation has ceased. The `textbook' SN~Ia progenitor consisting of a near-Chandrasekhar mass white dwarf and a hydrogen-rich donor is a natural candidate, as these systems do not explode at very short delay times according to some studies \citep[see Fig.~2 of][]{hillebrandt2013a}.  These Mn producing SNe Ia would have still exploded in these dwarf galaxies, however, their chemical signature would not be seen today in the atmospheric spectra of stars that formed at an earlier time. This is corroborated by the fact that [Mn/Fe] vs [Fe/H] in dwarf galaxies with early burst-like star formation histories, such a Sculptor \citep{kirby2019a, delosreyes2022a} or Ursa Minor \citep{McWilliam2018a}, differs qualitatively from dwarf galaxies with more extended star formation histories, such as Fornax or Leo I, which show significantly higher (about 0.2 dex) [Mn/Fe] at a given [Fe/H] \citep[see e.g.][]{delosreyes2020a}. To be clear, we are not saying that (Mn-producing) long-delay time SNe~Ia do not explode in dwarf galaxies like Sculptor. However, the nucleosynthetic signature of the long-delay time SNe~Ia is not found in the atmospheric spectra of stars observed in such dwarfs today since star-formation had largely ceased at the time of the supernova explosion. The ejecta of any long-delay time SN~Ia would escape the gravitational well of the small host galaxy before having a chance to be recycled into new generations of stars. 

In the coming years, the 4MOST survey of dwarf galaxies and their stellar streams \citep[4DWARFS,][]{skuladottir2023a} is anticipated to greatly increase the number of stars with spectroscopic abundances in these and other satellite dwarf galaxies and provide us with an incredible dataset to further test and refine these conclusions. 

\subsubsection{Intra-cluster medium}

Measuring the abundances of elements in stellar atmospheres via spectroscopy is fundamental for constructing models of galactic chemical evolution. However, not all supernova ejecta remain within galaxies; both Type Ia and core-collapse supernova explosions drive outflows that enrich the intra-cluster medium (ICM) with synthesized elements, as noted by \citet{finoguenov1999a}. In fact, the majority of the baryonic matter is not in stars and gas in galaxies but rather in the ICM \citep{giodini2009a}. Since the hot ICM gas is near collisional ionization equilibrium, it is possible to obtain precise determinations of abundance ratios by modelling X-ray spectra, as discussed in the review by \citet{mernier2018a}. Initially, XMM-Newton-derived data suggested large enhancements of Mn, Cr, and Ni relative to Fe, which led to the interpretation that near-Chandrasekhar mass primary WDs should have a dominant contribution to Type Ia supernovae \citep[e.g.][]{mernier2016a}. A few years later, improvements to atomic models and much higher spectral resolution data from the Hitomi satellite revised these to near-solar ratios. Nonetheless, in the Perseus cluster, revised values still suggest significant contributions from near-Chandrasekhar mass white dwarfs to Type Ia supernovae \citep{hitomi2017a, simionescu2019a}. Similarly, for the Centaurus cluster, \citet{fukushima2022a} conclude that significant contributions from near-Chandrasekhar mass Type Ia supernovae are still required to explain near-solar [Mn/Fe] and [Cr/Fe] ratios. Another recent study by \citet{batalha2022a} also investigates the problem of which  supernova explosion models are most likely to have contributed to the ICM metal enrichment. 
The authors present a non-parametric probability distribution function analysis to assess the likelihood of different SN yields from a large set of explosion models from the literature by making a comparison against observations of galaxy groups and galaxy clusters observed with Subaru. Regarding the role of Type Ia supernovae, \citet{batalha2022a} conclude that that 3D near-Chandrasekhar mass delayed-detonation models outperformed other tested combinations of supernova models.

\subsection{Radioactive nickel and ejecta masses from lightcurves and spectra } 
\label{sec:ejecta-masses}

\begin{figure*}[ht]
 \centering
    \includegraphics[scale=0.6]{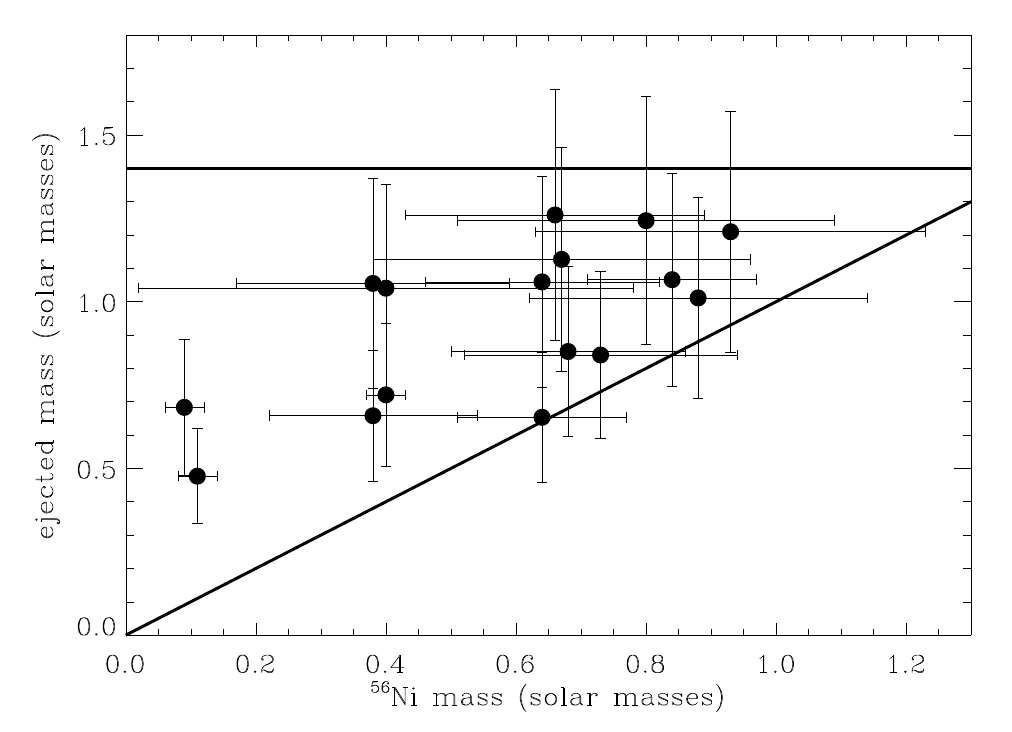}
\caption{Range of derived ejecta masses for 16 Type Ia supernovae as a function of estimated $^{56}$Ni mass. Note the data points show that the inferred amount of synthesized $^{56}$Ni varies within the sample by a factor of ten and the ejecta mass by more than a factor of two. Image reproduced with permission from \citet{stritzinger2006a}, copyright by ESO.}
\label{fig:Mej_stritzi}      
\end{figure*}

Over the last few decades, a number of works were important in making progress toward understanding the structure of white dwarfs prior to explosion based on studies of the SN ejecta. 
As mentioned in Sect.~\ref{sec:1}, consideration of the radioactive decay law coupled with diffusion of light as it travels through the SN ejecta then allow us to make an estimate of $^{56}$Ni mass from measuring SN~Ia lightcurves around their maximum. The associated ``Arnett's rule'' states that the instantaneous radioactive decay energy input (which is proportional to the $^{56}$Ni mass) at the time of the lightcurve maximum is proportional to the peak luminosity \citep{arnett1979a,arnett1982a}.
From UVOIR lightcurves, \citet{stritzinger2006a} derived ejecta masses and applying Arnett's rule (see their Sect.~3.2 and references therein) also derived $^{56}$Ni masses for a set of 16 SNe~Ia (see Fig.~\ref{fig:Mej_stritzi}). \citet{stritzinger2006a} thereby provided some of the first observationally derived evidence that the ejecta mass of SNe Ia varies by at least a factor of 2, and they suggested that sub-Chandrasekhar mass progenitors could be the reason.

In terms of sub-luminous SNe~Ia, \citet{mazzali1997a} found that outer parts of ejecta in the sub-luminous SN1991bg were under-abundant in Fe-group, indicating the presence of some unburned material. 
They found that this supernova likely had an ejecta mass 0.4--1.0 \msun, with a total synthesized Ni mass of just 0.07 \msun. This ejecta mass is rather small in comparison to what is expected of SNe~Ia at more typical peak luminosities. 
In terms of ejected nickel masses from the SN1991bg-like supernovae, \citet{stritzinger2006b} estimated a $^{56}$Ni ejecta mass of 0.09 \msun\ for SN1991bg itself and 0.05 \msun\ for SN1999by from UVOIR data (see their Table 2). A $^{56}$Ni ejecta mass in the range of 0.04--0.06 \msun\ was estimated for the fast-evolving SN2021qvv \citep[][Table~3]{graur2023a}, 
and a notably large 0.22 \msun\ for SN2022xkq \citep{pearson2024a}, a transitional/91bg-like. 

In the absence of conversion of kinetic energy to radiation via shock interaction, the energy source for the observed light output of the supernova are thus radioactive decays, and not the exothermic fusion reactions that power the explosion. One should keep in mind of course that there are complicated physical effects that may significantly affect the observed light curve evolution, including: magnetic field dependent positron transport and escape 
\citep{chan1993a,penney2014a,kushnir2020a}; time-dependent delayed recombination \citep{fransson1993a}, an effect that depends on non-thermal excitation of ions \citep{fransson2015a}; or light echoes \citep[e.g.][]{schmidt1994a,patat2006a}. 
Effects like these complicate the derivation of radioactive isotope masses from late-time light curves \citep[e.g.][]{kerzendorf2014a,tucker2022a}. 

Ejecta masses of ``superluminous Type Ia supernovae" -- stellar explosions assumed to be thermonuclear but exhibiting more luminous than standard peak luminosities and typically slower-declining lightcurves -- have been estimated to be as high as 2.8 \msun, with an inferred nickel mass of at least 1.2 \msun\ for SN2009dc \citep{taubenberger2011a}. Though, arguably, this is not a very precise way to estimate the ejecta mass in all cases, which renders the label of ``Super-Chandrasekhar mass Type Ia supernova" misleading. In particular, it is very plausible that a significant fraction of such events may derive from a conversion of kinetic energy (via interaction) to light, which can explain ``Super-Chandra SNe~Ia" without an actual need for super-Chandrasekhar mass progenitors \citep{hachinger2012a}. However, other scenarios for super-Chandrasekhar mass explosions have indeed been proposed: spin-up / spin-down, or the merger of two white dwarfs resulting in a(n initially rapidly-spinning) white dwarf with a new mass above the Chandrasekhar mass limit (see \citet[][section 2.4]{neunteufel2022a}).  
Rapidly-rotating explosion models of super-Chandrasekhar mass explosions were investigated by \citet{fink2018a}, and it was found their predicted observables did not match those of the observationally-labelled `super-Chandrasekhar' events.

 An approach to determining $^{56}$Ni masses from the time evolution of the [Co $\textsc{iii}$]5893 line during the nebular phase was pioneered by \citet{childress2015a}. This method exploits the fact that the line flux of this nebular Co-line is proportional to two powers of the mass of $^{56}$Co (and thus $^{56}$Ni) as a function of time. The reason for this is that the observed line flux both scales with the number of available Co atom targets and the probability of exciting the line. As long as the ionization state of the ejecta does not significantly evolve, the flux of this Co-line thus decays much more rapidly than surrounding Fe-lines, which do not face the same problem of an exponentially decreasing target abundance. \citet{childress2015a} find that radioactive $^{56}$Ni masses of SNe Ia fall into two regimes: i) SNe~Ia with narrow lightcurves have low $^{56}$Ni masses around $0.4~\msun$ with a weakly increasing yield as ejecta mass goes from $1.0~\msun$ to $1.4~\msun$, and ii) SNe~Ia with broad lightcurves show a scatter in $^{56}$Ni masses from around 0.6 to $1.2~\msun$, with ejecta masses clustering around $1.4~\msun$. 
 \citet{scalzo2014a} find a similar bimodal distribution of ejecta and $^{56}$Ni  masses in their lightcurve analysis of 337 SNe~Ia, corroborating the evidence for at least two distinct explosion mechanisms for normal SNe Ia. 

\subsection{Supernova abundance tomography}

Supernova abundance tomography is a method to infer the abundance profiles (the radial distribution or distribution in velocity space) of chemical elements in the ejecta of supernovae. The technique was pioneered by \citet{stehle2005b}. In essence, the abundance tomography method takes advantage of the gradual shift of the photon emitting layer into deeper and deeper layers of the ejecta over time. In other words, as the supernova expands, the photosphere recedes (in a Lagrangian sense) deeper into the ejecta. Since spectral features form in the layers above the photosphere, the earliest spectra are sensitive to the chemical abundances in the outermost layers, whereas post-maximum light spectra probe much deeper layers of the ejecta. Therefore, by modelling a time series of spectra taken in succession, abundance profiles of chemical elements that best reproduce the observed spectral evolution can be reconstructed from the outside in.
Abundance tomography has been performed for several well-observed SNe~Ia, including SN~2002bo \citep{stehle2005a, obrien2021a}, SN~2004eo \citep{mazzali2008b}, 
SN~2003du \citep{tanaka2011a}, SN~2009dc \citep{hachinger2012a}, SN~2010jn \citep{hachinger2013a}, SN~1991T \citep{sasdelli2014a}, 
SN~2011fe \citep{mazzali2015a}, SN~1986G \citep{ashall2016a}, and SN~1999aa \citep{aouad2022a}. 
For most of these studies, however, density profiles corresponding to a mass of the exploding white dwarf near the Chandrasekhar limit were a priori assumed. \citealt{mazzali2015a} also used sub-Chandrasekhar mass profiles for SN~2011fe, \citealt{hachinger2012a} also explored super-Chandrasekhar mass profiles for SN~2009dc, and \citealt{obrien2021a} used a larger variety of models in their Bayesian probabilistic modelling approach. Assuming a Chandrasekhar mass profile pre-emptively confines the derived white dwarf mass to 1.4 \msun, which is currently not considered to be a valid assumption \citep{hillebrandt2013a}. 
In spite of this caveat, meaningful abundance profiles for individual SNe can be derived in this way. Such inferred abundance profiles can then be directly compared to the modelling predictions of different explosion models. To give one example, \citet{aouad2022a} performed abundance tomography for 3 (similar) density profiles (all Chandrasekhar mass models) for SN~1999aa and they found that the innermost 0.3\,\msun\ consist of mostly stable IGE, then about 0.65\,\msun\ radioactive nickel, then a thin layer of IMEs and then just over 0.2\,\msun\ of an oxygen rich outer layer.

\subsection{Potential surviving companions} 
\label{sec:survivors}

In addition to efforts to look for SN~Ia companions in present-day, nearby supernovae (Sect.~\ref{sec:companion_search}), searching for companions of historical supernovae in the vicinity of supernova remnants has been ongoing \citep[see][]{ruiz-lapuente2019a}. A star that survives the blast wave by its nearby stellar companion may appear to be observationally peculiar. Additionally, geometric effects on the supernova ejecta imparted by the surviving star have also been explored \citep{kasen2004a}. 
Certain observational signatures might be present even well after the remnant has entered its Taylor--Sedov phase, such as heavy-element enrichment \citep{liu2013a}, bloating leading to higher luminosity for both hydrogen-rich \citep{marietta2000a,shappee2013a} and helium-rich \citep{pan2013a} companions, and supernova kicks that result in higher-than-usual proper motion \citep{kerzendorf2009a}. We refer the interested reader to \citet{liu2023a}, Sect.~5.2.1, for details regarding ejecta-companion interaction searches.  

The search for surviving companions of SNe~Ia has in the past mostly focused on the donor stars in single degenerate systems; for a recent review on the topic see \citet{ruiz-lapuente2019a}. However, double-degenerate systems can have their own surviving companions if only one of the WDs explodes. For dynamically-driven mergers (see Sect.~\ref{sec:dyn}), the secondary star might survive the explosion \citep[e.g.][]{burmester2023a}. For these double-degenerate systems, the orbital velocities are high (several hundred to a few thousand km/s) when the compact stars interact. As a result, the explosion of the primary leads to the expulsion of the secondary WD at high-velocity. Although these surviving companions are ejected largely unscathed, the expectation is that their composition is heavily altered by the interaction with the supernova ejecta \citep[e.g.][]{tanikawa2018a}. The runaway stars thus carry a signature of the nucleosynthesis of the supernova explosion and they should present as chemically peculiar.

Modern spectroscopic surveys such as Gaia \citep{gaiaDR2} have revealed very fast-moving chemically peculiar dwarf stars; some of these `runaway' stars may be the surviving companions of SN~Ia explosions from double degenerate binaries \citep{shen2018a, igoshev2023a}. However, such runaways, if they originate from SN~Ia exploding binaries -- even if many are lurking in the Galaxy but eluding detection -- are predicted to constitute only a small fraction of the total Galactic SN~Ia population \citep{neunteufel2022a,pakmor2022a}. 

Chemically peculiar runaway stars carrying the nucleosynthetic signature of thermonuclear supernova explosion could also arise from near-Chandrasekhar mass explosions \citep{raddi2018a}, in particular from a pure deflagration scenario where the primary is not fully unbound by the thermonuclear runaway (cf. Sect.~\ref{sec:faileddef}). Several candidate stars heavily enriched in O/Ne and showing signatures of high-density thermonuclear burning (e.g. super-solar Mn/Fe) that could be the stellar remnants of such thermonuclear supernovae have now been identified \citep{raddi2019a, igoshev2023a}.

\subsection{Supernova-remnant archaeology}

Some SN~Ia formation scenarios require extended periods of accretion from a secondary companion star onto the primary WD, growing the mass of the primary and increasing the central density leading up to the ignition. The required mass accretion rate to obtain stable burning on the surface and hence to grow the primary white dwarf mass results in a hot surface of the primary  emitting strongly in soft X-rays and UV in a time window some 10 million years prior to explosion \citep{kahabka1997a}. The paucity of such soft X-ray emission from nearby elliptical galaxies and bulges has been used by \citet{gilfanov2010a} to limit the contribution of the H-accreting single-degenerate progenitor channel to less than 5\% in these old stellar populations. 

The ionizing radiation from SN~Ia progenitor systems involving stably accreting white dwarfs would naturally also have an imprint on the interstellar medium surrounding the explosion site, and one could expect to detect signatures of ionized gas up to 100,000 years post-supernova from such progenitors. Despite efforts to reveal such signatures, thus far any strong evidence for such a progenitor through this detection method has not been found  \citep{woods2017a,kuuttila2019a,souropanis2022a}. Specifically, the lack of a Strömgren sphere surrounding Tycho's SNR (the remnant of SN~1576) has been used
\citep{woods2017a,woods2018a} to argue that this famous Galactic supernova more likely originated in a double degenerate merger. 

\subsection{Rates and delay time distributions} 
\label{sec:Rates}

The observed rate of Type Ia supernovae, barring numerous observational biases, will be a convolution of the SN~Ia delay time distributions and the star formation histories of the supernova-hosting environments.
The empirically estimated Galactic SN~Ia rate is ${\sim}0.5$ per century \citep[][section 1, for a description of various methods used in deriving the SN~Ia rate from observations]{li2011c,wiseman2021a}.

\begin{figure*}[ht]
\centering
\includegraphics[scale=0.5]{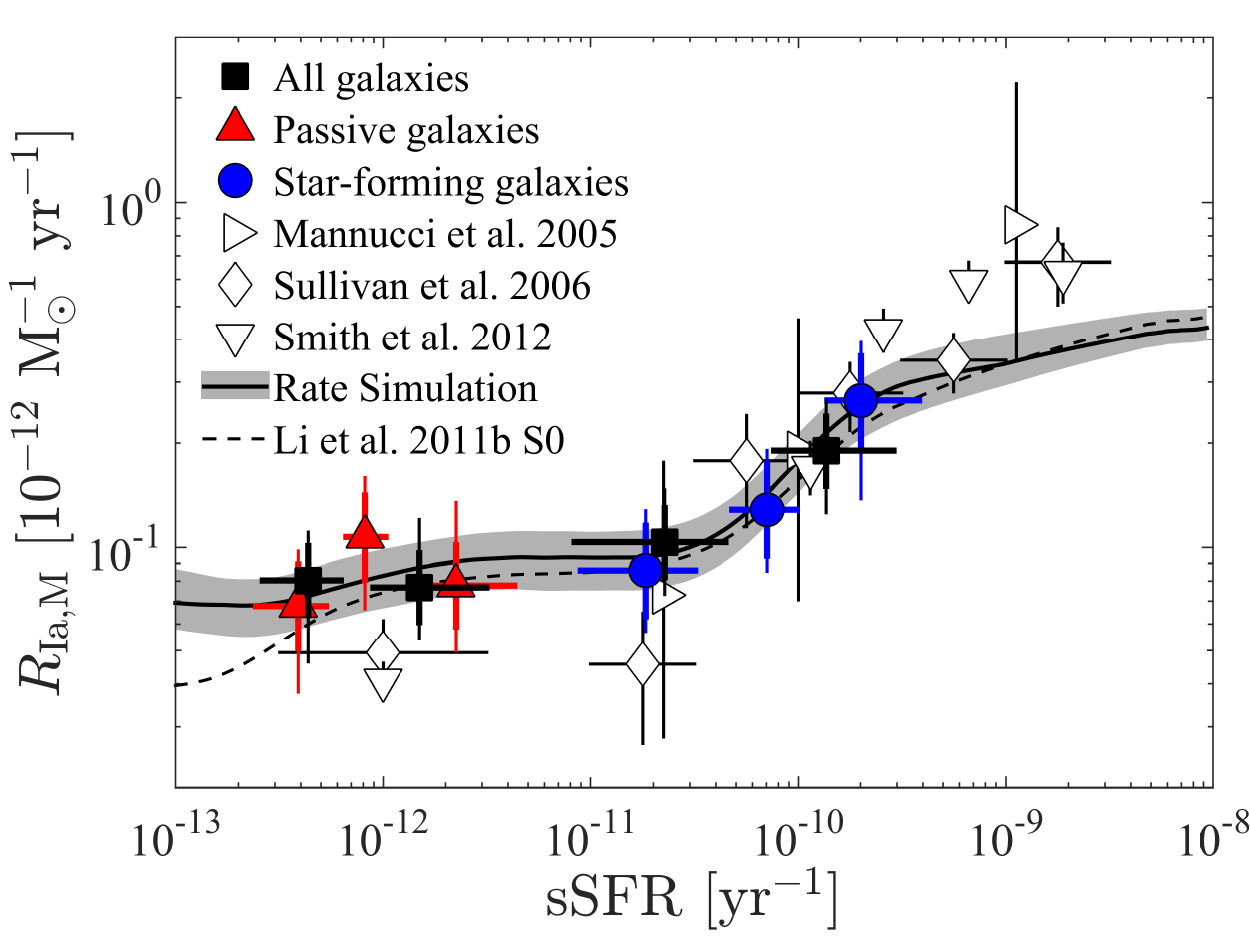}
\caption{SN~Ia rate as a function of specific star formation rate (per unit stellar mass). Coloured data points have been mapped to galaxy types, either passive or star forming: A clear progression in SN~Ia birthrate from passive galaxies to those that actively form stars is confirmed across a number of works. Image reproduced with permission from \citet{graur2015a}, copyright by the author(s); (see also their Table~2).}
\label{fig:graur-fig}      
\end{figure*}

The number of works focused on determining rates of SN~Ia has grown substantially, and necessarily, as instrumentation has greatly improved, resulting in increased capability to detect events at higher redshift. Type Ia SN rates have been observationally-derived using various techniques, from data taken with different telescope facilities, with a range of sample sizes and host properties \citep[][to highlight a number of papers from the last quarter century]{cappellaro1999a,tonry2003a,dahlen2004a,mannucci2005a,sullivan2006a,botticella2008a,li2011a,graur2011a,maoz2012b,graur2014a,rodney2014a,friedmann2018a,brown2019a,strolger2020a,wiseman2021a,toy2023a}. A relationship between the SN~Ia rate and specific star formation rate out to a redshift of ${\sim}$0.25 is shown in Fig.\ref{fig:graur-fig}.

As mentioned already in Sect.~\ref{sec:hosts}, it has been well-known for decades that, without taking into account standardization for cosmological studies, the more luminous SNe~Ia are those that occur among younger, star-forming galaxies, while dimmer SNe~Ia are found among older stellar populations. The rate of SNe~Ia is also much higher among young stellar populations, with the SN~Ia rate in late-type (e.g. spiral) galaxies being larger by a factor of 20 or so when compared to early-type galaxies \citep{mannucci2005a}. Often, the supernova rate is expressed in SNuM: ($\frac{{\rm number \, of \, SN~Ia}}{{{100}{\rm yr}\,{10^{10}}}{\msun}}$), though sometimes simply expressed in SNe \msun$^{-1}$ yr$^{-1}$, where the mass often represents the mass (born) in stars. The rate can also be expressed per unit volume \citep[e.g.][]{cappellaro2015a}, or a volumetric rate that excludes mass but factors in the dependence of an assumed Hubble constant \citep[SNe yr$^{-1}$ Mpc$^{-3}$ $h_{70}^3$, e.g.][]{frohmaier2018a}. 

\begin{figure*}[ht]
 \centering
      \includegraphics[width=\textwidth]{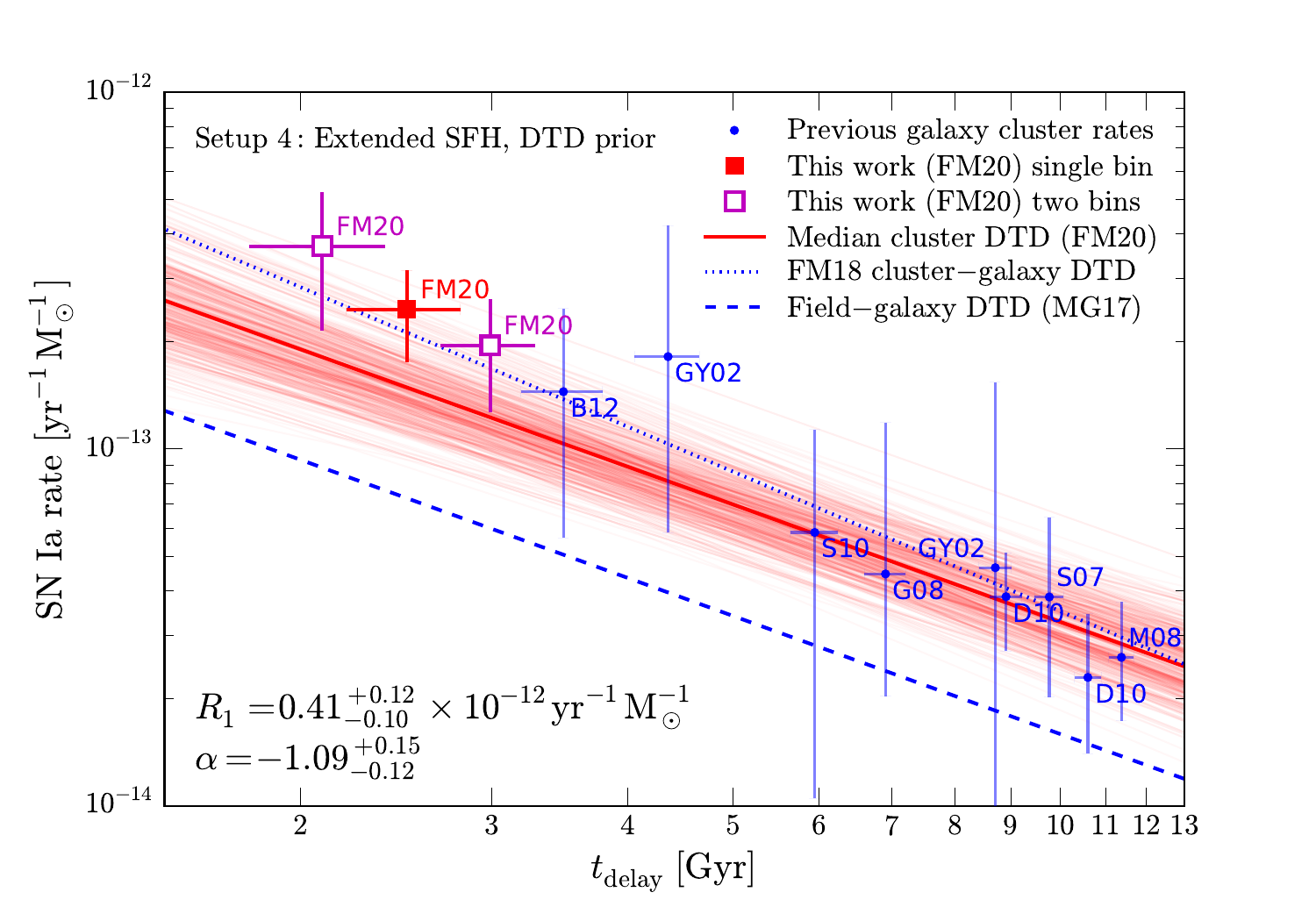}
\caption{Observationally-derived power-law DTD fits showing supernova rate per unit time per mass born in stars. Over-plotted are supernova rates from various works. The lower dashed blue line represents a fit based on field galaxies in the local universe, while the blue dotted line represents a best-fit based on the sample of higher-redshift galaxy clusters. Red lines represent DTD models from MCMC Bayesian inference from \citet[][see paper for details]{freundlich2021a}. According to this work, for a given supernova age (or delay time), the overall supernova rate is higher in galaxy cluster environments compared to the local Universe by a factor of nearly 2 or more, with the discrepancy being higher at short delay times (see text). Figure from \citet{freundlich2021a}.}
\label{fig:FM21}      
\end{figure*}

Observations of SNe~Ia in the mid-2000s indicated that SNe~Ia were likely comprised of two main populations: a `prompt' component with short delay times and a `tardy' component with longer delay times that exploded over a wider age range \citep{mannucci2006a}. Motivated by the apparent existence of two populations of SNe~Ia, \citet{scannapieco2005a} came up with a simple 2-component analytical expression for the SN~Ia rate as a function of both instantaneous star formation rate and total stellar mass: The ``A $+$ B'' model. This model however does not capture the full picture because observationally-inferred delay time distribution shapes clearly show a significant number of events at intermediate delay times (see e.g. Fig.~\ref{fig:FM21}).  

In terms of more recent works, \citet{friedmann2018a} found that the rate of SNe~Ia in galaxy clusters (at a mean redshift $z = 1.35$) is higher than that observed for field galaxies, with a cluster SN~Ia rate of $2.6 \pm ^{3.2}_{1.5}$ $\times$ $10^{-13}$  yr$^{-1}$ \msun $^{-1}$, though the reason(s) for this higher efficiency are not clear. 
Already from observational studies of SN~Ia rates and delay times over a decade ago \citep{maoz2012a}, it was apparent that the rate of SNe~Ia per unit stellar mass in galaxy clusters is higher than in the local Universe. This finding was also recently re-confirmed by \citet[][see Fig.~\ref{fig:FM21}]{freundlich2021a}; 
a similar figure can be found in \citet{strolger2020a}, their Fig.~9.

\begin{figure*}[ht]
\centering
\includegraphics[scale=0.60]{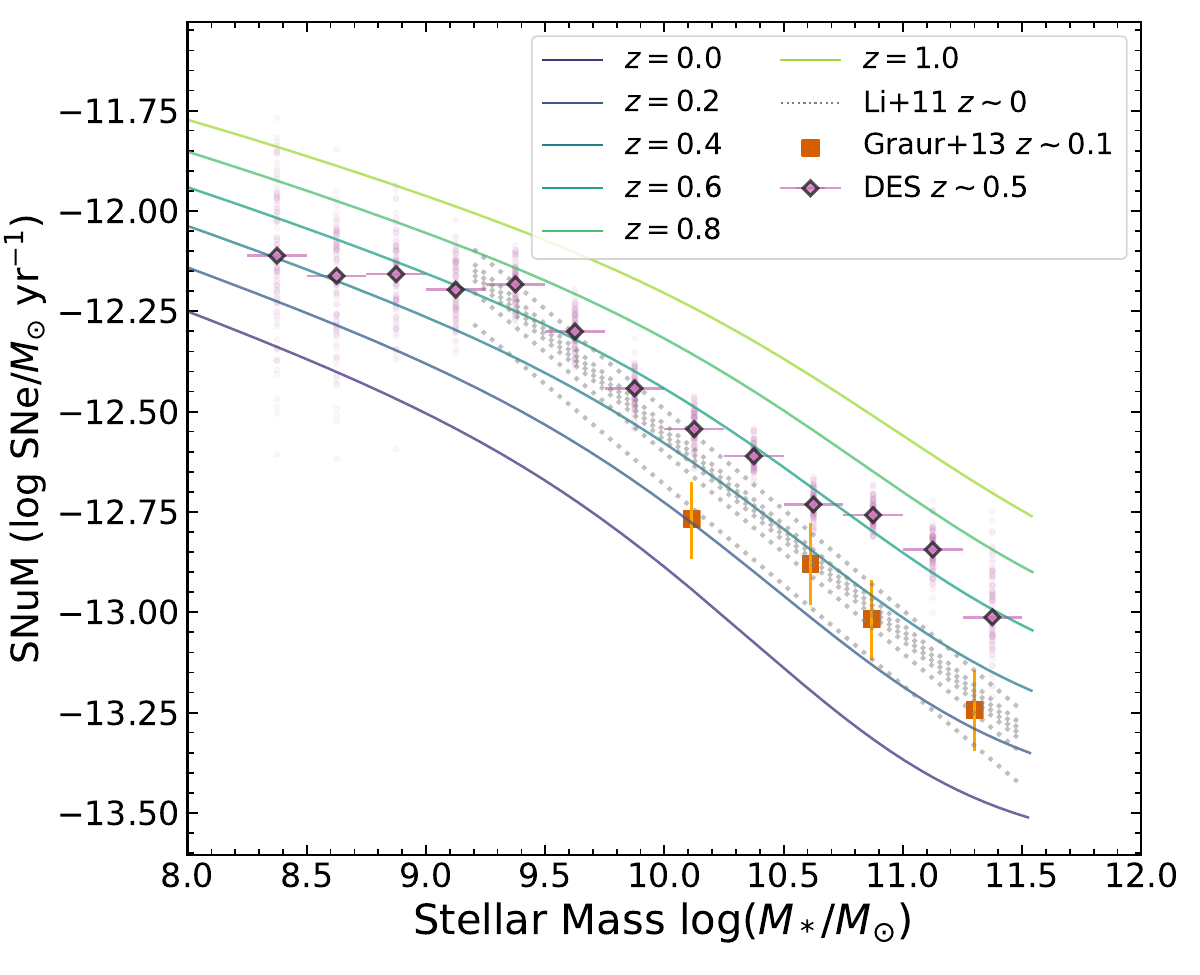}
\caption{Purple diamonds show the rate of SNe~Ia per unit stellar mass as a function of galaxy stellar mass from \citet{wiseman2021a}. Dotted lines and orange squares show previous rate estimates from earlier works. Image reproduced with permission from \citet{wiseman2021a}, copyright by the author(s); see their Fig.~11 for details.}
\label{fig:DES}      
\end{figure*}

Having said this, \citet{toy2023a} found the rate of SNe~Ia in field galaxies to be quite similar to that of galaxy clusters (see their Fig.~8), plausibly at odds with the trend found by \citet{friedmann2018a}. \citet{toy2023a} cite possible reasons for the difference (see their Sect.~4.2.1), however we note that the redshift range of cluster SNe in the \citet{toy2023a} study is significantly smaller ($0.1 \leq z < 0.7$) compared to the redshifts of galaxy clusters in \citet{friedmann2018a}, so one could naively speculate that metallicity plays some role. Employing a completely different method, \citet{wiseman2021a} derive a SN~Ia rate from the Dark Energy Survey of 
$2.6 \pm 0.05 \times 10^{-13}$ yr$^{-1}$ \msun $^{-1}$ (see their Sect.~5, and see Fig.~\ref{fig:DES}, for the SN~Ia rate as a function of galaxy stellar mass).  

There have also been several works dedicated to modelling the time from progenitor binary system birth to explosion: the delay time \citep[i.e.][]{yungelson1998a}. When this time is calculated for a number of supernovae, we uncover the delay time distribution (DTD). While it is possible to estimate the DTD of SNe~Ia through analytic formulations \citep{greggio2005a},  
calculating delay time distributions from binary population synthesis models takes into account binary evolution physics in greater detail and thus enables the inclusion of a wider variety of possible outcomes \citep{han2004a,ruiter2009a,bogomazov2009a,wang2010a,mennekens2010a,toonen2012a,liu2015a,yungelson2017a}. In BPS codes, the delay time is simply found by examining the time between when the binary system is formed (usually assumed to be formed through field evolution, not through dynamical interactions in dense stellar environments) to when the supernova event occurs based on pre-set conditions \citep[see][for a parameter study]{claeys2014a}. The DTD is rather powerful in that it sets a limit on SN~Ia progenitor ages. 

Since certain progenitor channels are predicted to only produce `prompt' events with very short (e.g. well under 1 Gyr) delay times, we can then exclude such progenitors from explaining events occurring among old stellar populations, and vice-versa for any explosions that may be predicted to kick in only at long delay times. 
When considering the Chandrasekhar mass explosions from the single degenerate scenario, binary population synthesis calculations indicate there are two distinct age populations that emerge from the DTD: prompt events that stem from SNe~Ia with hydrogen-stripped, helium-burning star donors, and the more canonical events where a WD accretes hydrogen-rich material from an unevolved or slightly evolved star, which have a longer average delay time distribution (see \citealt{ruiter2011a} Fig.~1; see also \citealt{wang2009a}). Unlike white dwarf mergers whose DTD is primarily governed by the timescale associated with gravitational wave radiation, the DTD of single-degenerate scenario SNe~Ia is highly dependent on the evolutionary timescale of the (future) donor star. The H-stripped, He-burning donor stars that contribute to the prompt events derive from more massive progenitors than their H-rich donor counterparts, thus have shorter main sequence lifetimes \citep[see][Sect.~4.1]{ruiter2011a}. 
We note however that, as with any result based on modelling assumptions, predicted DTDs may not be a true reflection of what actually happens in nature. Interestingly, varying certain parameters, such as common envelope efficiency, may not only affect the overall birthrates, but could uncover new evolutionary channels -- and thus a differently-shaped DTD \citep[][so-called AM~CVn channel in their Fig.~1]{ruiter2009a}. 

Observational delay time distributions on the other hand must be reconstructed, and there are a number of sources of uncertainty. One major assumption to be made is the star formation history of the stellar population associated with the supernova explosion, which is nearly impossible to know with any high accuracy, though is generally more straightforward for old stellar populations \citep{totani2008a}. Nonetheless, some clever methods have been developed to re-construct delay times -- in other words, ages -- of the stars responsible for the supernovae \citep{maoz2010b,maoz2011a,graur2013a,heringer2017a,wiseman2021a}. While several studies agree that the $s$ parameter is on the order of $1$ in the $t^{-s}$ power-law fit to observational DTDs \citep[e.g.][]{maoz2012a,sand2012a,castrillo2021a}, there is no reason to a priori assume that all SNe~Ia must obey such behaviour \citep[see recent study by][]{palicio2024a}. Some studies have demonstrated that, if indeed a single continuous power-law is assumed, $s$ has some (unsurprisingly) apparent dependence on stellar environment \citep[e.g.][]{chen2021a}. The power-law fit was found to be slightly shallower (steeper) if the DTD was derived from field (cluster) galaxies \citep[][see also Fig.~\ref{fig:FM21}]{maoz2017a}. Further, \citet{heringer2017a} found that supernovae with faster lightcurves were well-fit by a steeper DTD power-law ($t^{-1.27}$) whereas SNe~Ia with broader lightcurves could not be fit by any single continuous power-law.  
  
Figure 5 in \citet{wang2012b} gives an overall summary of model DTDs from BPS studies compared to observations. The model DTDs are separated into two groups: single-degenerate and double-degenerate \citep[originally presented in][]{nelemans2013a}. The summary is that various BPS codes roughly agree in terms of producing DTDs from double white dwarf mergers (see Sect.~\ref{sec:dd} for more discussion). However, DTDs of single degenerate scenario Chandrasekhar mass SNe calculated with different codes greatly differ from one another, and in general do not match the observed DTD shape, and do a poorer job at coming close to the actual rate numbers, compared the DTD predicted for double degenerate\footnote{The various authors were agnostic regarding the explosion mechanism for the DWD mergers in the DTD study.} systems. As discussed, building up the mass of a white dwarf toward the Chandrasekhar mass limit is a challenge for binary evolution. Other differences in the various code results are found to simply arise from different assumptions on the treatment of binary evolution physics \citep{toonen2014a}. While not a cause for concern in terms of numerical accuracy, the lack of agreement between different codes does makes it hard for the community to agree on which evolutionary channel(s) is/are most representative of single degenerate SNe~Ia that would be found in nature. 

\section{Near-Chandrasekhar mass SN Ia explosion models} 
\label{sec:Chandra}
\subsection{The central ignition problem}
\label{subsec:central_ignition}

The Chandrasekhar mass limit is the maximum mass limit (a mathematical singularity where the radius approaches zero) of an idealized (zero-temperature) equation of state for a non-rotating white dwarf star supported by electron degeneracy pressure of a relativistic Fermi gas \citep{chandrasekhar1931a}. The exact value of this mathematical mass limit depends on composition, with the mean molecular weight per electron fraction $\mu_e$ being the key parameter. The textbook value is $M_{\rm Ch} = \frac{5.80}{{\mu_e}^2} \msun\ $ \citep[see e.g.][]{clayton1968}. We note that for self-conjugate (nuclei with the same number of protons and neutrons) nuclear matter, such as $^{4}$He, $^{12}$C, $^{16}$O, or $^{24}$Mg,  $\mu_e$ is very close to 2, leading to the familiar $M_{\rm Ch} \approx 1.45$ \msun. Inclusion of neutron rich isotopes, such as $^{22}$Ne or $^{56}$Fe, leads to larger $\mu_e$ and thus to a smaller Chandrasekhar mass limit. 

Since this is a review, we note here that the sometimes portrayed story-line that \textit{a massive white dwarf accretes mass from a companion and when it reaches its limit of stability (the Chandrasekhar mass) it explodes} is incorrect. As a mathematical singularity of an idealized equation of state, an accreting white dwarf will not reach or exceed the Chandrasekhar mass, and if anything it should collapse to form an even more compact object, most likely a neutron star, which would result in a faint transient \citep{darbha2010a}. 

What is true however is that the realistic equation of state of massive white dwarfs becomes very soft. What this means is that the star is more easily compressed, thus
a small amount of additional mass leads to a large increase in compactness, and therefore a large increase in central density. For a composition rich in $^{12}$C, realistic conditions lead to ignition of carbon burning (in the intermediate thermo-pycnonuclear regime) for densities around the range of $2 - 6 \times 10^9\,\mathrm{g} \, \mathrm{cm}^{-3}$, see e.g. Fig.~5 of \citet{gasques2005a}. In this intermediate regime, where uncertainties surrounding the Coulomb screening of the reaction rate only add to the complexity, the nuclei contributing most to the fusion rate are either slightly bound or slightly unbound to the crystalline lattice (which is about to melt). 

In addition to the inherent uncertainties in the nuclear reaction rate, the ignition problem is further complicated through heating and cooling processes. For example, heat produced by surface accretion can have a feed-back effect on the core-temperature, whereas cooling (dominated by neutrino losses) can delay the ignition \citep[e.g.][]{lesaffre2006a}. Additionally, during the ``simmering" phase lasting some 100 -- 1000 years, nuclear reactions that lead up to the thermonuclear runaway already modify the initial composition driving it more neutron-rich \citep{piro2008c}. Furthermore, the impact of convective flows and turbulence on the birth of a thermonuclear deflagration flame \citep{hoeflich2002a, nonaka2012a} add considerable variance to the models as well. These effects all contribute to the complexity of the ignition problem of the deflagration. As a result, important parameters of the ignition remain uncertain, such as the central density when ignition occurs or the exact location of the ignition spot(s).  Over the last few decades, many groups have explored the implications of the (stochastic) ignition mechanism, including ignition conditions, configuration, and whether and how the initial subsonic deflagration flame can transition to a supersonic detonation, on the observables of SNe~Ia. 

In what follows, we first give an overview of the characteristic nucleosynthesis occurring in such near-Chandrasekhar mass white dwarfs. Then we describe different types of binary systems in which central ignition of near-Chandrasekhar mass WDs can be achieved as described above. We conclude this section with a range of different possible explosion mechanisms that have been proposed to occur following the (near central) ignition of the deflagration flame.

\subsection{Nuclear burning in Chandrasekhar mass white dwarfs}
\label{sec:nuclear_burning}

The explosions that are Type Ia supernovae are powered by the nuclear binding energy released that results from the transmutation of lighter, less-bound nuclear ``fuel" species, such as $^{12}$C, $^{16}$O, $^{22}$Ne, into heavier, more tightly-bound nuclear ``ash" species, such as $^{32}$S, $^{55}$Co, or $^{56}$Ni. The binding energy of $^{12}$C is 7.68 MeV per nucleon (MeV/nuc), the binding energy of $^{16}$O is 7.97 MeV/nuc, and the binding energy of $^{56}$Ni is 8.64 MeV/nuc. Hypothetical fusion of a 50/50 mix (by mass) of $^{12}$C and $^{16}$O to $^{56}$Ni thus would release about 0.82 MeV/nuc. Using this simple ``back-of-the-envelope" calculation, we see that completely burning a 1.4 \msun \, 50/50 CO WD into $^{56}$Ni would release:

\begin{equation}
\frac{1.4 \times 2 \times 10^{33} {\rm g \, (WD \, mass)}}{1.66 \times 10^{-24} {\rm g \, (mass \, per \, nuc)}} \rightarrow 1.67 \times 10^{59} {\rm nuc}\times 0.82 \,{\rm MeV/nuc} ,
\end{equation}

\noindent which is about $1.37 \times 10^{57}$ MeV, or $2.2 \times 10^{51}$erg of nuclear energy, which is less than the approximately $3\times10^{51}$erg of gravitational binding energy of such a massive WD. So one may ask -- how can the white dwarf actually become unbound from such an explosion? 

\begin{figure*}[ht]
\centering
    \includegraphics[scale=0.9]{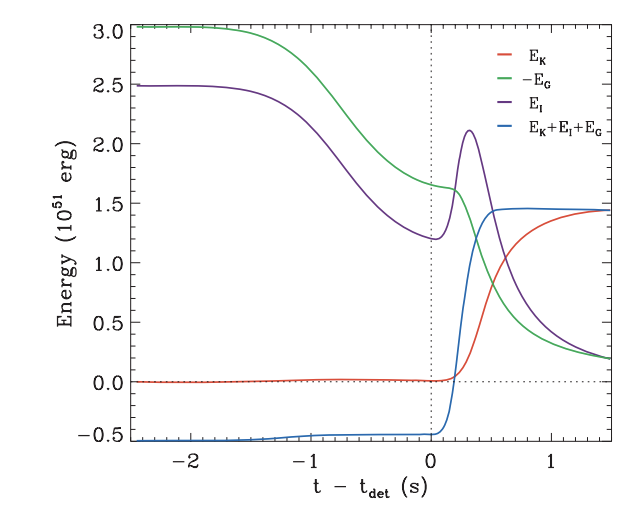}
\caption{The evolution of kinetic (red), internal (purple), and gravitational potential (green)
energies for a gravitationally-confined detonation model (see Sect.~\ref{sec:GCD}) of a Type Ia supernova explosion. For this model, the flame bubble was ignited 25 km from the centre of the star. Post-detonation (dotted vertical line), nuclear binding energy gets rapidly converted to internal energy, which in turn converts to kinetic energy, and the star blows up. Image reproduced with permission from  \citet{meakin2009a}, copyright by AAS.}
\label{fig:meakin}      
\end{figure*}

To get the full picture, we must also account for the internal energy, which is dominated by the internal energy of the degenerate electron gas and already accounts for a large fraction of the gravitational binding energy. The nuclear binding energy released by the explosive nuclear fusion reactions only needs to overcome the difference in initial gravitational binding energy and the internal energy. Kinetic explosion energies around the canonical value of $10^{51}$erg (1 Bethe) are thus readily achieved, even if only some fraction of the star is burned to the most tightly-bound iron group nuclear species, see e.g. Figure~\ref{fig:meakin}. As an aside, we note that the energy released by the thermonuclear fusion only powers the explosion but not the energy emitted by radiation -- the lightcurve -- which is powered by the radioactive decay of long-lived nuclear species produced in the explosion, such as $^{56}$Ni and $^{56}$Co at early and intermediate phases, and $^{57}$Co and $^{55}$Fe at late phases \citep{seitenzahl2009d}.

We have summarized above that it is the difference between the nuclear binding energy of the ``fuel" and the ``ash" that powers the explosion. But what determines the composition, and hence nuclear binding energy, of the ``ash"? For a fixed fuel composition, the fuel density at the time of ignition is the key parameter that determines the outcome of the explosive nuclear burning, since it sets the peak temperature reached. Broadly, we can categorize the outcomes into different burning regimes. At high densities, nuclear fusion reactions are so rapid (compared to the expansion timescale) that the composition can reach a state of nuclear statistical equilibrium (NSE) \citep[see][]{clifford1965a, woosley1973a, seitenzahl2009a}, dominated by iron-group elements. At lower densities, expansion and cooling leads to a ``freeze-out" of nuclear reactions before the burning is complete, leading to partially burned ash states dominated by intermediate mass elements, such as Si, and S \citep[e.g.][]{iwamoto1999a}. For fuel mostly composed of carbon and oxygen, we distinguish five different burning regimes (see e.g. figure 1 of \citealt{seitenzahl2017a}):
\begin{enumerate}
    \item at the highest densities, $\rho \gtrsim 10^9\mathrm{g}\,\mathrm{cm}^{-3}$: High-density allows efficient neutronization via electron captures, resulting in neutron-rich NSE, characterized by relatively high abundance of stable iron-group nuclei, such as $^{54}$Fe, $^{56}$Fe, or $^{58}$Ni (stable nickel). 
    \item at the next lower densities, but $\rho \gtrsim 2 \times 10^8\mathrm{g}\,\mathrm{cm}^{-3}$: ``normal" (low-entropy) freeze-out from NSE composition, dominated by $^{56}$Ni, and characterized by relatively high abundance of e.g. $^{55}$Co, which ends up as stable manganese. 
    \item at the next lower densities, but $\rho \gtrsim 2 \times 10^7\mathrm{g}\,\mathrm{cm}^{-3}$: ``alpha-rich" (high-entropy) freeze-out from NSE composition, dominated by $^{56}$Ni, and characterized by a paucity of e.g. $^{55}$Co and increased relative abundance of nuclei past $^{56}$Ni that are produced during the alpha-rich freeze-out, such as $^{64}$Ga and $^{66}$Ge (both decay to stable zinc), see e.g. Fig.~6 of \citet{lach2020a}.
    \item at the next lower densities, but $\rho \gtrsim 3 \times 10^6\mathrm{g}\,\mathrm{cm}^{-3}$: Carbon and oxygen burn, but the silicon-burning is incomplete. This density range is therefore characterized by large abundances of intermediate mass element isotopes such as $^{28}$Si and $^{32}$S, and in addition some iron group isotopes characteristic for incomplete silicon-burning, such as $^{55}$Co and $^{56}$Ni.
    \item at the next lower densities, but $\rho \gtrsim 3 \times 10^5\mathrm{g}\,\mathrm{cm}^{-3}$: At such low densities, carbon and neon still burn, leading to a net production of oxygen, which is already largely inert, and slightly heavier intermediate mass elements.
\end{enumerate}

We note that the transitions are not sudden and there is overlap between the different regimes. The densities given are only indicative and will vary somewhat for different explosion models (see e.g. figure 2 of \citealt{lach2020a}). For a review on how the detailed explosive nucleosynthesis is typically calculated from multi-dimensional hydrodynamical simulations via tracer particle methods see \citet{seitenzahl2023a}. 

Detonations in helium-rich fuel are still viable at even lower density, and they carry a different characteristic nucleosynthetic signature. Most conspicuously, $\alpha$-isotopes, such as  $^{36}$Ar, $^{40}$Ca, $^{44}$Ti, $^{48}$Cr or $^{52}$Fe are typically produced in much greater abundance compared to explosive carbon and oxygen burning. For introductions to the literature on explosive helium burning as it applies to thermonuclear supernovae \citep[see][] {timmes2000b,moore2013a}.

\subsubsection{Recent comparison of observational iron-group element masses and isotope ratios with explosion model predictions (Chandrasekhar vs. sub-Chandrasekhar)}
\label{subsubsec:fegroup}

 \begin{figure*}[ht]
    \includegraphics[width=\textwidth]{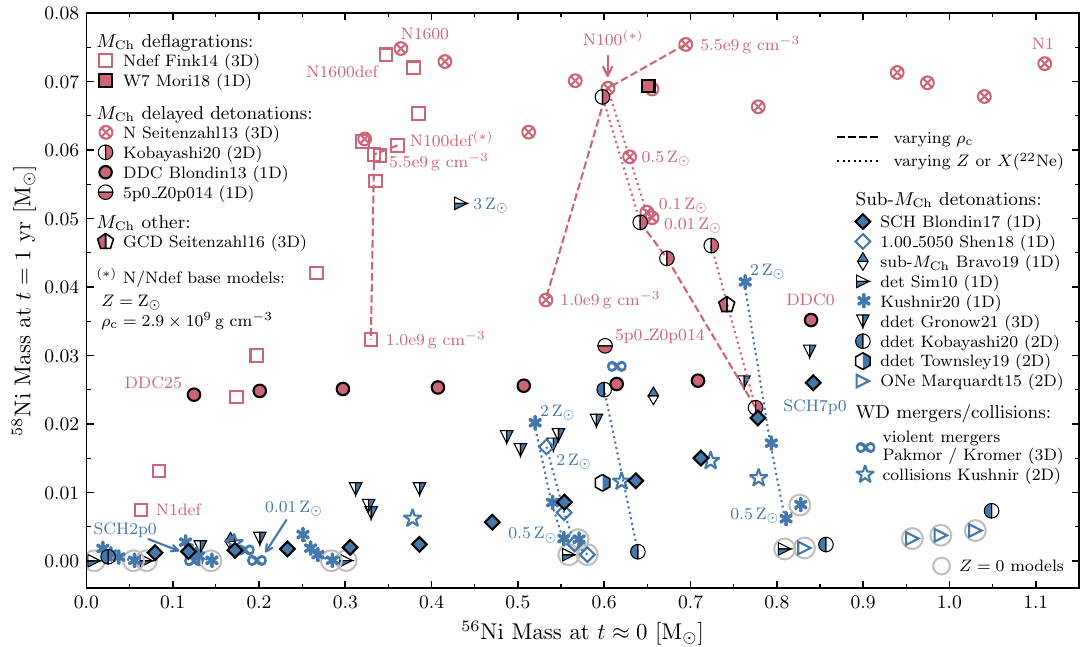}
\caption{Range of $^{56}$Ni (radioactive nickel produced in explosion) vs. $^{58}$Ni (decayed after 1 year; stable) for a number of simulated SN~Ia explosions from the literature. Chandrasekhar mass models are shown in red while sub-Chandrasekhar mass models are shown in blue. Image reproduced with permission from \citet{blondin2022a}, copyright by the author(s).}.
\label{fig:blondin}      
\end{figure*}

A recent study by \citet{blondin2022a} presented the amount of stable nickel produced (a year after SN explosion) as a function of the amount of radioactive nickel produced in the explosion for a wide variety of 1D, 2D and 3D models, including Chandrasekhar- and sub-Chandrasekhar mass simulations (encompassing multiple assumed progenitor channels). A general trend is found such that Chandrasekhar mass SNe~Ia tend to produce more stable nickel overall (with some exceptions), which would make for easier detection of forbidden nickel lines like [Ni~\textsc{ii}] in late time spectra of SNe~Ia for these systems (see Fig.~\ref{fig:blondin}). 

In another study, \citet{tiwari2022a} analyze trends in A=57 to A=56 and A=55 to A=56 production ratios for a diverse selection of explosion models from different research groups. They find clear trends of these ratios as a function of metallicity and primary WD mass. Notably, the model-to-model variance of the sub-Chandrasekhar model category is insufficient to account for the full range of observationally-inferred production ratios, even when allowing for significantly super-solar initial metallicities. 
\citet{tiwari2022a} conclude both near-Chandrasekhar and sub-Chandrasekhar models are required to explain the observationally inferred production ratios of these iron group isobars.

\subsection{Progenitor scenarios of near-Chandrasekhar mass explosions}
\label{sec:prog-MCh}

\subsubsection{Merging white dwarfs leading to near-Chandrasekhar mass WD ignition}

The double degenerate channel of two merging carbon-oxygen white dwarfs has been considered a plausible progenitor scenario for many years \citep{webbink1984a,iben1984a,han1995a,postnov2014a}. Until around 2010, it was assumed that in such a scenario the less massive (larger) degenerate dwarf would disrupt upon filling its Roche lobe and form a torus of material around the smaller, more massive dwarf, and this material would be accreted at a relatively high rate \citep[up to $10^{-5}$ \msun\, per year,][]{yoon2007a}. If enough mass is present in the donor white dwarf, the more massive white dwarf -- if allowed to continually accrete -- would approach the Chandrasekhar mass limit. What happens at this stage depends on several factors, namely the accretion rate behaviour over the course of mass transfer evolution and subsequent burning of the freshly-accreted material.  

One possibility is that the accreting white dwarf is `transformed' from a carbon-oxygen white dwarf to one that contains heavier elements, i.e. an oxygen-neon-magnesium white dwarf \citep{miyaji1980a,saio1985a,isern1991a,shen2012a}. This change in chemical make-up has a very important consequence: as an ONe(Mg) WD approaches the Chandrasekhar mass limit, it is more prone to electron captures on $^{24}$Mg and $^{20}$Ne which are produced in oxygen burning \citep[see][for more details]{jones2016a}. With electron capture rates increasing, degeneracy pressure loses its influence, and the ONe(Mg) white dwarf likely collapses to form a neutron star in what is called an accretion-induced collapse\footnote{Electron-capture supernovae \citep[ECSNe][]{nomoto1984c} are often likened to accretion-induced collapse supernovae because both involve a stellar core accreting mass leading to a core-collapse neutron star, and both explosions produce relatively low energies compared to SNe~Ia or core-collapse SNe \citep{dessart2006a}. ECSNe occur in stars with masses in the range between those that produce degenerate ONe white dwarfs and those that produce iron core-collapse neutron stars, and ECSN progenitors are referred to as Super-AGB stars \citep{doherty2017a}.} or more specific to this case -- a merger-induced collapse \citep{ivanova2008a,ruiter2019a,schwab2021a}. To be more confident of the outcome as to whether a massive ONe-rich white dwarf would definitely collapse to a neutron star or possibly explode, even partially, requires further study with multidimensional hydrodynamical simulations \citep[see][]{jones2019a}. 

On the other hand, if the CO white dwarf remains for the most part chemically unchanged and manages to approach the Chandrasekhar mass limit through accretion of its disrupted companion, it will explode, or possibly will accrete slightly above $1.4$ \msun\ given the appropriate physical conditions \citep{piersanti2003a,distefano2011a}. In the case where the Chandrasekhar mass is approached, it is believed that the central ignition would take place in the manner as outlined above in Sect.~\ref{subsec:central_ignition}, in essence in the same way as the ignition would occur in the core-degenerate or the single-degenerate Chandrasekhar mass scenarios discussed below. As mentioned in Sect.~\ref{sec:ejecta-masses}, explosion models involving super-Chandrasekhar mass white dwarfs have yet to show significant promise in reproducing the observed properties of the so-called super-Chandrasekhar SNe~Ia sub-class. 

Over the last dozen years or so, a number of sophisticated three-dimensional white dwarf merger simulations from different research groups have revealed that robust thermonuclear explosions can occur in white dwarfs that are well-below the Chandrasekhar mass limit (see Sect.~\ref{sec:subChandra}). Further, these explosions exhibit physical behaviour (ejecta velocities and nucleosynthesis) that are in agreement with those of observed SNe~Ia -- at least they are no worse than those models of exploding Chandrasekhar mass white dwarfs \citep{roepke2012a}. Since it has become clear in recent years that white dwarfs in mergers readily explode well before the more massive star has accreted much mass -- through the `double-detonation' rather than delayed detonation mechanism -- for the rest of this section we do not discuss double degenerate white dwarf mergers in the context of Chandrasekhar mass explosions; we return to mergers in Sect.~\ref{sec:subChandra}. Though our current understanding disfavours the Chandrasekhar white dwarf mass model in white dwarf mergers, we acknowledge that this could indeed be an incorrect assumption. 

\subsubsection{The core-degenerate scenario} 

It should be noted that the `core-degenerate scenario' \citep{sparks1974a,livio2003a,kashi2011a} is another potential, though less-studied \citep[but see][]{aznar-siguan2015a} progenitor scenario of SNe~Ia that involves the merger between a white dwarf and a relatively massive (non-degenerate) core of an evolved star during or following a common envelope episode. A central idea is that during the common envelope phase, not all material is ejected and ${\sim}$1--10\% of the envelope remains bound \citep{ilkov2012a} to the binary system. The resulting circumbinary disc can decrease the merger timescale between the AGB core and the white dwarf, which results in the (still hot) core becoming a relatively massive (possibly super-Chandrasekhar) object. We note here that binary star mergers have also been proposed as a channel for producing highly-magnetic, rather massive white dwarfs \citep{wickramasinghe2000a,bogomazov2009a}. In terms of explaining a large fraction of SNe~Ia, the core degenerate scenario has drawn a lot of criticism owing to the fact that SN~Ia explosions occurring during (or shortly after) a common envelope phase would likely exhibit tell-tale hydrogen features, and such features are only rarely seen in SNe~Ia spectra \citep[though there are exceptions, see][]{dilday2012a,kollmeier2019a}. It has been argued though that if a white dwarf $+$ AGB core star merger does result in an eventual explosion, there could be a range of plausible delay times (between merger and explosion) spanning up to ${\sim} 1$ Gyr because magneto-dipole radiation torques could delay the spin-down time of the newly-formed ${\sim} 1.4-1.5$ \msun\ white dwarf \citep{ilkov2012a}. Though this scenario is potentially promising \citep[see also][]{soker2023a}, and we indeed do expect these types of mergers to occur in nature, more numerical simulations of this particularly challenging scenario are required before we can say more with a high degree of certainty. 

\subsubsection{Single-degenerate near-Chandrasekhar mass scenario}
\label{sec:SDChandra}

For decades, it was assumed that all SNe~Ia occur in carbon-oxygen white dwarfs that have managed to approach the Chandrasekhar mass limit \citep{thielemann2004a} due to accretion from a stellar companion. The companion is often assumed to be a hydrogen-rich main sequence or evolved star (either in the Hertzsprung gap or a red giant). However, though less common, the companion could instead be a hydrogen-stripped, helium-burning star (see Sect.~\ref{sec:SD}).

The single-degenerate near-Chandrasekhar mass progenitor scenario for SNe~Ia -- regardless of how the explosion actually occurs -- is the one that was made popular in numerous textbooks. It seemed attractive at first glance by offering a simple explanation as to how to evolve a binary system to ignition of an otherwise inert white dwarf, via mass accretion from a Roche-lobe filling non-degenerate companion onto a massive white dwarf primary. This scenario always leads to ignition of a more or less bare primary white dwarf just below the Chandrasekhar mass. This common mass was further seen as a natural explanation for the observation that many SNe~Ia (at least historically) look similar to each other in terms of their lightcurve behaviour and general spectral properties \citep[see][for some discussion on formation scenarios and birthrates]{han2004a}. 

One well-known issue with the canonical single-degenerate scenario of SNe~Ia is the challenge of building up the mass of the white dwarf to the Chandrasekhar mass through accretion of hydrogen \citep{sutherland1984a,wheeler1990a}. 
The issue is that the mass transfer rate has to be just right for the WD to gain mass, and this changes as the WD grows in mass, which appears as a fine-tuning problem \citep{han2004a,nomoto2018a}.
There is a small parameter space over which fully efficient hydrogen-rich burning, and thus sustained mass gain, on a CO WD can occur. This results in a rather narrow region in {\em white dwarf mass -- mass accretion rate} parameter space over which fully efficient burning is possible. Outside of this parameter space the WD will experience thermally unstable burning resulting in novae for low accretion rates, or significant expansion of the WD atmosphere and optically thick winds for high accretion rates \citep[the so-called `red giant' configuration,][]{prialnik1995a,nomoto2007a}. Potentially, the accreting WD may even experience mass erosion through nova eruptions in which case it would never attain near-Chandrasekhar mass \citep{patterson2017a}.

While the symbiotic channel has previously been investigated -- in which a white dwarf accretes material from the wind of an evolved star \citep{munari1992a}, it was determined that such a binary configuration was unlikely to lead to the formation of a Chandrasekhar mass white dwarf over the course of the short lifetime of the symbiotic \citep{kenyon1993a,hachisu2012a}, though such systems might contribute some small fraction of SN~Ia events to the sub-Chandrasekhar mass scenario. However, given the existence of systems like RS~Oph, which is estimated to host a white dwarf close to the Chandrasekhar mass limit, perhaps the jury is still out on the symbiotic question.

Now that we have broadly introduced the main binary star configurations (see also Sect.~\ref{sec:progenitors}) and discussed pathways to central ignition of near-Chandrasekhar mass models, we next give a summary of the different explosion models in the context of explosions in modern Chandrasekhar mass WD models.  

\subsection{Explosion mechanisms} 

The near-Chandrasekhar mass models discussed in this section all start with the ignition of a deflagration (sub-sonic burning) near the centre of the WD (see Sect.~\ref{subsec:central_ignition}). In the next Sect.~\ref{sec:faileddef}, we present the characteristics of models where all of the nuclear burning occurs only in a deflagration. As discussed there, predictions of state-of-the-art 3D pure deflagration models proved to be irreconcilable with many features of normal SNe~Ia, although they remain good candidates for certain sub-luminous sub-types of SNe~Ia, in particular the 2002cx-like SNe. Both pure detonation models, which burn way too much of the WD to IGEs (see Sect.~\ref{sec:20thcent}), and pure deflagration models, which, for realistic ignition conditions, burn far too little of the WD to IGEs, are therefore ruled out for the bulk of SNe~Ia. Phrased in this way, it is not surprising that the attention turned to models that include a combination of the two: an initial deflagration phase that in one way or another transitions to a later detonation phase.  

It was \citet{khokhlov1991a} that showed that an initial expansion of the white dwarf during a deflagration phase, followed by a transition of the mode of burning to a detonation at a time when the density of the unburned fuel had decreased to allow for production of intermediate mass elements, resulted in better agreement with observations. We refer to such models that pre-expand the star through an initial deflagration phase before detonating it when the central density has been lowered as ``delayed-detonation models". A few different main variants of delayed-detonation models exist: deflagration to detonation transition (DDT), gravitationally confined detonation (GCD), and pulsating reverse detonation models (PRD). We discuss these different variants in turn in the Sects.~\ref{sec:DDT} -- \ref{sec:PRD}, following immediately the description of pure deflagration models in the next section.

\subsubsection{Pure deflagrations (failed detonations)} 
\label{sec:faileddef} 

It became very clear in the late 1960s that if the near-Chandrasekhar mass models were to work, the initial ignition had to occur as a subsonic deflagration (see Sect.~\ref{sec:20thcent}). The successes of parameterized 1D fast-deflagration models -- most prominently the W7 model of \citet{nomoto1984a} -- in reproducing key observable features of normal SNe~Ia, led to a focus of the thermonuclear modelling efforts on pure deflagration models for the next 20 -- 30 years.

1D models however are un-physical, in that they do not capture the all-important effects of buoyancy, turbulence, and the Rayleigh--Taylor instability on the evolution of the flame front. 
In 1D, a deflagration ignited in a small central region simply burns outward in mass-coordinates and there are no pockets of unburned fuel left at high density. This produces an artificially large fuel consumption rate and leads to more energetic and luminous explosions than what is physical for the initial model. In 3D by contrast, a deflagration ignited in a small volume near the centre of the WD will be buoyant, since the energy release of the nuclear fusion reactions resulted in a small expansion and lower density of the hot ignition spot compared to the colder surrounding material. The bubble starts to float radially outwards in one direction towards the surface, accelerating and growing in size driven by the interaction of the flame front with the turbulent fluid flows and hydrodynamical instabilities \citep[see e.g. the review by][]{roepke2018a}. Consequentially, it was found that single spot ignition models only produced weak and faint explosions, unable to explain normal SNe Ia \citep[e.g.][]{reinecke1999b}.

To obtain more energetic and luminous explosions in 3D models without suppressing buoyancy and hydrodynamical instabilities, central and multi-spot ignition models were introduced, which lead to a faster initial growth of the flame surface and a more complete incineration of the high density interior. This results in explosion models that produce a few tenths of a solar mass of $^{56}$Ni and completely unbinds the white dwarf \citep[e.g.][]{gamezo2003a, gamezo2004a, roepke2005b, roepke2006a, roepke2007c}. 

It is worth noting that pure deflagration models produce high ratios of stable iron group isotopes (e.g. $^{54}$Fe, $^{58}$Ni) relative to radioactive $^{56}$Ni and also high [Mn/Fe]. The reason for this is that the contribution of the detonation, which in DDT models produces several tenths of a solar mass of $^{56}$Ni and intermediate mass elements, is missing. This occurs since the detonation in DDT models predominantly burns at densities where neutronization via electron captures is largely negligible and NSE is established in the high-entropy regime 
(see Sect.~\ref{sec:nuclear_burning}).

There was mounting evidence that the mixed ejecta of pure deflagration models are not in agreement with the more layered ejecta profiles inferred for normal SNe~Ia \citep[see e.g.][]{stehle2005a, mazzali2007a} and that in any case the ignition of the deflagration is most likely occurring in a single, off-centre spot \citep{zingale2009a, nonaka2012a}. Moreover, it was soon demonstrated that the turbulently mixed profiles of such pure deflagration models exhibit characteristics (e.g. colours, spectra) that are not reconcilable with observations of normal SNe~Ia \citep[see e.g.][]{fink2014a}. The pure deflagration model thus ceased to be a contender as a viable explosion model for normal SNe~Ia, and the focus for the near-Chandrasekhar mass ignition paradigm shifted to models where the sub-sonic deflagration transitions to a super-sonic detonation, in one way or another (see the following sub-sections).

However, \citet{jordan2012b} and \citet{kromer2013a} independently realized that, while too mixed and insufficiently energetic and luminous to explain the normal SNe~Ia, single-spot, off-centre ignited pure-deflagration models are excellent candidates to explain the sub-luminous sub-class of 2002cx-like SNe of the SN~Iax class. Notably, these models fail to unbind the whole white dwarf and they leave compact, high-velocity WD remnants behind that are polluted with some of the ashes of the thermonuclear fusion reactions \citep[see e.g.][]{fink2014a}. 

Off-centre, single-spot ignited pure-deflagrations in `hybrid' CONe WDs can eject even less mass and produce even fainter events, with the explosion model of \citet{kromer2015a} producing only $3.4\times10^{-3} \msun$ and providing a good match to SNe~Iax at the faint end of the SN~Iax luminosity range, such as SN~2008ha. Taking this even further, such partial explosions may even be possible in ONeMg WDs \citep[see e.g.][]{jones2016a}, where the high central densities near $10^{10}\,\mathrm{g}\,\mathrm{cm}^3$ lead to high neutronization rates and an over-production of neutron-rich Fe-group isotopes \citep[e.g.][]{jones2019a}. 

The chemically peculiar remnants that are left behind by such incomplete explosions or ``thermonuclear eruptions" may already have been identified; see e.g. \citet{raddi2019a} for plausible high-velocity stellar remnants and \citet{zhou2021a} for a plausible Galactic supernova remnant.

\subsubsection{Deflagration-to-detonation transition models}
\label{sec:DDT}
 Often referred to as DDT models, deflagration-to-detonation transition (DDT) models used to be the most widely-accepted explosion models that can possibly explain normal (and maybe some abnormal) SNe~Ia \citep{hillebrandt2000a}. In such DDT models, the ignition of the deflagration occurs as described above. The key ingredient for DDT models is then the spontaneous transition of the subsonic mode of nuclear burning (deflagration) to a supersonic mode of burning -- a detonation. DDTs are frequently observed on Earth, such as the ``knocking" in the cylinders of combustion engines. In these terrestrial cases, the DDTs are enabled by the walls of the confinement vessel. Stars do not have walls, however, and the problem of DDTs in an unconfined medium is more complicated. The basic idea, which goes back to \citet{zeldovichbook}, is that a gradient in the induction times (auto-ignition delay-times to thermonuclear runaway) leads to a shock-formation and a detonation. \citet{lee1978a} refined this basic picture and the Shock-Wave-Amplification-through-Coherent-Energy-Release (SWACER) mechanism was born: sound waves created by small volumes of fuel that ignited propagate outwards, these sound waves reach other neighbouring volumes of fuel that by the time of their arrival are also igniting. With an appropriate spatial-gradient in these ``induction" or auto-ignition timescales, the sound waves amplify and the pressure pulse can steepen into a shock, that for the right gradients can transform into a detonation, see e.g. \citet{seitenzahl2009b} for a determination of critical induction time gradients relevant to the SN~Ia problem.
The mechanism of how a suitable induction-time gradient can be set up in a deflagration involves mixing hot ash with cold fuel, when the deflagration enters the distributed burning regime, and strong turbulence, which intermittently produces small volumes with relatively large velocity (and temperature) fluctuations \citep[e.g.][]{khokhlov1991c,niemeyer1999a,lisewski2000a, woosley2007a, schmidt2010a, ciaraldi2013a}. For alternative mechanisms to transition to a detonation see also \citet{fisher2019a} and \citet{poludnenko2019a}.

\citet{khokhlov1991b} showed that the expansion of the star prior to the detonation results in the desired nuclear burning at densities where IMEs such as Si and S are synthesized. Early one-dimensional simulations \citep[e.g.][]{hoeflich1995a, hoeflich1996a} gave promising comparisons of the models against observations, however, key physical processes in the DDT explosion mechanism such as buoyancy, convection, and the turbulent cascade are inherently three-dimensional, requiring three-dimensional full-star simulations for a self-consistent approach to the problem \citep[see e.g.][]{pakmor2024a}. \citet{gamezo2005a} showed first in a three-dimensional simulation that a period of sub-sonic burning (a deflagration) in an expanding white dwarf followed by a supersonic detonation led to kinetic energies that were on par with what was observed for typical SNe Ia (${\sim} 1.3-1.6 \times 10^{51}$ erg), in contrast to the pure deflagration models which produced kinetic energies only about half of this range. Further, the resulting nucleosynthesis agreed more readily with observations compared to pure deflagration models, which would leave intermediate mass elements (i.e. oxygen, carbon) unburned near the white dwarf centre.

Parameterized model grids initially indicated that the width-luminosity relation may be recovered by DDT models \citep{kasen2009a}. However, \citet{sim2013a} demonstrated with 3D Monte Carlo radiative transfer calculations of the 3D model grid of \citet{seitenzahl2013a} that DDT models tend to lie orthogonal to the width-luminosity relation, in spite of giving relatively good spectral matches to some observed SNe~Ia. Varying secondary parameters, such as the carbon fraction, do not appear to change the orthogonality of the DDT models to the width-luminosity relation \citep{ohlmann2014a}.

\subsubsection{Gravitationally confined detonation models} 
\label{sec:GCD}

Another variant of near-Chandrasekhar mass explosion is the so-called ``gravitationally confined detonation'' (GCD) \citep{plewa2004a}. 
In GCD models, a deflagration ignites in a single bubble, typically some tens of km offset from the centre. Buoyancy drives the hot bubble towards the surface. As it grows in size and complexity due to the Rayleigh--Taylor instability, the rising plume accelerates to super-sonic speeds. Near the surface of the WD where the ambient pressure is sufficiently low, the hot ash can eventually spread laterally in all directions across the surface. While some material along the initial ignition axis reaches escape velocity, most of the hot ash material remains confined (bound) to the star as it rapidly spreads outwards, which is the origin of the somewhat misleading name ``gravitationally confined detonation'' scenario. The hot ashes then ``collide'' at the anti-podal point opposite to where the plume of ash first broke out of the star. Nuclear fuel in the collision region may be sufficiently compressed and heated to trigger a detonation \citep[see e.g.][]{seitenzahl2009b, seitenzahl2009c}. Since the detonation is triggered only after energy release of the deflagration has allowed the WD to expand, this is also a type of delayed-detonation model. 

 The weak deflagrations arising from single ignition point models only allow for moderate expansion of the WD before the detonation is triggered. The further out the initial ignition spot is, the faster the plume rises (weaker deflagration phase) and the more compact the white dwarf when it detonates, leading to more $^{56}$Ni and bright events. Moving the ignition spot closer to the centre reduces the initial buoyancy, allowing the deflagration to burn more mass. This increased energy release leads to more expansion and a weaker collision of the ash after break-out. For bubbles ignited close to the centre, a detonation becomes increasingly unlikely, limiting GCD models to the brighter end of the SN~Ia distribution \citep{fisher2015a, byrohl2019a}. The WD is thus still rather compact (relatively high density) when it detonates, which is the reason why these models are expected to be rather bright \citep{fisher2015a}, producing a lot of $^{56}$Ni and relatively small amounts of IMEs. \citet{meakin2009a} calculated the first detailed nucleosynthesis for single ignition point (2D) GCD models, which indeed exhibit a large IGE/IME ratio. However, since only very little material is burned in the deflagration at the highest densities where electron captures drive the material neutron rich, compared to DDT models these GCD models also produce lower Mn/Fe and relatively fewer neutron-rich IGE. With a 3D simulation, detailed nucleosynthesis, and 3D Monte Carlo radiative transfer calculations, \citet{seitenzahl2016a} demonstrated that their GCD models do not match to normal SNe~Ia. Further, although there are intriguing similarities with the 1991T-like SNe~Ia, their models have a little too much high-velocity IGE and not enough low-velocity stable iron.

\citet{jordan2012a} first introduced the pulsationally assisted gravitationally confined detonation variant of this class of explosion mechanisms. They ignited the deflagration more vigorously in multiple overlapping spots. As expected, the greater energy release leads to more expansion and the colliding ashes initially fail to produce detonation conditions. However, the energy release is too small to unbind the WD. After reaching a maximum radius, the WD contracts again and the additional compression and heating during the contraction phase leads to a detonation.

\citet{lach2022a} looked further into pulsationally assisted gravitationally confined detonations for a range of different ignition conditions. They produce models with $^{56}$Ni masses from 0.257 to 1.057 \msun. Comparing their model spectra with observations, they conclude that although the models cover a range of $^{56}$Ni masses all the way from sub-luminous to super-luminous SNe~Ia, only the 1991T-like sub-type provides a good match to the bright end of the model distribution.

\subsubsection{Pulsational reverse detonation models} 
\label{sec:PRD}
The pulsating reverse detonation (PRD) model \citep{bravo2006a} is another variant of a delayed-detonation. Similar to the GCD models discussed in the previous section, the explosion begins with a (weak) deflagration that fails to unbind the star. The energy input however excites a pulsational mode. When the white dwarf contracts again, an accretion shock forms as the expanded outer layers fall back onto the CO core. The conversion of the pulsational kinetic energy may sufficiently compress and heat the underlying material to initiate a detonation \citep[see][]{bravo2009a}.
Three-dimensional explosion simulations of the PRD variant show that energetic explosions with radioactive $^{56}$Ni masses that match the brighter SNe~Ia can be obtained \citep{bravo2009b}. However, this subclass of explosion models has deflagration ash at high velocity, and therefore large amounts of iron-group elements in the outer layers. The location of these iron-group elements in the ejecta leads to quite red colours and the predicted spectra are largely not in agreement with normal SNe~Ia \citep[e.g.][]{baron2008a}, although some viewing angles are better than others \citep{bravo2009b}.
Compared to e.g. DDT models, there is also a lot more unburned carbon, which leads to a C~II feature not typically seen in normal SNe~Ia near maximum light \citep{baron2008a, dessart2014a}. Moreover, the relatively small mass burned at high density by the deflagration means that for the \citet{bravo2009b} model even [Mn/Fe] is sub-solar. Overall, the PRD models are therefore not great candidates for SNe~Ia.

\subsubsection{Alternative models} 
There are some alternative explosion models in the literature that are less main-stream than the models discussed above, but nevertheless worth mentioning here. For example, it was pointed out by \citet{horowitz2021a} that actinides, including radioactive uranium, should preferentially crystallize in cooling massive white dwarfs first, owing to their high melting points. The authors entertain the idea that critical masses of radioactive actinides nuclei could condense, giving rise to a chain-reaction and ultimately fission-ignited thermonuclear supernovae,  \citep[see][]{deibel2022a, horowitz2022a}. 
In a modification of the near-Chandrasekhar mass explosion models, \citet{leung2015a} and \citet{chan2021a} considered the effect of white dwarfs with dark-matter cores on the explosion properties, which generally lead to fainter but slower declining events. The energy released by annihilating dark matter particles in collapsing dark matter cores has even been hypothesized as a possible alternative ignition scenario for the nuclear fuel \citep{janish2019a}. 

\section{Sub-Chandrasekhar mass SN Ia explosion models} 
\label{sec:subChandra} 

\subsection{Historical context}

As discussed in the previous chapter, the problem of building up the mass of a white dwarf toward the Chandrasekhar mass via mass-transfer is a well-known one. An advantage of sub-Chandrasekhar (\msub) WDs is that nature finds it easier to make them \citep[see e.g.][]{torres2021a}. 
The idea that Type Ia supernovae could originate through a double-detonation mechanism in \msub\ WDs was explored many decades ago \citep{woosley1980a,nomoto1982b}, and significant work with hydrodynamical simulations involving helium shell detonations on \msub\ mass WDs started to become more common-place throughout the 1990s. Explosive nucleosynthesis calculations and synthetic lightcurves indicated that \msub\ mass explosions should be taken seriously as viable SN~Ia progenitors. Though it was found that lower-mass WDs that explode via double-detonation are not capable of producing very large amounts of iron, they indeed produce large amounts of V, Cr, and Ti \citep{woosley1994a,livne1995a}. Two-dimensional models showed fairly good agreement with four real SN~Ia explosions: the well-known super-luminous and sub-luminous SN~1991T and SN~1991bg, respectively, as well as SN~1989B and SN~1992A \citep[see also][]{woosley1995a}. A main issue however was that these sub-Chandrasekhar mass explosion models from the 1990s involved rather massive helium-rich shells (${\sim 0.15 - 0.2}$ \msun) acquired via accumulation from a stellar companion, sitting on the WD surface. The second detonation deeper within the WD -- occurring at higher density -- was necessary to achieve any reasonable nucleosynthetic outcome resembling SNe~Ia. But when the first detonation occurs in a massive helium shell, helium-burning leads to a rather large amount of iron-group elements being synthesized, with too much iron-group material at high expansion velocities. A key issue with an abundance of IGEs in the outer layers is that fluorescence in elements like titanium and chromium redistributes blue photons to redder wavelengths and consequently, synthetic B-V colours are generally too red at peak and do not satisfy observational constraints \citep[e.g.][]{kromer2010a}. It was confirmed with the advent of more sophisticated numerical models including 3D hydrodynamics, that such massive (or `thick') helium shells were not actually necessary to achieve a double-detonation, putting the \msub\ mass double-detonation scenario back on the table as a promising progenitor candidate \citep{sim2010a,kromer2010a,fink2010a,woosley2011b,shen2014a}. 

In this section, we discuss \msub\ mass explosions in two parts: those where mass transfer is dynamically-driven (i.e. unstable), and those where the binary is undergoing stable mass transfer at the time of explosion; binaries involving a \msub\ mass white dwarf that gathered mass either through RLOF, or even through winds, can potentially produce Type Ia supernovae. Both types of double-detonation configurations -- dynamically unstable or stable mass transfer -- require different modelling approaches. 

`Dynamically-driven' (dynamically unstable mass transfer) implies that the rate at which mass is transferred between the stars leading up to the supernova proceeds on a dynamical timescale (rather than on a longer nuclear or thermal timescale). In the case of two white dwarfs of similar mass, mass transfer will be unstable and proceed rapidly and the stars will most certainly merge quickly \citep[see][]{marsh2004a}. 
On the other hand, for binaries undergoing sustained RLOF, the mass gain rate, which is dependent upon the orbital configuration of the binary, determines whether the material captured by the accretor will undergo nuclear burning on the white dwarf or whether material will simply accumulate on the WD surface, building up the mass of the helium `shell'. In a nut`shell', if enough helium-rich material is allowed to avoid burning and simply accumulate on the \msub\ WD at a relatively low rate \citep[${\sim}10^{-8}$ \msun yr$^{-1}$, i.e.][]{hashimoto1986a,piersanti2014a}, a detonation will be initiated in the recently-accumulated helium-rich shell.  
This first detonation will subsequently trigger a second detonation in the central regions of the white dwarf, thus destroying the star. Such a mechanism is referred to as a `double-detonation', and for about 4 decades has been the principle explosion mechanism thought to be responsible for \msub\ mass SNe~Ia.\footnote{Some recent simulations have explored the possibility of each star experiencing if not two, a minimum of one detonation each for a total of 3 or 4 detonations in the binary \citep[][see their table 2 for subsequent abundance computations]{tanikawa2019a,pakmor2022a}, but we do not discuss these systems here.} In this context, semi-detached binaries undergoing non-dynamical mass-transfer have canonically been assumed to potentially undergo double-detonations under the right conditions, and we will discuss these systems first. 

\subsection{Non-dynamical mass transfer} 

Population synthesis calculations have indicated that thermonuclear supernovae triggered from helium-shell detonations in mass-transferring binaries in which a white dwarf steadily accumulates mass from a helium-rich star\footnote{Steady here does not mean fully efficient. During stable mass transfer some fraction of the mass lost by the donor will be captured by the companion, either through an accretion disc or via direct accretion or accumulation onto the star, and this fraction of material that is burned (or accumulated) need not be close to 1, and likely will change with time as the system evolves.} could match the empirically-derived SN Ia rate of the Galaxy, which is $0.54 \pm 0.12$ per century; 
see \citet{li2011c} \citep[see also][]{tutukov1996a,cappellaro1997a}. 
While symbiotic binaries involving a white dwarf accreting hydrogen (which burns to helium) have been considered as potential candidates for the \msub\ scenario \citep{kenyon1993a}, such a configuration -- an accreting white dwarf with both a thin helium and hydrogen shell -- leads to frequent unstable burning and the probability for shell detonations leading to a WD detonation before the evolved donor has left the giant phase is unlikely \citep[see also][in the context of hydrogen and helium novae]{kemp2021a}.   
When considering binaries with WD accretors and corresponding predicted delay time distributions and rates for double-detonation supernovae \citep{ruiter2011a,wang2013a,ruiter2014a}, favoured binary star configurations include a CO white dwarf gaining material via RLOF from either a helium-rich white dwarf or the helium-rich core of a previously `regular' star that has had its envelope removed through binary interactions\footnote{Such objects can encompass a range of level of degeneracy and are often referred to in various ways in the literature: naked helium star, helium main sequence star, hydrogen-stripped helium-burning star, (semi-)degenerate helium core, etc.}.  
The AM~CVn systems are a well-known sub-population of cataclysmic variables in which a (CO) white dwarf accretes from a small helium-rich star \citep{warner1995a}. For some time AM~CVn systems were considered as plausible progenitors of some fraction of the SN~Ia population \citep{solheim2005a,bildsten2007a}. However, theoretical estimates of AM~CVn birthrates seem to be over-estimated in population synthesis studies, since predictions of the AM~CVn population numbers were typically an order of magnitude above those observationally-determined values, albeit with limited sample size \citep{nelemans2001a,roelofs2007a}.

Galactic AM~CVn binaries, of which more than 55 are known \citep{ramsay2018a}, will be crucial targets for future space-based gravitational wave detectors like {\em LISA} \citep{kupfer2018a,amaro-seoane2023a}, given their exceptional capability as multi-messenger (GW and electromagnetic) sources. Though earlier works concluded the donor star was likely a degenerate object and thus the `double white dwarf' formation scenario for AM~CVn systems was the most popular of the three main proposed scenarios, recent, detailed studies have demonstrated that at least some (if not the majority) of AM~CVn donor stars are actually non-degenerate, or only partially degenerate \citep[see][]{green2019a}. The majority of AM~CVn systems appear to harbour rather `run of the mill' CO WD accretors in terms of mass, with very low-mass companions, thus they are not deemed as particularly likely SN~Ia progenitors, even for the sub-Chandrasekhar mass events. However, a sub-set of AM~CVn-like binaries involving more massive WDs accreting from subdwarf B/O stars \citep{heber2016a} could potentially reach sufficient physical conditions to produce double-detonations, though the SN~Ia birthrate of this channel is expected to be rather low \citep[see also][for discussion of HD265435, a binary consisting of a white dwarf and a hot subdwarf]{pelisoli2021a}. It was recently estimated that such systems could be responsible for a rather small fraction ($\lesssim 1$ \%) of Galactic SNe~Ia \citep{zhou2014a,neunteufel2022a}.

\subsubsection{Double-detonation models in binaries undergoing non-dynamical mass transfer} 

In binary evolution physics, the inter-connection between changing orbital quantities (masses and mass ratio, separation, eccentricity) and evolving stellar properties (phase of stellar evolution e.g. size and density; composition) make predicting final states from initial states extremely challenging -- even for a simple stellar population having one metallicity! Rapid binary population synthesis studies, simulating between ${\sim}10^{5} - 10^{9}$ stars in one go, have greatly improved our ability to set limits on which binary star configurations plausibly produce a Type Ia supernova in the first place, thereby guiding 
computationally expensive hydrodynamical and radiative transfer simulations that focus on the most crucial part of the evolution: pre-explosion and/or explosive phase \citep[see also][]{neunteufel2016a}. 

In detailed hydrodynamical simulations of double-detonations in binaries undergoing stable mass transfer, the donor itself is generally not simulated. 
Such simulations offer an approach about the binary configuration and thus progenitor scenario. Despite the lack of information about binary interactions preceding the detonations in these models, such simulations can be indicative of what might occur in the case of a WD merger (see next sub-section). 

\begin{figure*}[ht]
\centering
    \includegraphics[width=\textwidth]{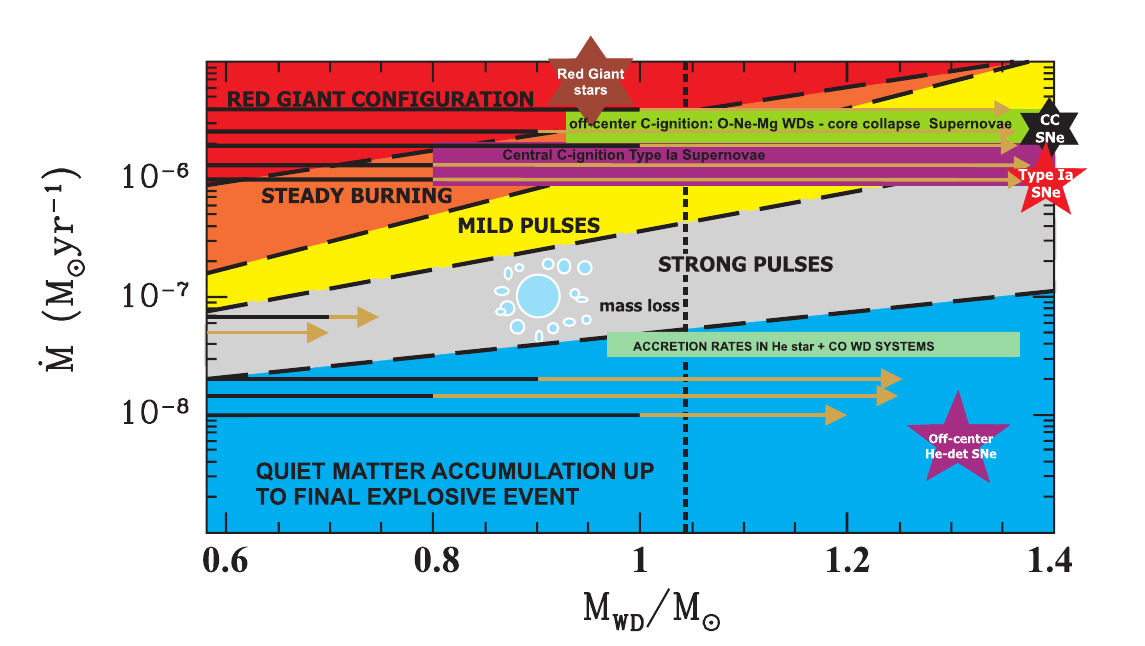}
\caption{Rate of accretion of helium as a function of white dwarf mass with various accretion regimes. In the Red Giant regime (red) the white dwarf cannot accept the large amount mass transferred and thus develops a red giant-like envelope \citep[i.e.][]{nomoto2007a}. For slightly lower accretion rates, nuclear burning on the white dwarf becomes fully efficient (orange region), and the star may become massive enough to explode as a near-Chandrasekhar mass WD. For even lower accretion rates (yellow and grey), mild or strong helium flashes develop \citep{kato2003a}, thereby decreasing mass retention efficiency; it is not fully clear whether accreting white dwarfs in this regime will erode in mass or ultimately gain mass \citep[see][for the hydrogen-accretion context]{starrfield1985a}. Finally, for rates below ${\sim}$ few$\times 10^{-8}$ \msun\ yr$^{-1}$, the build-up of a helium shell (no flashes) is possible, which may lead to a detonation in the helium shell once sufficient conditions are met \citep{fink2010a}. Image reproduced with permission from \citet{piersanti2014a}, copyright by the author(s).}
\label{fig:acrreg}      
\end{figure*}

The amount of helium that is collected on the WD surface before the initial helium shell ignition is dependent on the mass of the underlying white dwarf \citep{iben1989a,shen2009a}. If the helium shell ignites but fails to drive a second detonation deeper in the star, the result would be a so-called .Ia  \citep[\emph{`dot one Eh',}][]{bildsten2007a}. A potential observed counterpart of such a transient could be SN2002bj \citep{poznanski2010a}. As mentioned in Sect.~\ref{sec:SDChandra}, a white dwarf accreting helium-rich matter can on the other hand approach the Chandrasekhar mass limit, thus surpassing the opportunity to explode as a \msub\ event. The overall outcome (either Chandra or sub-Chandra) will depend heavily on the accretion rate, with higher rates leading to hotter envelopes that facilitate nuclear burning of helium, either stably or via a series of helium-shell flashes \citep{kato2004a}. \citet{piersanti2014a} carried out a comprehensive investigation of helium-accreting white dwarfs, taking into account the thermal response of the accretor using a modified version of the FRANEC evolutionary code \citep{chieffi1998a}. Figure~\ref{fig:acrreg} depicts a summary of their findings, which breaks down He-accreting WD systems into various regimes. Chandrasekhar mass white dwarf explosions may be achieved for relatively high accretion rates when the white dwarf undergoes stable burning, while \msub\ explosions are only achievable for accumulation rates on the order of ${\sim}$ few$\times 10^{-8}$ \msun\ yr$^{-1}$ or less. 

The overall picture is that the initial detonation is ignited in a helium-rich layer that has been convectively burning helium for a few days. For vigorous convection, conditions can be such that at the inflow of a convective shell a He-detonation can ignite \citep{glasner2018a}. This helium detonation in turn creates a propagating shock wave that travels through the core, closes in on itself, and ultimately ignites a detonation in the core. 
Where exactly the helium shell detonation takes place, in terms of scale height and geometry, and then how and precisely where the second detonation in the white dwarf occurs, are still not completely clear \citep{moll2013a}. When a helium shell detonation, which could arise by compressional heating during mass accumulation, leads directly to a second detonation at the shell-WD core interface, the mechanism is often referred to as an `edge-lit' detonation \citep{livne1990b,sim2012a}. However, another, possibly more favoured model occurs when the initial helium detonation shock wave does not immediately initiate a detonation near the shell-core boundary, but rather travels through the star and converges somewhere off-centre on the far side of the WD (on the opposite side of the initial ignition spot), resulting in a second detonation close to the WD centre \citep{livne1990a}, where densities are rather high ($\gtrsim 10^{7}$ g cm$^{-3}$). Such a `convergent shock scenario' for double-detonations with rather small helium shell masses have shown to be extremely promising when comparing synthetic lightcurves and spectra of those with observed SNe~Ia \citep{townsley2012a}. Using high-resolution 3D hydrodynamical simulations with Arepo \citep{pakmor2016a}, \citet{gronow2020a} found that even before the convergent shock mechanism occurs, double-detonations can already occur via the `scissor mechanism' \citep[see also e.g.][]{livne1995a,garcia1999a,forcada2006a}, in which a carbon detonation is ignited during the convergence of the detonation wave at the base of the helium shell. Such a mechanism highlights the importance of mixing in (3D) simulations: helium shells that contain some fraction of carbon not only produce synthetic observables that are in better agreement with observations \citep{kromer2010a,shen2014a}, but enhance the He burning rate thus leading to stronger shocks.

\begin{figure*}[ht]
\centering
    \includegraphics[width=0.9\textwidth]{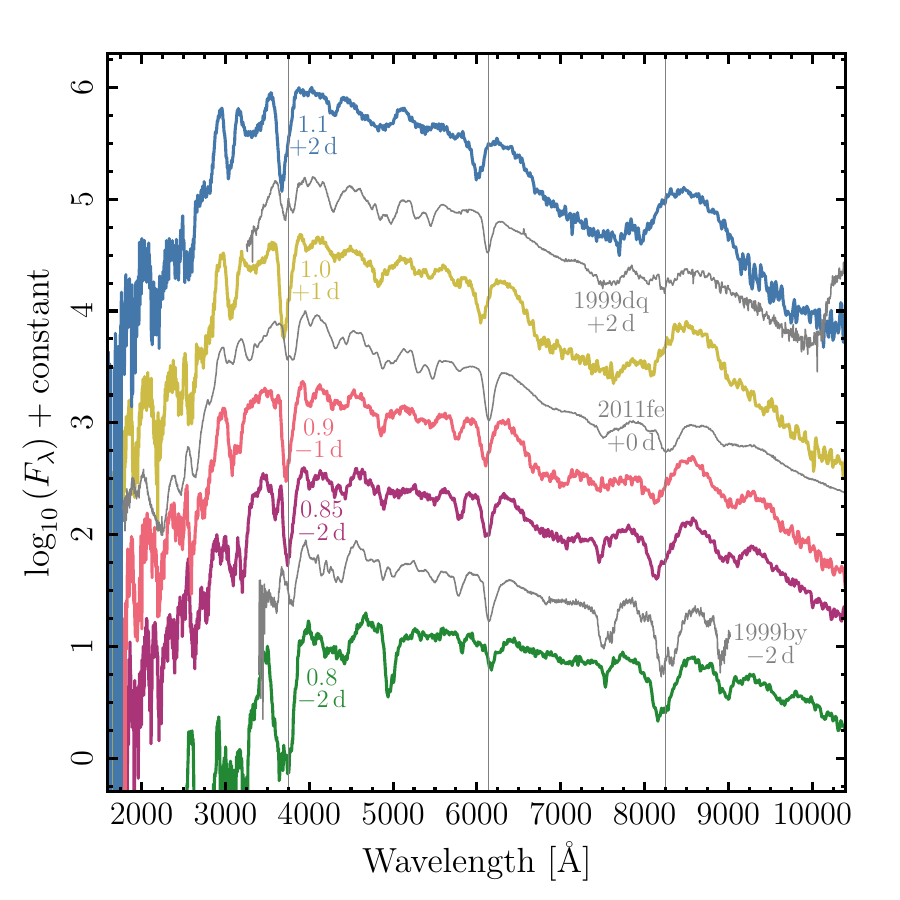}
\caption{Observed (grey) and synthetic (colours; 50/50 C/O fraction assumed)  spectra near B-band maximum light. Exploding WD masses are indicated along with time in days before or after maximum. The Si~\textsc{ii} 6355 and Ca~\textsc{ii} H\&K and near-infrared triplet regions are indicated by vertical lines. All spectra are 
offset on the y-axis by arbitrary constants. Image reprinted with permission from \citet{shen2018a}, copyright by AAS.}  
\label{fig:shen2018}      
\end{figure*}

\citet{townsley2019a} studied double-detonations in \msub\ mass white dwarfs (2D: 1 \msun \, WD with a 0.02 \msun \, helium shell) confirming that these explosions can produce normal SNe~Ia spectra, with higher ejecta velocities being found in the simulations where the supernova is analyzed at higher latitudes. It was noted however that the lightcurve decline rate in this particular study was somewhat lower than those of observed SNe~Ia like SN2011fe. 
Computing detonations in \msub\ WDs with an extensive reaction network, \citet{shen2018a} found that exploding WDs with mass ${\sim} 1$ \msun\ produced nucleosynthetic observables in very good agreement with observations of normal SNe~Ia (see Fig.~\ref{fig:shen2018}), even following the Phillip's relation. \citet{shen2021a} found WDs in the mass range ${\sim}0.9-1.1$ \msun\ exhibit features very similar to normal SNe~Ia, while less massive WDs ${\sim}0.85$ \msun\ fared rather well in explaining the sub-luminous events (see their Figs.~8 and 20). 

\citet{boos2021a} performed a parameter study of 2D full star detonations using {\sc FLASH} and a 55-species reaction network \citep{townsley2019a}. In their study they included both the preferred thin-shell models as well as thick-shell models, which are plausibly good representatives of other exotic thermonuclear transients \citep{polin2019a,de2019a}. A notable point of the \citet{boos2021a} study is that the secondary detonation did not always occur at a consistent location relative to the WD centre, with the thickest shell (0.1 \msun\ helium shell model on a 1.0 \msun\ WD) resulting in a detonation most off-set from the symmetry axis, i.e. closer to the core-shell boundary. The different models exhibited a range in nucleosynthetic yields ranging a factor of ${\sim}2$ in intermediate mass elements and a factor of ${\sim}7$ in $^{56}$Ni (see their Table~2).  

\begin{figure*}[ht]
\centering
    \includegraphics[width=\textwidth]{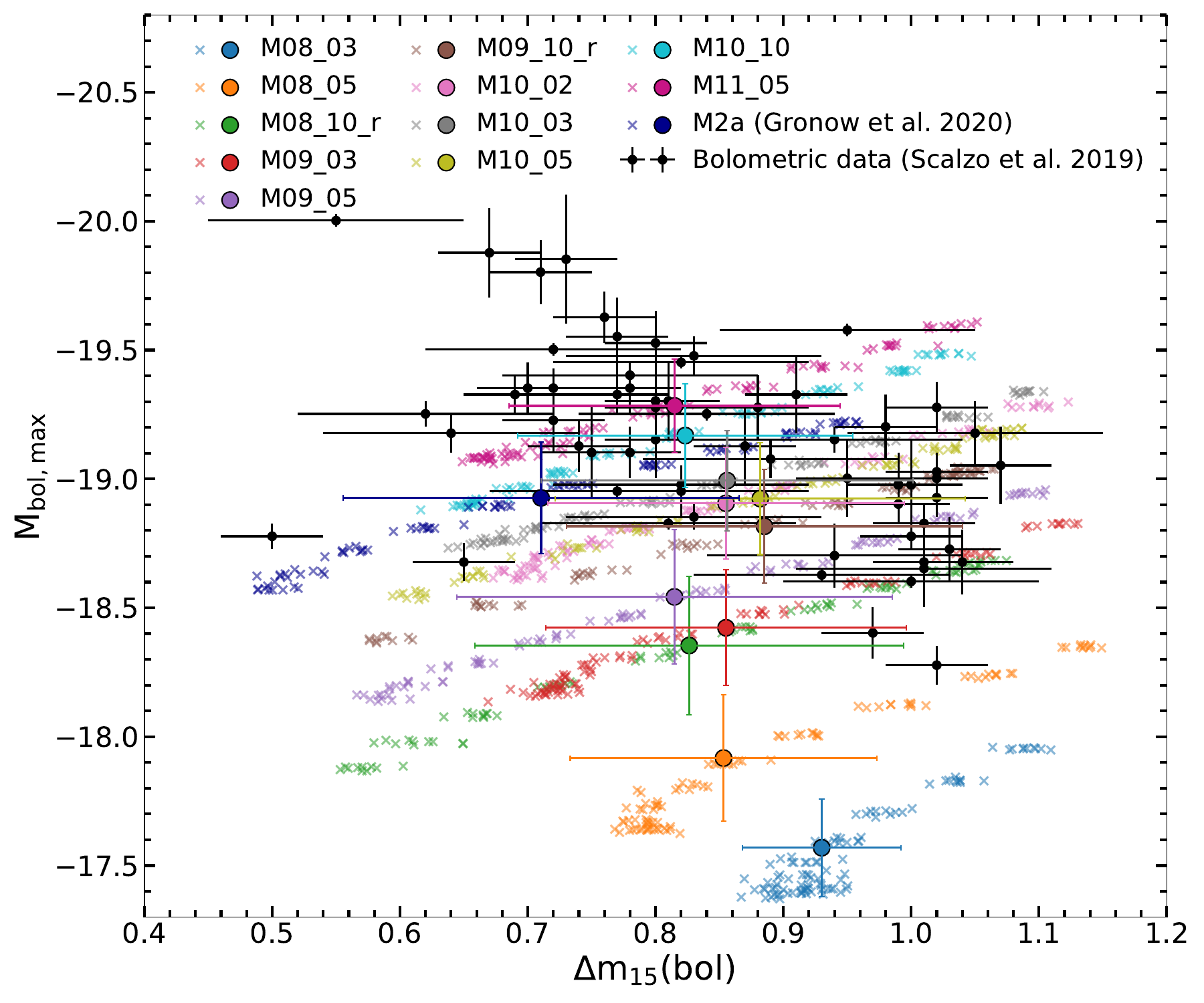}
\caption{Coloured dots represent angle-averaged (over 100 viewing angles) $\Delta$m$_{15_{\rm {bol}}}$ vs. peak bolometric magnitude for various double-detonation explosion models. In the legend, M11\_05 means the white dwarf `core' mass was 1.1 \msun\ while the helium shell had a mass of 0.05 \msun. The models with more massive white dwarfs and less massive shells more readily cover the range of observed SNe~Ia in this parameter space. The sample of nearby SNe~Ia from \citet{scalzo2019a} are represented by black cross symbols. Image reprinted with permission from \citet{gronow2021a}, copyright by ESO.}  
\label{fig:gronow2021}      
\end{figure*}

While 3D hydrodynamical studies of double-detonations are now feasible at high-resolution, much of the parameter space remains to be explored. One emerging trend appears to be that at least some \msub\ explosion models (and their associated radiative transfer calculations and predictions of nucleosynthetic structure) are promising for explaining some SNe~Ia of the more peculiar sub-classes \citep{collins2022a}. 
 On the other hand, in a parameter study of 3D double-detonations of varying shell (0.02--0.1 \msun) and core mass (0.8--1.1 \msun), \citet{gronow2021a} found that bolometric properties of simulated double-detonations were a fairly good match to faint and normal Type Ia supernovae (see Fig.~\ref{fig:gronow2021}). 

Finally, while canonical models assume a CO WD as the exploding star, other situations have been explored for accretors of different composition. White dwarfs with significant amounts of helium are generally not massive enough (thus do not harbour the right density) to readily synthesize  elements near the iron peak \citep{hashimoto1986a}. On the other hand, \msub\ oxygen-neon white dwarfs possess sufficient densities, and could plausibly account for some small fraction of events \citep{marquardt2015a}. However, ONe WDs are likely harder to ignite in the first place \citep[see][for an overview of challenges and various works associated with achieving core detonations in double-detonation simulations]{shen2014a}. Other more exotic means of producing double-detonation SNe~Ia have also been explored, for example tidally-induced double-detonations in white dwarfs that pass close to intermediate-mass black holes and become tidally disrupted \citep{tanikawa2018b}. One can be hopeful that in the near-future, we will be able to learn more about these types of transient events with dedicated deep surveys such as Rubin \citep{ivezic2019a}, in particular if such transients can be optimized for multi-messenger follow-up over the electromagnetic (and in some cases gravitational wave) spectrum. 

\subsection{Dynamically-driven}
\label{sec:dyn}

When the orbital separation of two white dwarfs becomes too small for both stars to remain within their Roche lobes, the larger (less massive) white dwarf will fill its Roche lobe and matter will be transferred away from it. In the ideal (easily to treat numerically) case, matter flows from the donor through the inner Lagrangian point between the stars toward the accretor, which then accepts the new material. However, in the case of dynamically-driven mass transfer, the transfer of material is too fast: the donor is unable to stay within its Roche lobe and structurally re-adjust itself in a short-enough timescale, and the accretor is unable to accept this influx of material. 
When the mass-losing star (donor) has a relatively distinct core-envelope structure, its envelope engulfs both stars in a common envelope \citep[see][for a recent review on numerical techniques]{roepke2023a}. However, in the case of two stellar cores with no envelope(s), the stars merge. The precise timescale on which the merger occurs depends on the properties of the system, as there are many competing processes to consider \citep{piersanti2003a,shen2012a}. These merging white dwarfs are the systems we discuss next in the context of SN~Ia progenitors. 
 
\subsubsection{Mergers of white dwarfs} 
\label{sec:wdm}

\begin{figure*}[ht]
\centering
\includegraphics[scale=0.65]{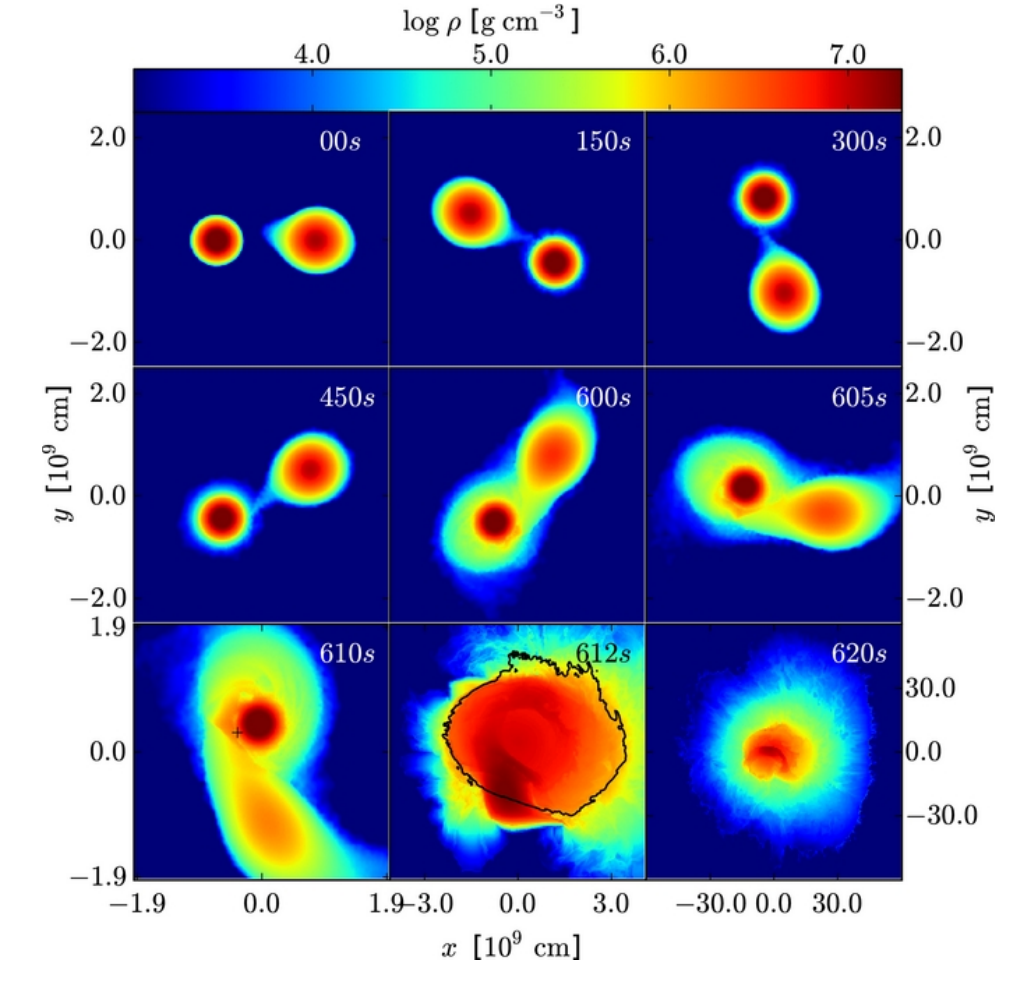}
\caption{A white-dwarf merger simulated with {\sc GADGET} resulting in a carbon-ignited detonation (marked with black cross) of the primary white dwarf; the {\sc LEAFS} code is used to model the detonation flames \citep{reinecke1999b}. Mass transfer proceeds stably initially but after ${\sim}10$ minutes, mass transfer becomes dynamically unstable. The masses of the primary and secondary CO WDs are $1.1$ and $0.9$ \msun, respectively, and the primary remains mostly fully intact throughout the interaction, thus retaining much of its initial density profile, until after the detonation. The black line in the bottom row (middle panel) shows the detonation front, which engulfs both stars, leaving no surviving companion in this case. Image reproduced with permission from \citet{pakmor2012a}, copyright by AAS.}
\label{fig:merger}      
\end{figure*}

Simulations of white dwarf mergers are very computationally demanding, and have only become possible relatively recently at a resolution where length scales relevant to the detonation physics are starting to be resolved \citep[e.g.][]{moran2024a}.  
Smoothed Particle Hydrodynamical simulations of white dwarf mergers initially indicated that conditions necessary for a carbon detonation were not likely to be met \citep{guerrero2004a}, but exceptions were indeed possible \citep{yoon2007a}. 
\citet{pakmor2010a} demonstrated with 3D hydrodynamical simulations coupled with radiative transfer modelling that two CO white dwarfs with roughly equal mass ($\sim 0.9$ \msun) would merge violently to create an explosion through direct carbon ignition near the core of the primary white dwarf without the requirement for a double-detonation.\footnote{In the carbon-ignited merger scenario, the merger occurs violently enough such that the primary WD suffers sufficient compressional heating in its central regions and undergoes a prompt detonation there, completely unbinding the star.} In addition, synthetic spectra and lightcurves matched those of 1991bg-like (sub-luminous) SNe extremely well. The \citet{pakmor2010a} study also concluded that such mergers could occur in young and old galaxies alike, thus discounting the perception that such mergers are rare, `niche' events. Additional simulations of carbon-ignited WD mergers soon followed \citep{pakmor2011a,pakmor2012a,sato2015a}, demonstrating that some of these explosions could plausibly reach peak luminosities comparable to normal SNe~Ia (see Fig.~\ref{fig:merger} for snapshots of the merger process). While successful SN~Ia explosions are mostly expected to only occur in double white dwarf systems above a critical mass ratio \citep{sato2016a} and only for systems above a critical mass threshold for the primary star \citep[but see][]{vankerkwijk2010a}, some of the first detailed WD merger simulations \citep{guillochon2010a} demonstrated that the primary white dwarf density, and thus primary WD mass, plays the dominant role in determining the amount of iron-group elements that are synthesized in the explosion. In other words: {\em The primary white dwarf mass offered an elegant explanation for the observed diversity in SN~Ia peak luminosity}, with more massive primaries giving rise to a higher production of radioactive nickel, resulting in more luminous events at peak \citep{ruiter2013a,shen2017a}. 

It was soon realized that explosion of the primary white dwarf is much more readily achieved if the binary system contained some (even small) amount of helium, for example on the white dwarf surface. How would the helium have gotten there? This could be achieved either through rapid mass transfer from a helium or hybrid white dwarf \citep[see][]{guillochon2010a}, or could be a thin layer of helium that is left over from previous stellar evolution, pre-existing well before the final binary interaction phase  \citep{pakmor2013a}. Such `helium-ignited' mergers then quickly became the poster child as the favoured sub-Chandrasekhar mass progenitor channel of SNe~Ia. Over the last dozen years, simulations of white dwarf mergers, and studies investigating post-merger configurations in the context of SNe~Ia, have been active areas of research by different groups employing a variety of numerical techniques: \citep{schwab2012a,ji2013a,dan2014a,moll2014a,dan2015a,raskin2014a,kashyap2015a,zhu2015a,kashyap2018a,pakmor2021a,roy2022a,pakmor2022a,zenati2023a,burmester2023a}. Despite the increasing number of works in this area, much of the parameter space of 3-dimensional white dwarf mergers still remains largely unexplored. One challenge to the field is that the learning curve associated with understanding how to use and additionally correctly interpret output from a 3D hydrodynamical codes is quite steep and time-consuming. 
We note that white dwarf mergers nonetheless are increasingly an extremely promising progenitor scenario, potentially for normal SNe~Ia as well as other sub-classes, and we strongly encourage further numerical studies in this area. Such studies will be useful in the larger context of white dwarf merger research (not limited to SNe~Ia progenitors) with prospective work supported by future space-based gravitational wave observatory missions that are projected to come online starting in the 2030s. 

\subsection{WD+WD collisions} 
\label{sec:coll}

Distinct from white dwarf mergers where gravitational wave radiation emission is what drives the stars together -- white dwarf collisions refer to white dwarfs crossing each other's path before total orbital decay. Such coalescing of white dwarfs that may occur in dense stellar environments, such as in the cores of globular clusters. In a white dwarf collision (similar to the case of mergers), the combined stellar mass need not be above the Chandrasekhar mass limit: white dwarf densities can rise significantly due to shock compression, in particular if the collision occurs head-on \citep[see][who performed a parameter study in WD mass and composition]{rosswog2009a}. The range of impact conditions results in a range of $^{56}$Ni that comfortably encompasses the expected range of normal SNe~Ia, though \citet{raskin2009a}, who conducted 3D smoothed-particle hydrodynamical simulations, found a more modest (${\sim} 0.4$ \msun; i.e. sub-luminous) amount of nickel produced for their explored (thought to be most probable) case of two $0.6$ \msun\ white dwarfs. 

Using 2D hydrodynamical simulations, \citet{kushnir2013a} found that collisions of white dwarfs with a range of masses and impact parameters bracket the observed properties of SNe~Ia, and argue that this channel could be relatively common if triple star systems are considered \citep[see also][]{iben1999a,thompson2011a,hamers2022a,rajamuthukumar2023a}. 
However, no population synthesis nor dynamical studies were performed in that work to confirm this. A population study is needed to properly quantify the probability of such systems being successfully produced in nature; stars increase dramatically in size as they evolve, and so avoiding a merger prior to the double white dwarf phase could be a major hurdle for the triple scenario. 
\citet{piro2014a} found that 
colliding white dwarf masses would need to be relatively large to produce sufficient amounts of $^{56}$Ni in an explosion. Thus, we should only expect normal SNe~Ia via collisions to be found among young stellar populations (i.e. with short delay times), which does not fit the general trend of SNe~Ia, with normal events occurring among a large range of stellar environments and ages. 

\citet{toonen2018a} explored a range of models using triple star evolution coupled with dynamical secular evolution and found the rate of WD+WD collisions plausibly leading to SNe~Ia is on the order of 0.1\%. A recent parameter study by \citet{rajamuthukumar2023a} that incorporated stellar evolution and dynamical interactions of triple star systems found that WD+WD collisions could account for 0.4--4\% of SNe~Ia.

\section{Final remarks}

The current state of the field is that white dwarf mergers leading to sub-Chandrasekhar mass explosions are an extremely promising channel that seem to be able to account for a large fraction, if not the majority, of Type Ia supernovae. In terms of various model predictions, double white dwarf mergers fare extremely well in terms of delay time distribution and rates (see Sect.~\ref{sec:Rates}), and perhaps more importantly, the expected range in exploding white dwarf mass provides a natural, simple explanation for the observed diversity among SN~Ia peak luminosity (Sect.~\ref{sec:wdm}). Whether the companion explodes too, or not, is not clear. However, some modern 3D hydrodynamical simulations indicate that the predicted differences between the two cases would not be immediately apparent, and any differences would not be extreme \citep{pakmor2022a}. The fact that deeper searches for potential hyper-velocity `fly away' white dwarf companions did not recover a large amount of candidates might give credence to the idea that, in the case of WD mergers leading to SNe~Ia, both WDs are likely destroyed more often than not. However, it could be that a large fraction of such high-velocity ex-donors are still eluding detection \citep{el-badry2023a}. 
On the other hand, taking into account constraints set by stellar abundance data from the solar neighbourhood, it is likely that at least some SNe~Ia must have had to originate in massive white dwarfs around the Chandrasekhar mass to have produced sufficient amounts of IGEs, in particular Mn (Sect.~\ref{sec:chem}), which is more supportive of a single-degenerate origin.
Taken together, the over-arching theme has been that sub-Chandrasekhar mass SNe~Ia, in particular coming from white dwarf mergers, are slowly emerging as the most-favoured progenitor scenario, but they cannot be the only progenitor. 

We are entering a phase in which the rate of the discovery of new transients will continue to increase even more drastically in the coming years. Rubin will detect about 10 million transients per night; an overflow of data unlike any we have encountered before. Since it is not feasible for humans to efficiently sift through this rapid data-deluge in real time using traditional methods -- 
separating the useful data into distinct categories (asteroids/fireballs, exoplanets, core-collapse and thermonuclear supernovae, gamma ray bursts etc.) -- several stream alert ‘brokers’ have been in development by a variety of teams. The brokers\footnote{\url{https://www.lsst.org/scientists/alert-brokers}}, many of which are actively receiving and processing alerts from the Zwicky Transient Facility 
\citep{bellm2019a,patterson2019a}, each have their own unique set of priorities, objectives, and methods for identifying and rapidly prioritizing diverse transient data for the scientific community, who can then initiate further, detailed (e.g. spectral) follow-up. Among the Rubin brokers, Fink \citep{moeller2021a} has the capability of SN~Ia classification even several days before peak magnitude, using a deep learning framework based on SuperNNova classifier scores \citep{moeller2020a,leoni2022a}. 

Other than refining rate estimates for various SN sub-classes, gathering numbers and statistics of SN explosions as a function of host environment (metallicity, galaxy Hubble type) will be useful particularly for high-redshift supernovae where, currently, it is not always possible to resolve a host. The deep (over 10 year) Legacy Survey of Space and Time that Rubin will conduct will offer an unprecedented opportunity to drill the environments of high-redshift supernovae. Combined with data from the Euclid mission, it will be possible to increase the sample of events even further \citep{bailey2023a}. Going a few years into the future, Nancy Grace Roman Telescope NIR data will play an important role in detecting SNe out to higher redshifts, which will be important for future studies in precision cosmology \citep[e.g.][]{dettman2021a}.  

One question to ask is: how will increasing the number of SN lightcurves from deep surveys enable us to learn more about their physics? We are already at a stage where the acquisition of observational data far exceeds (and will continue to exceed) the development of mature models that describe the very phenomena we see through their light. Recent, nearby SNe (like SN~2011fe and SN~2014J) have enabled astronomers to set strict likelihood limits on progenitor scenarios for these individual events. Likewise, historic, nearby young supernova remnants such as Kepler, Tycho, and SNRs in the LMC have enabled astronomers to put some constraints on the nature of the explosion and the progenitor system \citep{decourchelle2017a,vink2017a,seitenzahl2019a}.

While nearby supernovae are rare, they offer an unprecedented opportunity to probe explosion physics which is not possible with the majority of events. 
High-quality SN~Ia spectra obtained during the nebular phase can give valuable and constraining information on SN physics, but as mentioned already (Sect.~\ref{sec:specandLC})  one must be careful when interpreting the observations to draw conclusions about progenitor structure. This is because while no model is flawless,
some simulations may not incorporate all the relevant physics required to fully capture intrinsic 3D behaviour \citep{shingles2020a,pakmor2024a}. Investment in 3D spectral modelling of the nebular phase, ideally arising from different explosion conditions, could be a promising approach toward making stronger connections between observed SN~Ia sub-classes and progenitor origin.


\begin{acknowledgements}
AJR acknowledges financial support from the Australian Research Council under award number FT170100243. This research was supported in part by the National Science Foundation grant PHY-1748958 to the Kavli Institute for Theoretical Physics (KITP). AJR is grateful for helpful discussions with A. Cikota, D. Hamacher, A. Karakas, A. Möller, and R. Pakmor, and thanks the anonymous referees for their comments and suggestions that helped to improve this manuscript.  
AJR also acknowledges her late PhD supervisor, Chris Belczynski, who helped to foster her interest in studying SN~Ia progenitors and actively encouraged her to participate in the first SN~Ia meeting of her career in Arcetri, Italy, in 2008. Chris was passionate about life and science, and embraced challenges with great enthusiasm and tenacity.  

\end{acknowledgements}

{\noindent \small
\textbf{Conflict of interest} The authors declare no competing interests.
}

%
%

\phantomsection
\addcontentsline{toc}{section}{References}
\bibliographystyle{spbasic-FS-etal}      
\bibliography{ashbibz2}

%
%

\end{document}